\documentclass[12pt]{article}
\pdfoutput=1
\usepackage{graphicx}
\usepackage{mathrsfs}
\usepackage{amssymb}
\usepackage{amsmath}
\usepackage{graphicx}
\usepackage{amsmath,amsfonts}
\usepackage{algorithm}
\usepackage{algorithmic}
\usepackage[round]{natbib}
\usepackage{amscd}
\usepackage[mathscr]{eucal}
\usepackage{lineno}
\usepackage{extarrows}
\usepackage{color}
\usepackage{nameref}
\usepackage{pdflscape}

\usepackage{pdfpages}

\usepackage{tikz}

\numberwithin{thm}{section}

\newcommand{\be}{\begin{equation}} \newcommand{\ee}{\end{equation}}
\newcommand{\bd}{\begin{displaymath}} \newcommand{\ed}{\end{displaymath}}
\newcommand{\ba}{\begin{align}} \newcommand{\ea}{\end{align}}
\newcommand{\baa}{\begin{align*}} \newcommand{\eaa}{\end{align*}}
\newcommand{\ben}{\begin{enumerate}} \newcommand{\een}{\end{enumerate}}
\newcommand{\bi}{\begin{itemize}} \newcommand{\ei}{\end{itemize}}

\newcommand{\ud}{\mathrm{d}}
\newcommand{\E}[1]{\operatorname{E}\left[ #1 \right]}

\newcommand{\var}[1]{\operatorname{Var}\left[ #1 \right]}
\newcommand{\cov}[2]{\operatorname{Cov}\left[ #1,#2 \right]}

\usepackage[normalem]{ulem}

\begin{document}


\title{The Ornstein--Uhlenbeck process with migration: evolution with interactions}

\author{{\sc Krzysztof Bartoszek}, {\sc Sylvain Gl\'emin}, \\ {\sc Ingemar Kaj}, and
{\sc Martin Lascoux}} 

\maketitle

\begin{abstract}
  
  The Ornstein--Uhlenbeck (OU) process plays a major role in the
  analysis of the evolution of phenotypic traits along
  phylogenies. The standard OU process includes drift and stabilizing
  selection and assumes that species evolve independently. However,
  especially in plants, there is ample evidence of hybridization and
  introgression during evolution.  In this work we present a
  statistical approach with analytical solutions that allows for the
  inclusion of adaptation and migration in a common phylogenetic
  framework.  We furthermore present a detailed simulation study that
  clearly indicates the adverse effects of ignoring migration.
  Similarity between species due to migration could be misinterpreted as
  very strong convergent evolution without proper correction for these
  additional dependencies.  Our model can also be useful for studying
  local adaptation among populations within the same species.
  Finally, we show that our model can be interpreted in terms of
  ecological interactions between species, 
providing a general framework
  for the evolution of traits between ``interacting'' species or
  populations.
\end{abstract}


\section{Introduction}

Comparing phenotypic traits or gene expression patterns across species
or populations is intrinsically difficult since species or populations
are inherently non--independent entities: species are related through
phylogenies and populations through gene genealogies
\citep{JFel1985}. This historical component of the variation within
and between species will also need to be accounted for if our aim is
to understand trait evolution and in particular test whether traits
evolved neutrally or under natural selection.

The first proposed model of continuous trait evolution was the
Brownian motion (BM) model \citep{AEdw1970,JFel1985}.  This is
essentially a neutral model of evolution with the trait
oscillating around the ancestral value with linearly increasing
variance over time. The lack of any stabilization in the process was
quickly noticed and hence the Ornstein --Uhlenbeck model was proposed
as an alternative \citep{JFel1988,THan1997}. Since the seminal papers
of Edwards and Felsenstein \citep{AEdw1970, JFel1985}, various forms
of the Ornstein-Uhlenbeck process have been used to assess the
relative importance of genetic drift and selection in phenotypic trait
evolution across species.  Initial versions incorporating only drift
and stabilizing selection, but recent forms of the OU process now
consider simultaneously genetic drift and both stabilizing and
directional selection \citep{JBeaetal2012, Held2014}.  Until recently,
however, species or populations were assumed to evolve independently,
neither interacting nor exchanging genetic material.  Since migration
between populations or hybridization between species will tend to
homogenize phenotypes directly by sharing the underlying genetic
information, the application of such models at the species level
\citep[i.e. between populations within species, see discussion
by][]{GStoSNeeJFel2011} could pose a serious problem. More generally,
hybridization between closely related species or divergence with gene
flow during species formation is likely more frequent than initially
thought \citep{RAbbetal2013}, leading to the same problem at the
phylogenetic scale where BM and OU models where classically applied.
Ecological interactions between species can also affect phenotype
evolution if the strength and/or direction of selection on the
phenotype of a focal species is affected by the phenotypes of other
species. For instance, competition can lead to character displacement
and diversification.  On the contrary, mimicry, as migration, will
tend to homogenize phenotypes. Two recent studies combined
phylogenetic structure and interactions during trait
evolution. \citet{SNuiLHar2015} described an interaction framework
combined with an OU evolution process. However the interactions
between the species occur only through ``the covariance among
population mean phenotypes for species $i$ and
$j$''. \citet{JDruJClaMManHMor2016} further related to this model but
also included species interaction in the ``deterministic part''
of the stochastic dynamics in their framework.  In our work we focus
on the migration problem and consider a selection---migration model
where different species or populations are allowed to exchange genes
thus affecting each others' primary optima \citep[in the spirit
of][]{THan1997}. We show that \citet{JDruJClaMManHMor2016}'s model is
a special subcase of ours, which can thus also be applied to the
species interaction problem.  We then used simulations to assess the
consequences of ignoring migration when testing for the presence of
selection.

\section{Phylogenetic models of selection with migration}
\subsection{Selection model}

The Ornstein-Uhlenbeck process is defined by the
following stochastic differential equation
(SDE) for the trait $X$,

\be\label{eqOU} \ud X(t) = -\alpha \left(X(t) - \theta \right) \ud t +
\sigma \ud W(t), 
\ee 
where $W(t)$ is the standard Wiener process. The
interpretation of the individual model parameters has been discussed
at length \citep[e.g.][]{MButAKin2004,THan1997,THanJPieSOrz2008}.
Shortly, $\theta$ is the optimal value for the trait, $\sigma$
determines the magnitude of genetic drift affecting the trait, and
$\alpha$ controls the speed of the approach to the optimum, i.e. the
strength of selection.  The neutral case in the sense of absence of
selection, $\alpha=0$, corresponds to the Brownian motion model.  The
expected value (mean) and variance of the process are 

\be
\begin{array}{rcl}
\E{X(t)} & = & \theta + e^{-\alpha t}\left(X_{0} - \theta\right) \to \theta \\
\var{X(t)} & = & \frac{\sigma^{2}}{2\alpha}\left(1-e^{-2\alpha t} \right) \to \frac{\sigma^{2}}{2\alpha}.
\end{array}
\ee 
This means that after a long time the trait will stabilize around
the optimal value $\theta$ with stationary oscillations of variance
$\sigma^{2}/(2\alpha)$.
A key parameter of this model is the half--life \citep{THan1997}, $t_{0.5} = \ln 2/\alpha$,
which quantifies how much time is needed to lose half of the ancestral effect.

\begin{figure}[!ht]
\begin{center}
\includegraphics[width=0.3\textwidth]{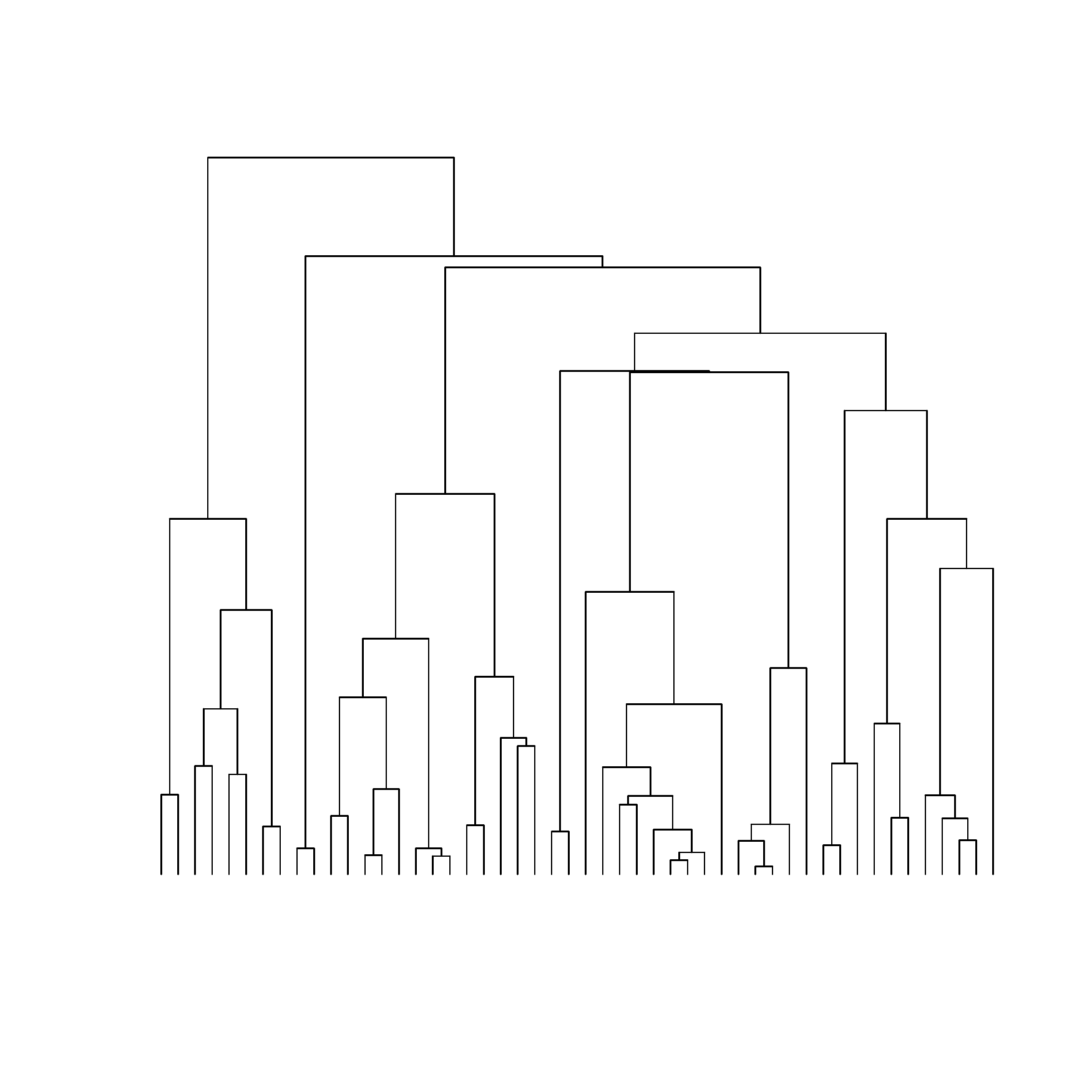}
\includegraphics[width=0.3\textwidth]{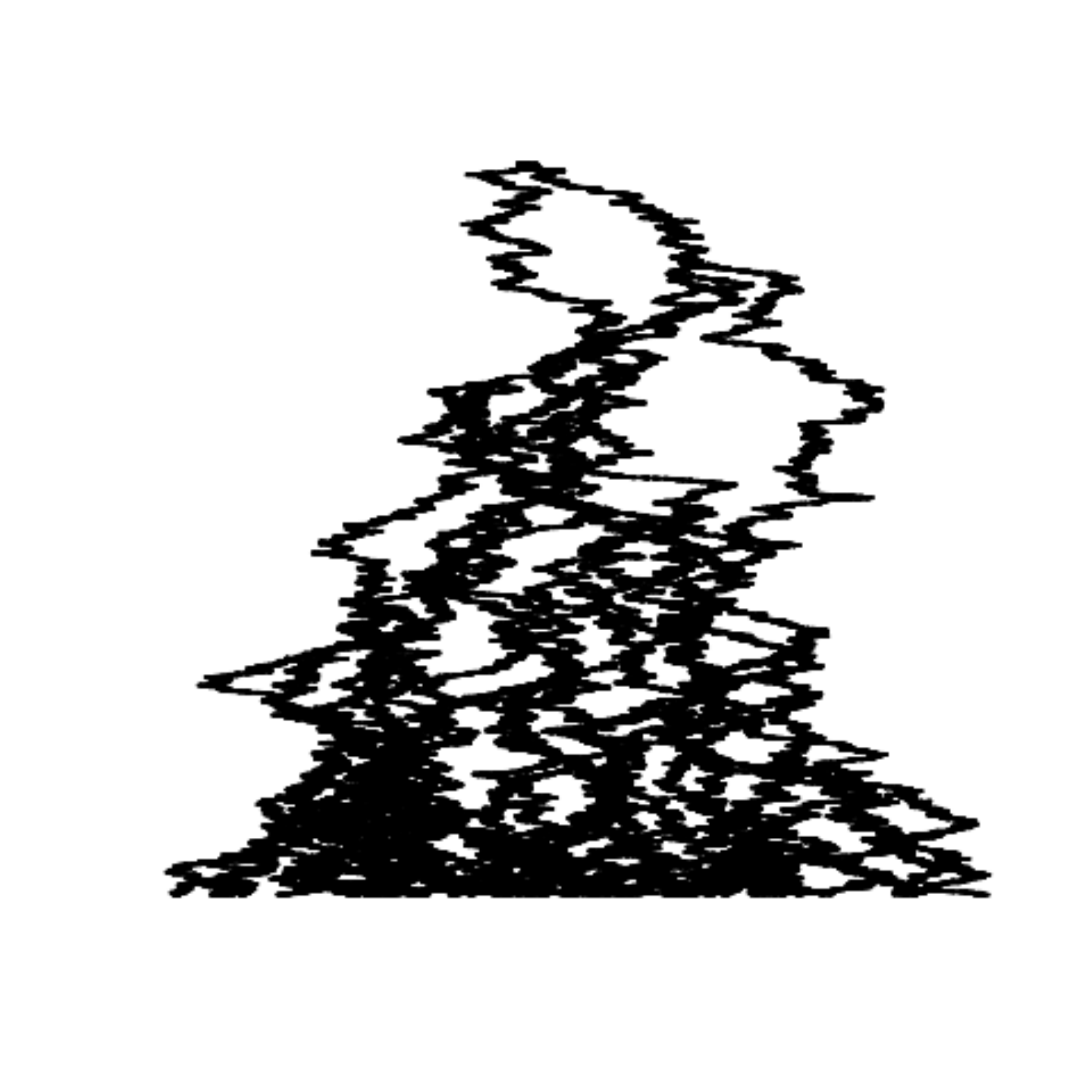}
\includegraphics[width=0.3\textwidth]{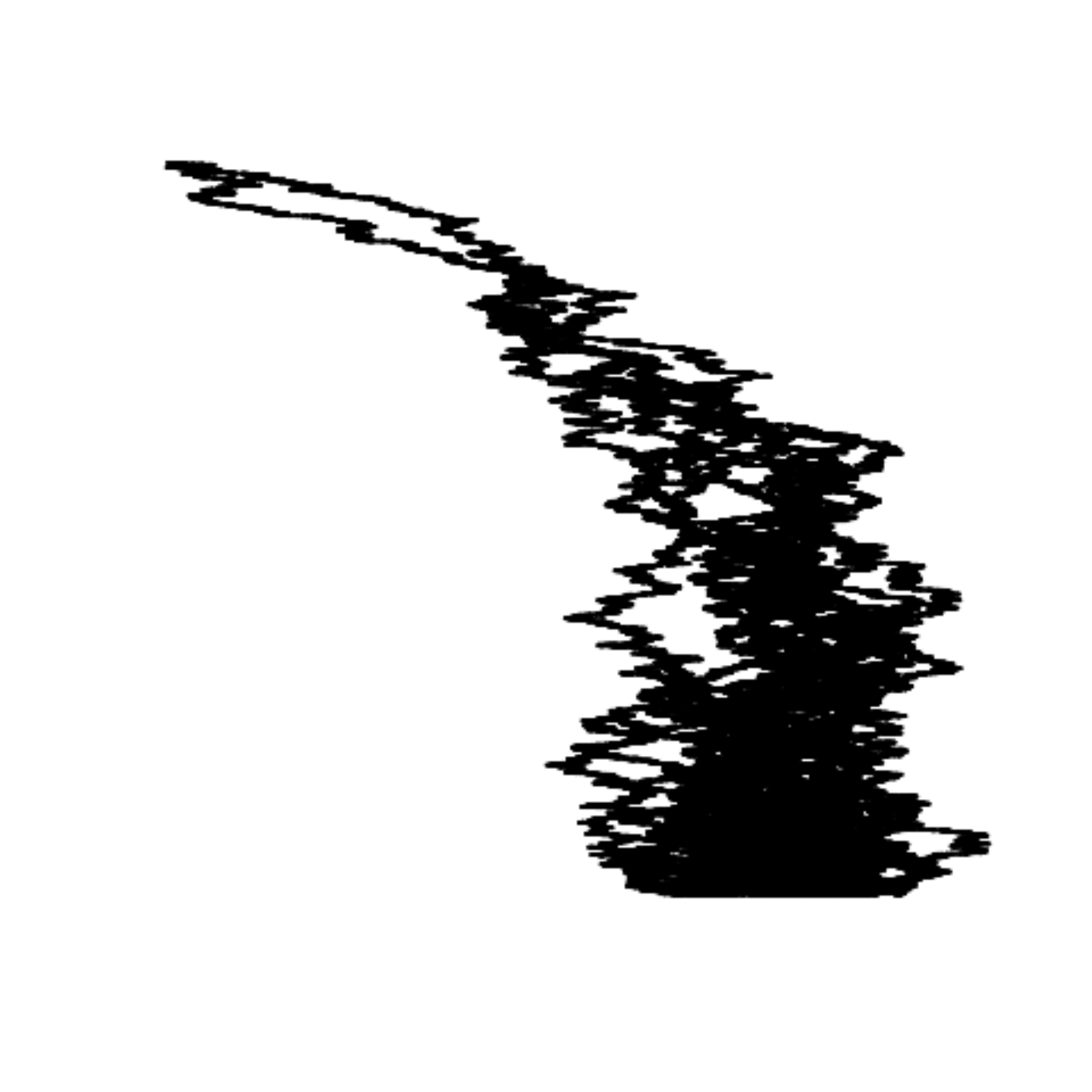}
\caption{Left: a simulated pure birth tree  on $50$ tip species
\citep[speciation rate $1$, TreeSim R package;][]{TreeSim1,TreeSim2,R}
, centre: Brownian motion simulated on
this tree , right: Ornstein--Uhlenbeck process simulated on this tree
\cite[BM: $X_{0}=0$, $\sigma^{2}=1$; OU: $X_{0}=0$, $\theta=5$, $\alpha=1.5$, $\sigma^{2}=1$; mvSLOUCH R package][]{KBaretal2012}.
\label{figBproc}}
\end{center}
\end{figure}

The trait follows the speciation pattern described by the
phylogeny. The trait changes along each branch according to the
diffusion process in \eqref{eqOU} and at speciation it splits into two
processes. Commonly these two daughter lineages evolve
independently. Nearly all current phylogenetic comparative methods
make this ``no exchange or interaction between species'' assumption.
However, one can easily argue that this is not realistic biologically:
species interact with each other due to e.g. ecological interactions,
gene flow or hybridization.

\subsection{Including migration}

Assume we have $n$ species or populations under selective pressure
present in the time interval $[0,T]$ and that there is gene flow
between some of them.  For simplicity and parallelism with other OU
models, we will use the term species in the rest of the manuscript,
though gene exchange is more likely to occur between populations of
the same species. The mathematics of the model is the same whether
one studies evolving species or populations, whereas the parameter
interpretations and model implications can be different.

If the state of the process at time $t$ is a
vector $\vec{X}_{t}$ consisting of the $n$ current trait-values, then the
state $\vec{X}_{t+\Delta t}$ after an additional time $\Delta t$ is

$$
\vec{X}_{t+\Delta t} =  \vec{X}_{t} - \alpha\mathbf{I}(\vec{X}_{t}-\vec{\theta})
\Delta t  +  \mathbf{M}\vec{X}_{t} \Delta t.
$$
The matrix $\mathbf{M}$ describes the migratory gene flow between the $n$
species while the other parameters are the standard OU process
parameters.  Taking $\Delta \to 0$, we obtain the ordinary
differential equation (ODE)

$$
\ud\vec{X}_{t} = -\left(\alpha\mathbf{I} - \mathbf{M}
 \right)\vec{X}_{t} + \alpha\mathbf{I}\vec{\theta}.
$$
Equivalently, letting $\mathbf{H}$ denote the matrix 

$$
\mathbf{H}=\alpha\mathbf{I} - \mathbf{M},
$$
in which all interactions between the species are coded, we have 

\be\label{eqOUMode}
\begin{array}{rcl}
\ud\vec{X}_{t} & = & -\mathbf{H}\left(\vec{X}_{t} -\alpha
  \mathbf{H}^{-1}\vec{\theta}\right).
\end{array}
\ee
Combined with the random noise we obtain an OU with migration stochastic model

\be\label{eqOUMsde}
\begin{array}{rcl}
\ud \vec{X}(t) & = & -\mathbf{H}\left(\vec{X}(t) - \alpha
\mathbf{H}^{-1}\vec{\theta}\right) \ud t + \sigma_{x}\mathbf{I}\ud\vec{W}(t),\\ 
\end{array}
\ee where $\vec{W}$ is a vector of $n$ independent Wiener processes.
In Eq.\ \eqref{eqOUMsde} we may now recognize $\vec{X}(t)$ as a
multivariate, $n$--dimensional, OU process with ``deterministic part''
the matrix $\mathbf{H}$ and optimum value $\alpha
\mathbf{H}^{-1}\vec{\theta}$.  This means that one can obtain the
moments of the process analytically. The knowledge of the mean and
variance--covariance function then characterizes its normal law, as
the OU model is a normal process.

We represent the  eigen decomposition of the matrix $\mathbf{H}$ by

$$
\mathbf{H} = \alpha \mathbf{I} - \mathbf{M} \equiv \mathbf{P} \mathbf{\Lambda} \mathbf{P}^{-1}.
$$
Under the assumption that $\mathbf{H}$ is non--singular the mean vector and covariance matrix will be \citep[]{KBaretal2012} 

\be\label{eqmeanvarOUM}
\begin{array}{rcl}
\E{\vec{X}(t)} &= & \mathbf{P}e^{-\mathbf{\Lambda}t} \mathbf{P}^{-1}\vec{X}_{0} +
\alpha\mathbf{P}\mathbf{\Lambda}^{-1}\left(\mathbf{I} -e^{-\mathbf{\Lambda}t} \right) \mathbf{P}^{-1}  \vec{\theta} 
\\
\var{\vec{X}(t)} &= & 
\frac{\sigma_{x}^{2}}{2} 
\mathbf{P}\mathbf{\Lambda}^{-1}
\left(\mathbf{I} - e^{-2 \mathbf{\Lambda} t}   \right)
\mathbf{P}^{-1}.
\end{array}
\ee 
The remaining case, when $\mathbf{H}$ is singular, amounts to some
of its eigenvalues being $0$.  Assuming that eigenvalue $i$ is $0$, in
Eq. \eqref{eqmeanvarOUM} we need to consider the limit

$$
\begin{array}{rcl}
\Lambda_{i}^{-1}\left(1 -e^{-x \Lambda_{i}t} \right) &
\xlongrightarrow{\lambda_{i} \to 0} xt. 
\end{array}
$$
Then the mean and variance formulae of Eq. \eqref{eqmeanvarOUM} hold
with $xt$ at appropriate diagonal entries of the diagonal matrix
$\mathbf{\Lambda}^{-1}\left(\mathbf{I} - e^{-x \mathbf{\Lambda} t}
\right)$.  In fact we will discuss an important submodel where this is
the case, namely the model of \cite{JDruJClaMManHMor2016}.

The model setting laid out above is quite general. Each species tends
towards its own optimum (which in particular may be the same as the
optimum of all/some other species) with $\alpha$ controlling the speed
of the approach.  Along its route it exchanges migrants (hence shares
trait values) or more generally
interacts 
with all/some other species as controlled by $\mathbf{M}$, hence
$\mathbf{H}$. As a result the model is high--dimensional: with $n$
species the matrix $\mathbf{H}$ is parametrized by $n^{2}+1$
parameters. This number increases further if $\mathbf{M}$ changes over
the phylogeny.  Usually in a phylogenetic comparative method (PCM)
setting there are only $n$ measurements -- all from contemporary
species.  Therefore one needs to add further structure on $\mathbf{M}$
to obtain a feasible framework for analysis, and furthermore make the
parameters interpretable.

Let us write out in detail the migration component for species 
(remembering that in Eq.\ \eqref{eqOUMsde} $\Delta t \to 0$)

$$
\begin{array}{ll}
X_{t+\Delta t,i} &  =   
X_{t,i} + \mathrm{~selection~} + \mathrm{~noise~}\\
&\; - \left(-m_{i,i}X_{t,i}+
m_{i,1}X_{t,1}+\ldots+m_{i,i-1}X_{t,i-1}+m_{i,i+1} X_{t,i+1}+\ldots+m_{i,n} X_{t,n}
\right)\Delta t.
\end{array}
$$
Now let us assume that migration is conservative, i.e. that the number
of immigrants and emigrants are equal.
This translates to the sum of entries in each row of $\mathbf{M}$
equalling $0$, i.e.

$$
m_{i,i} = m_{i,1}+\ldots+m_{i,i-1}+m_{i,i+1} +\ldots+m_{i,n}.
$$
Let us assume furthermore that the exchange of information is
symmetric between all species, i.e.  $m_{i,i}\equiv m$ and $m_{i,j}
\equiv m/(n-1)$ for $i\neq j$. This implies that the elements of
$\mathbf{H}$ are given by 

$$
[\mathbf{H}]_{ij} = \left\{
\begin{array}{cc}
\alpha+m & i=j \\
\alpha-\frac{m}{n-1} & i \neq j
\end{array}
\right. .
$$
Such a matrix $\mathbf{H}$ is eigen decomposable -- it has the simple
eigenvalue $\alpha$ and an eigenvalue of multiplicity $n-1$
equalling $\alpha+(1+1/(n-1))m$. The eigenvectors matrix and its
inverse are 

$$
\mathbf{P} = 
\left[
\begin{array}{cccccc}
1 & -1 & -1 & -1 & \ldots  & -1 \\
1 & 1 & 0 & 0 & \ldots & 0 \\
1 & 0 & 1 & 0 & \ldots & 0 \\
\vdots & \vdots & \ldots & \ddots & \ldots & \vdots \\
1 & 0 & \ldots & \ldots & \ldots & 1
\end{array}
\right],
~~
\mathbf{P}^{-1} = 
\left[
\begin{array}{cccccc}
\frac{1}{n} & \frac{1}{n} & \frac{1}{n} & \frac{1}{n} & \ldots  & \frac{1}{n} \\
-\frac{1}{n} & \frac{n-1}{n} & -\frac{1}{n} & -\frac{1}{n} & \ldots & -\frac{1}{n} \\
-\frac{1}{n} & -\frac{1}{n} & \frac{n-1}{n} & -\frac{1}{n} & \ldots & -\frac{1}{n} \\
\vdots & \vdots & \ldots & \ddots & \ldots & \vdots \\
-\frac{1}{n} & -\frac{1}{n} & \ldots & \ldots & \ldots & \frac{n-1}{n}
\end{array}
\right].
$$
From these representations one easily obtains the moments of the
process, Eq. \eqref{eqmeanvarOUM}. 

The above assumptions on the process may seem restrictive. They,
however, perfectly describe the ``mean field'' corresponding to the
Wright island model in population genetics or to a situation where a
number of exchangeable species constantly interacts with each
other. 
The magnitude of migration is measured by $m$.
At the species level, this corresponds to a model of local adaptation as 
analyzed by \citet{AHenTDayETay2001}.
However, they only considered two populations.

\subsection{Including the phylogeny}

So far we have considered the evolution of traits without any
branching structure -- a fixed number of species evolve and interact
within a given time frame.  Ideally we would like to model 
the phylogenetic or populations history setup -- where the trait evolves
with time and species not only interact but also speciate (or
populations split) and new lineages, that also may interact, arise.

In order to trace the evolution of traits in the species tree and to
derive the law of the contemporary sample we introduce some additional
notation.  By $t_{k}$ we denote the time (the origin of the tree is
$t_{0}=0$) of the $k$th speciation event, i.e.\ at time $t_{k}$ the
number of species changes from $k$ to $k+1$.  Hence
$T_{k}:=t_{k}-t_{k-1}$ is the duration of time when there were $k$
lineages present. In particular if the tree has no root branch but
starts at the first bifurcation event we will have
$T_{1}=t_{0}=t_{1}=0$.  We assume that in each of these intervals all
parameters are constant.  A parameter with subscript $k$, e.g.
$\mathbf{H}_{k}$, indicates the value of the parameter at the time
when there are $k$ lineages present in the phylogeny.  After each
speciation event the two daughter lineages inherit
their ancestral trait value and the traits in both lineages begin to
evolve over time following the assumed OU dynamics with possible
interactions.  The subscripted trait vector $\vec{X}_{k}(t)$ will be
used to underline that at time $t$ there are $k$ lineages present.
Just after speciation the two daughter lineages are identical, hence
they have to share the same mean, variance and covariance with other
lineages. In principle this causes a momentary singularity of the
process, which however immediately disappears. To formalize the mean
and covariance of the phylogeny traits process at this instance, as
$\vec{X}_{k}(t_{k}-)$ changes dimension to $\vec{X}_{k+1}(t_{k}+)$, we
introduce a branching operator ${\curlyvee}$, which increases the size
of a vector by one element and the size of a square matrix by adding a
row and a column.  This operator copies
the appropriate entry of the traits' mean vector and the appropriate
row and column of the variance--covariance matrix when speciation
happens, resulting in a recursive (in terms of speciation events)
formula for the mean and variance 

$$
\begin{array}{rcl}
\E{\vec{X}_{k}(t_{k})} &= &
e^{-\mathbf{H}_{k} T_{k}} \E{\vec{X}_{k-1}(t_{k-1})}^{\curlyvee}
+ \left(\alpha\mathbf{H}_{k}^{-1} - \mathbf{P}_{k}\left(\alpha
    \mathbf{\Lambda}_{k}^{-1}e^{-\mathbf{\Lambda}_{k}T_{k}} \right)
  \mathbf{P}_{k}^{-1} \right) \vec{\theta}_{k} \nonumber
\\ 
&= &
e^{-\mathbf{H}_{k} T_{k}} \E{\vec{X}_{k-1}(t_{k-1})}^{\curlyvee}
+ \alpha\mathbf{P}_{k}\mathbf{\Lambda}_{k}^{-1}\left(\mathbf{I}_{k} -e^{-\mathbf{\Lambda}_{k}T_{k}} \right) \mathbf{P}_{k}^{-1}  \vec{\theta}_{k},
\label{eqEX}
\end{array}
$$

$$
\begin{array}{rcl}
\var{\vec{X}_{k}(t_{k}) } &=  &
\frac{\sigma_{x}^{2}}{2} 
\mathbf{H}_{k}^{-1}\left(\mathbf{I}_{k} - e^{-2 \mathbf{H}_{k} T_{k}}   \right)
+ e^{-\mathbf{H}_{k} T_{k}}
\var{\vec{X}_{k-1}(t_{k-1})}^{\curlyvee}
e^{-\mathbf{H}_{k}^{T} T_{k}}  \nonumber
\\
&=&\frac{\sigma_{x}^{2}}{2}  \mathbf{P}_{k}
\mathbf{\Lambda}_{k}^{-1}
\left(\mathbf{I}_{k} - e^{-2 \mathbf{\Lambda}_{k} T_{k}}   \right)
\mathbf{P}_{k}^{-1}
+
e^{-\mathbf{H}_{k} T_{k}}
\var{\vec{X}_{k-1}(t_{k-1})}^{\curlyvee}
e^{-\mathbf{H}_{k}^{T} T_{k}}.
\label{eqVarX}
\end{array}
$$
In principle the operator should be indexed by the phylogeny and
speciation number but we suppress this notation to avoid clutter.

\begin{figure}[!ht]
\begin{center}
\includegraphics[width=0.4\textwidth]{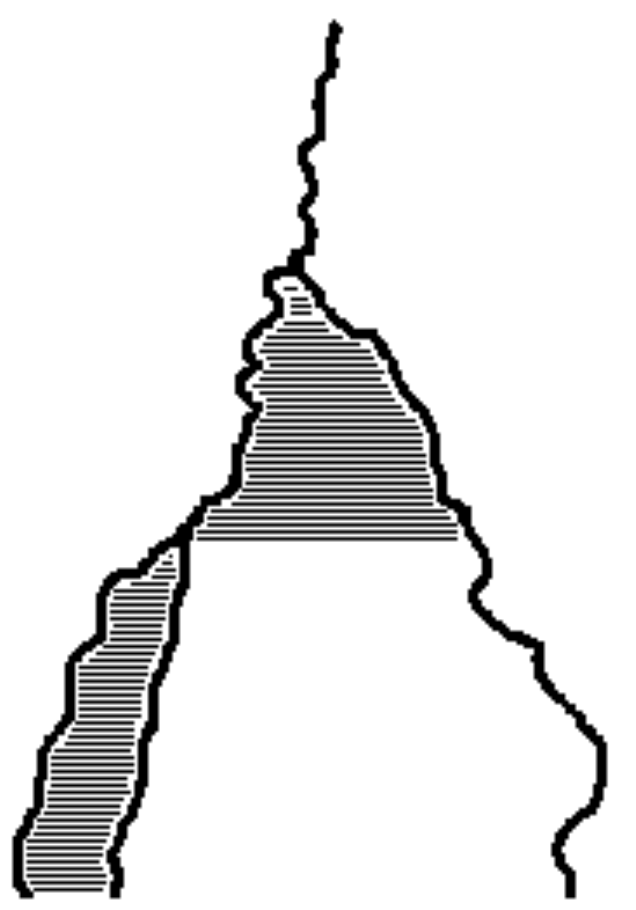}
\caption{An evolving trait on a phylogenetic tree with migrations indicated,  by horizontal lines.\label{figPhylM}}
\end{center}
\end{figure}

\subsection{Including measurement error and missing observations}

Our setting is targeted to maximum likelihood optimizations, which
allow for analyzing measurement errors and missing observations.

Trait measurements at the tips of the phylogeny are typically species'
means obtained as averages over many individuals. The resulting
intra--species variation adds a sample variance attached to each
observation, which is straightforward to include but could be highly
misleading to ignore \citep[e.g.][]{THanKBar2012}.
From the vector of observations,
$x_{i,1},\ldots,x_{i,n_{i}}$ for species $i$ we can estimate the
variance for the $i$--th species as

$$
\sigma_{i}^{2}=\frac{1}{n_{i}} \sum\limits_{k=1}^{n_{i}} \left(x_{i,k} - \bar{x}_{i} \right)^{2},
$$
where $\bar{x}_{i}$ is the average for the $i$--th species. These
variances are then added to the diagonal of the variance--covariance
matrix of the $n$ contemporary species, $\var{\vec{X}_{n}(t_{n}) }$,
given by Eq. \eqref{eqVarX}.  This is a common approach used in
phylogenetic comparative software that allows for measurement error
\citep[see e.g.][]{KBaretal2012}.

To correctly handle species with missing measurements, the rows of the
design matrix, expectation vector, Eq. \eqref{eqEX} and rows and
columns of the variance--covariance matrix, Eq. \eqref{eqVarX} that
correspond to measurements with missing values should be removed.  In
the case of each species being described by a single trait this is the
same as just removing the ``missing value'' species from the data set.
However, in the case of a multivariate analysis, when observations are
just missing from some traits, there is no need to remove a whole
species from the analysis. One just removes the missing components by
removing the appropriate rows and columns as described above
\citep{KBaretal2012}.

\section{Simulations --- effects of migration and selection}

To assess the effects of migration we simulated populations of
individuals with migration between them. We considered a fixed star phylogeny
of species with height equal to $5$ and $300$ tips. We do not
have a phylogenetically structured population but rather a population
descending from a common ancestor and all individuals are constantly
interacting.

We draw $n=300$ trait values from a normal distribution with mean
vector and variance--covariance matrix defined by Eqs. \eqref{eqEX}
and \eqref{eqVarX} for different parameter values. Time was always
assumed as $5$, the initial value $0$, $\sigma^{2}=1$. The optimum
$\theta$ was either constant at $0$ or $5$, or half the population had
$\theta=0$ or $\theta= -5$ while the other half had $\theta=5$.  The
adaptation parameter varied as $\alpha \in \{0.5,1,10\}$ and the
migration parameter $m \in \{0,0.1,0.5,1,10\}$.  Here, $m$ is an
arbitrary value on the trait scale.  For the special case $0\le m \le
1$ this is consistent with the common interpretation of $m$ as a
proportion of individuals coming from abroad. 

\begin{figure}[!ht]
\begin{center}
\includegraphics[width=0.3\textwidth]{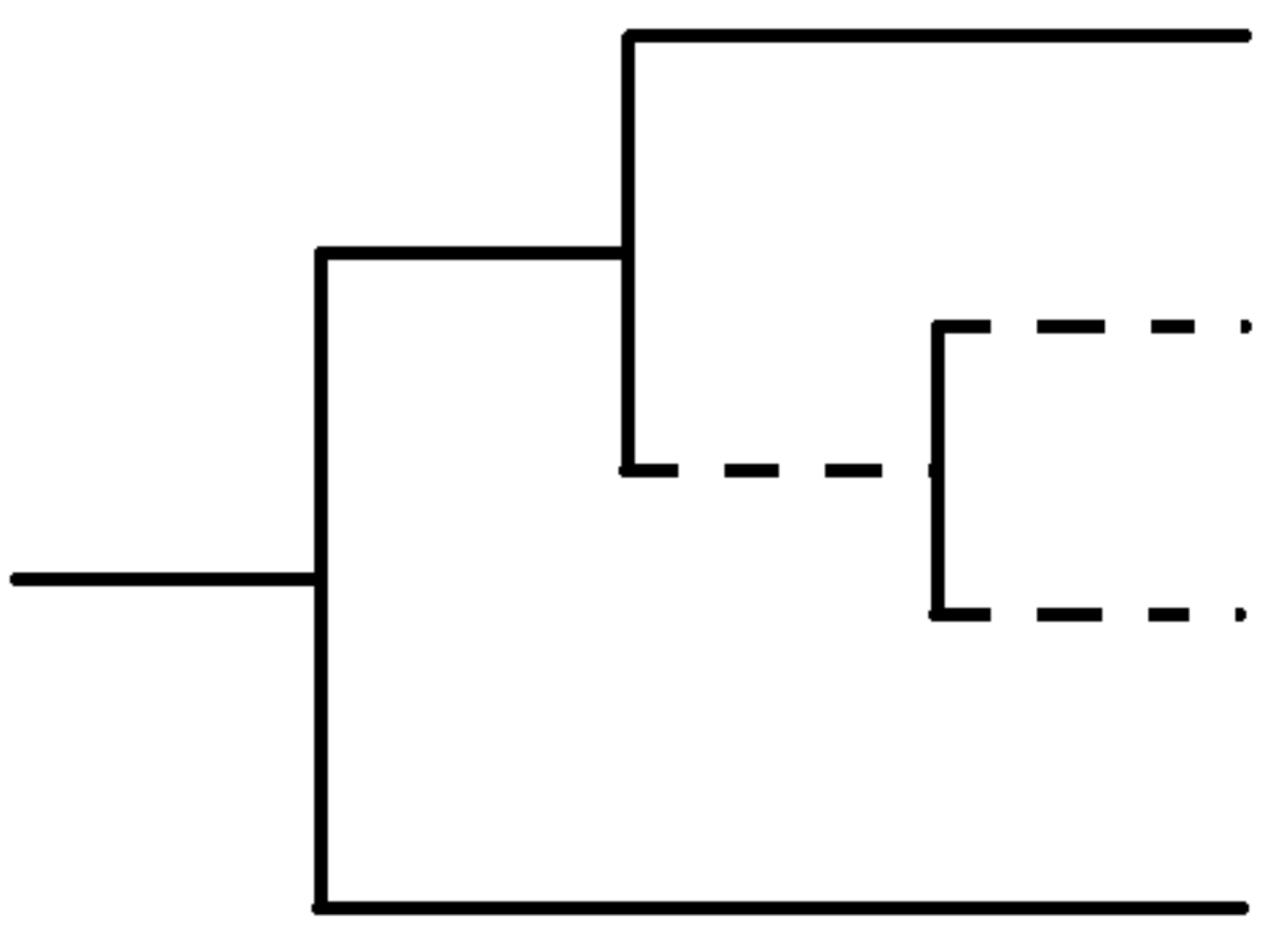}\\
\includegraphics[width=0.3\textwidth]{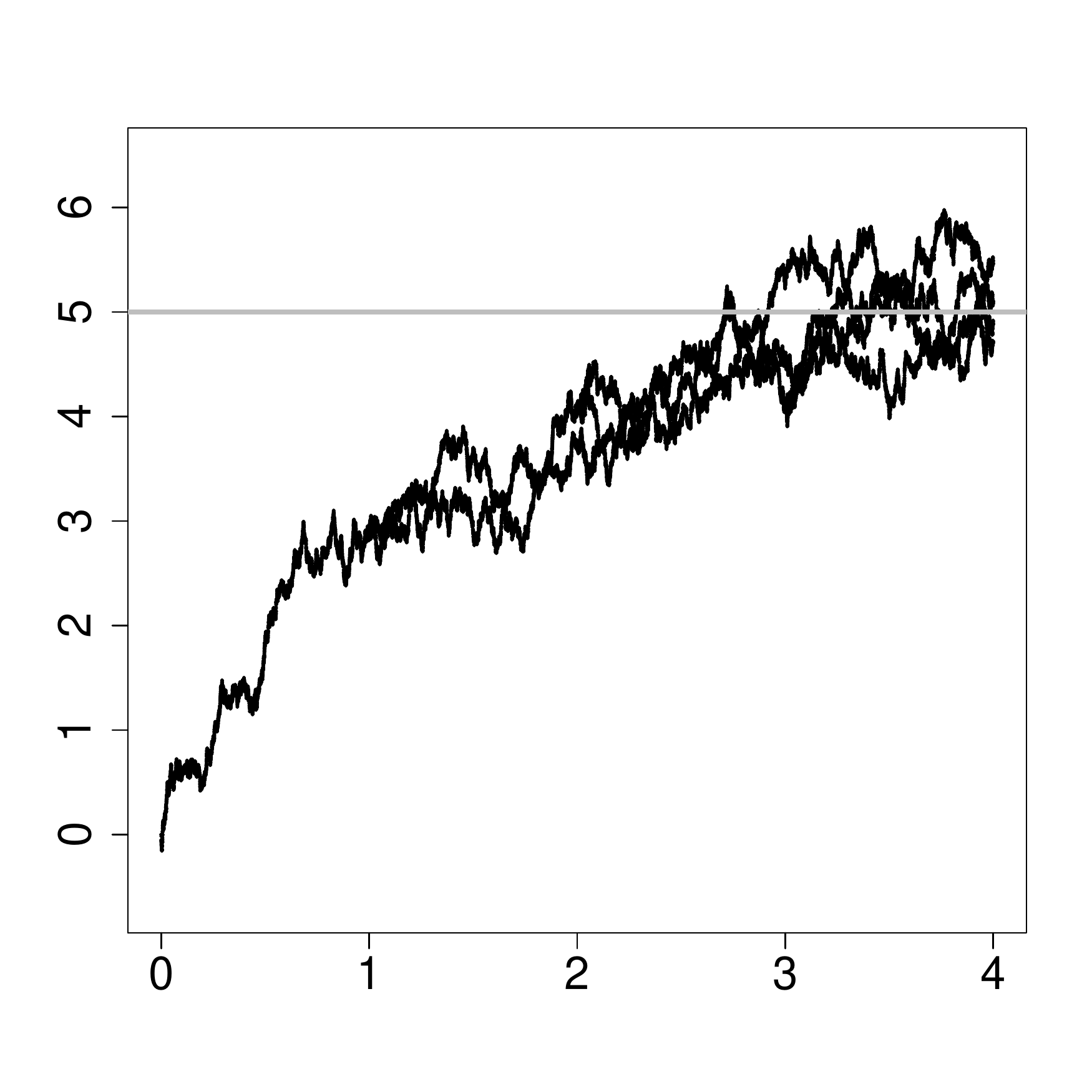}
\includegraphics[width=0.3\textwidth]{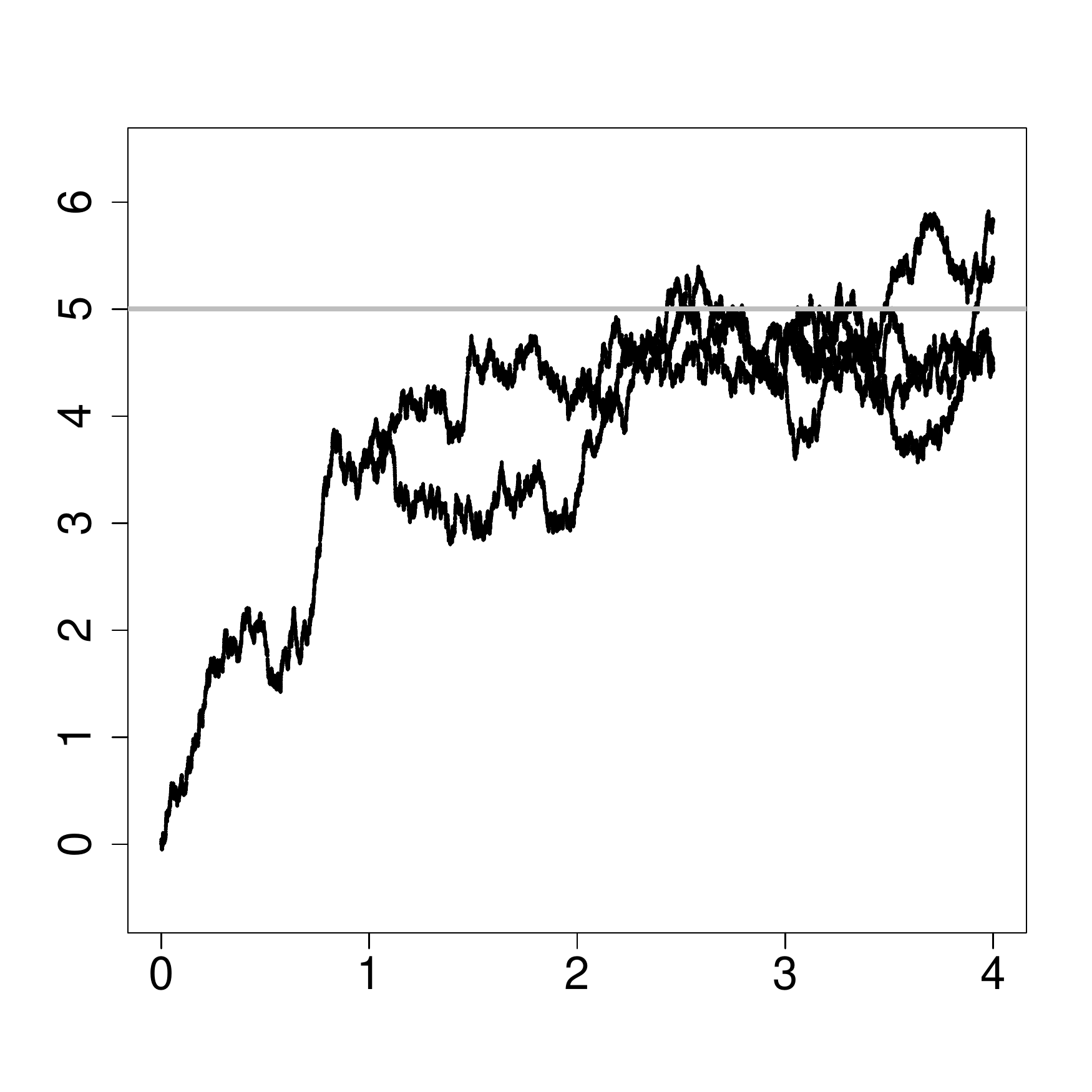}
\includegraphics[width=0.3\textwidth]{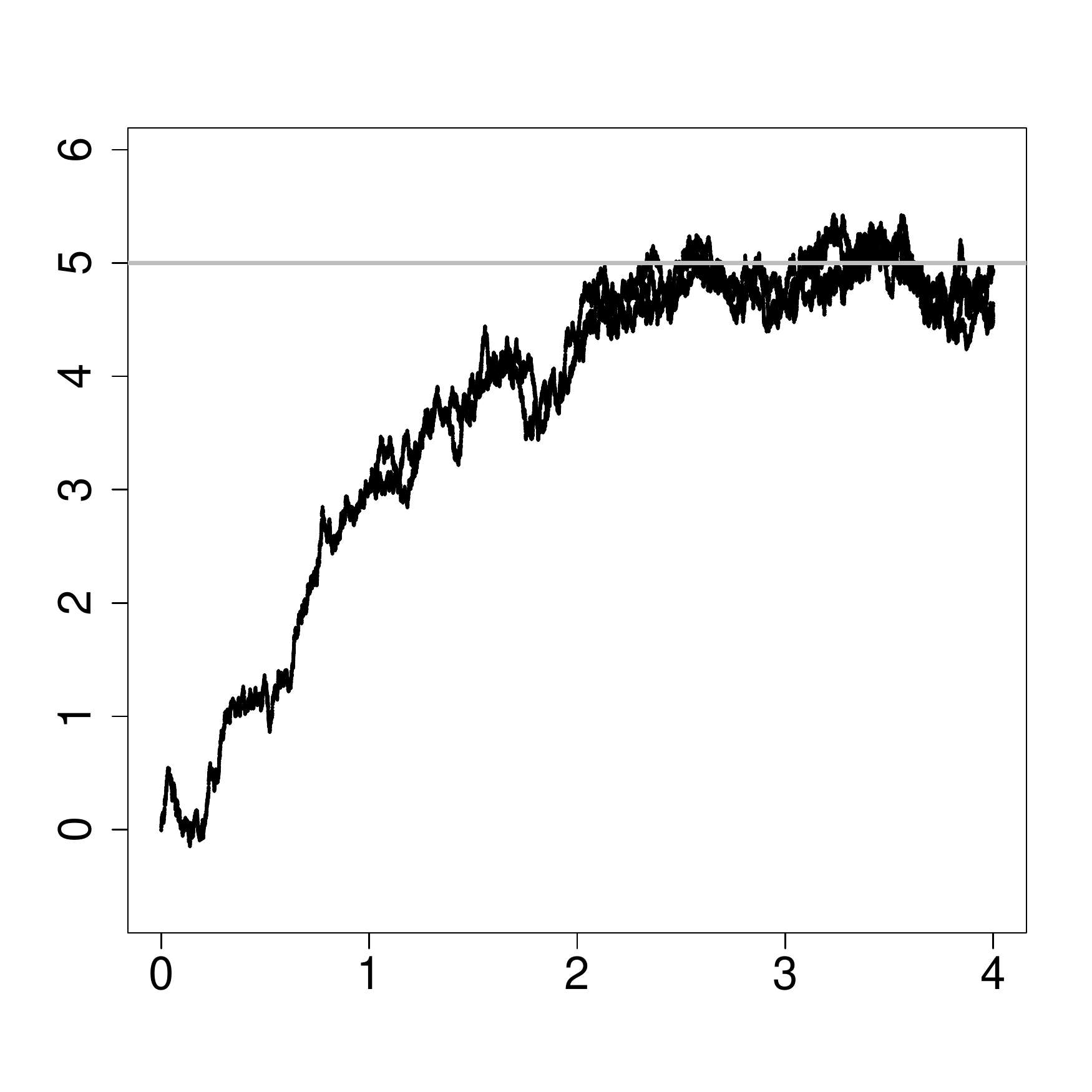} \\
\includegraphics[width=0.3\textwidth]{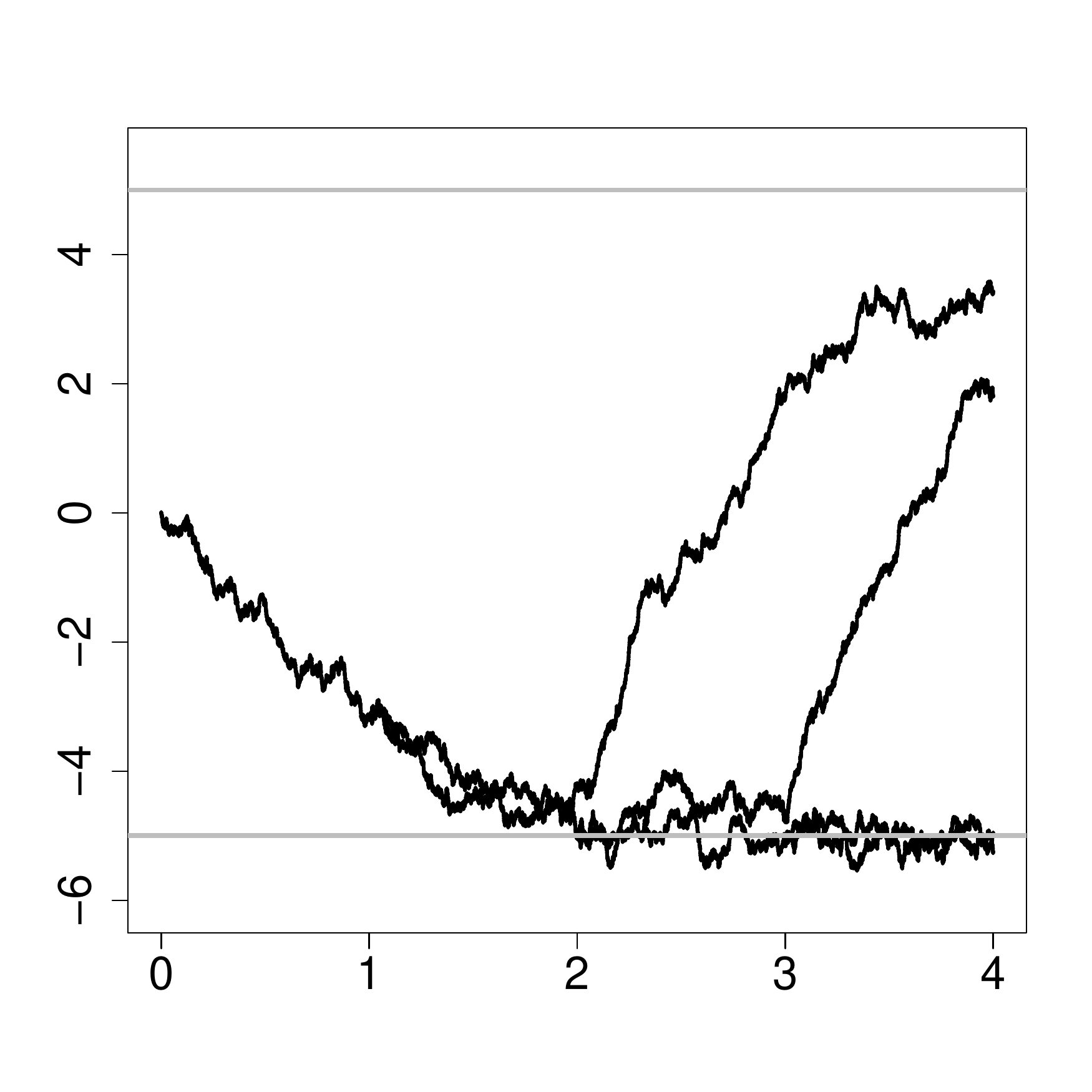}
\includegraphics[width=0.3\textwidth]{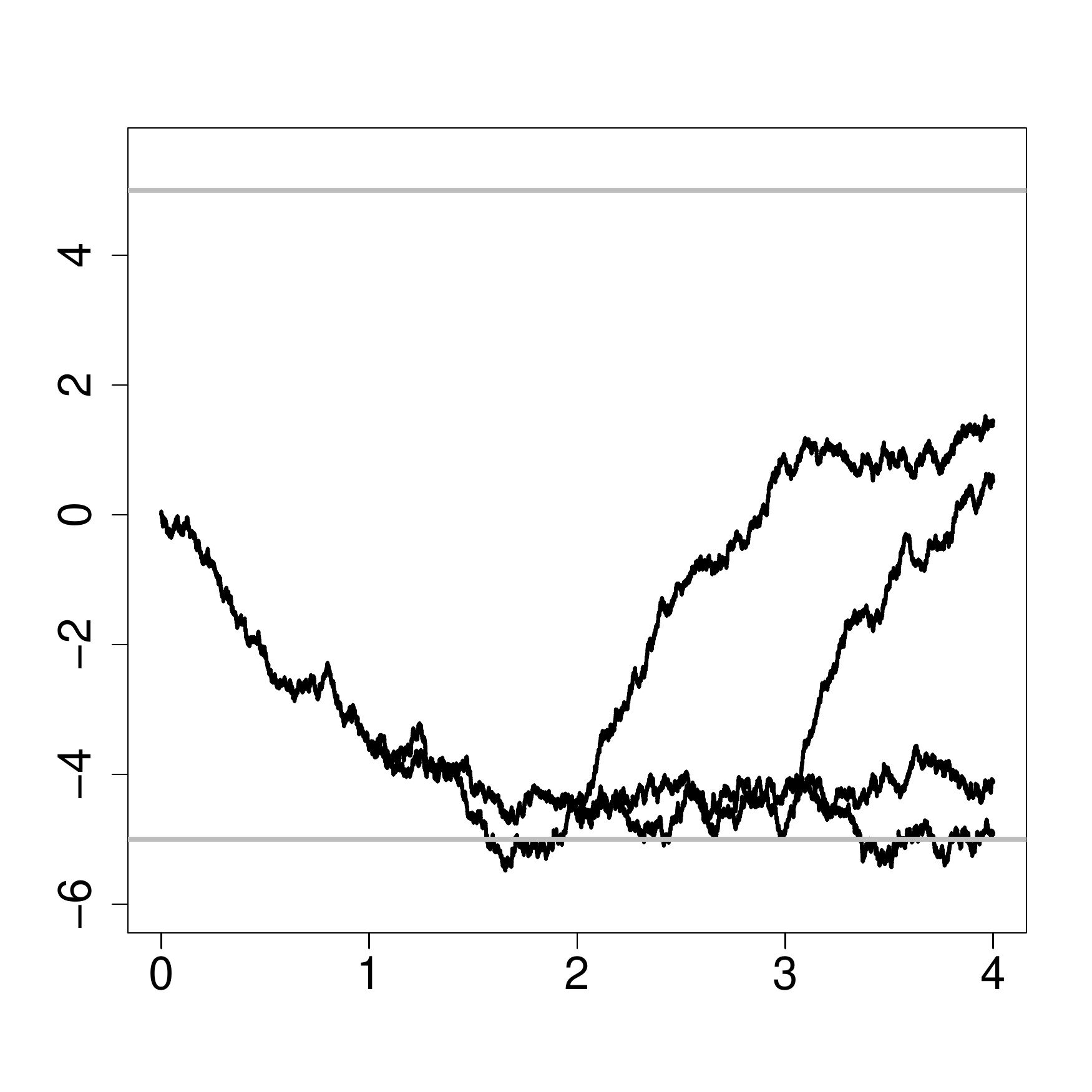}
\includegraphics[width=0.3\textwidth]{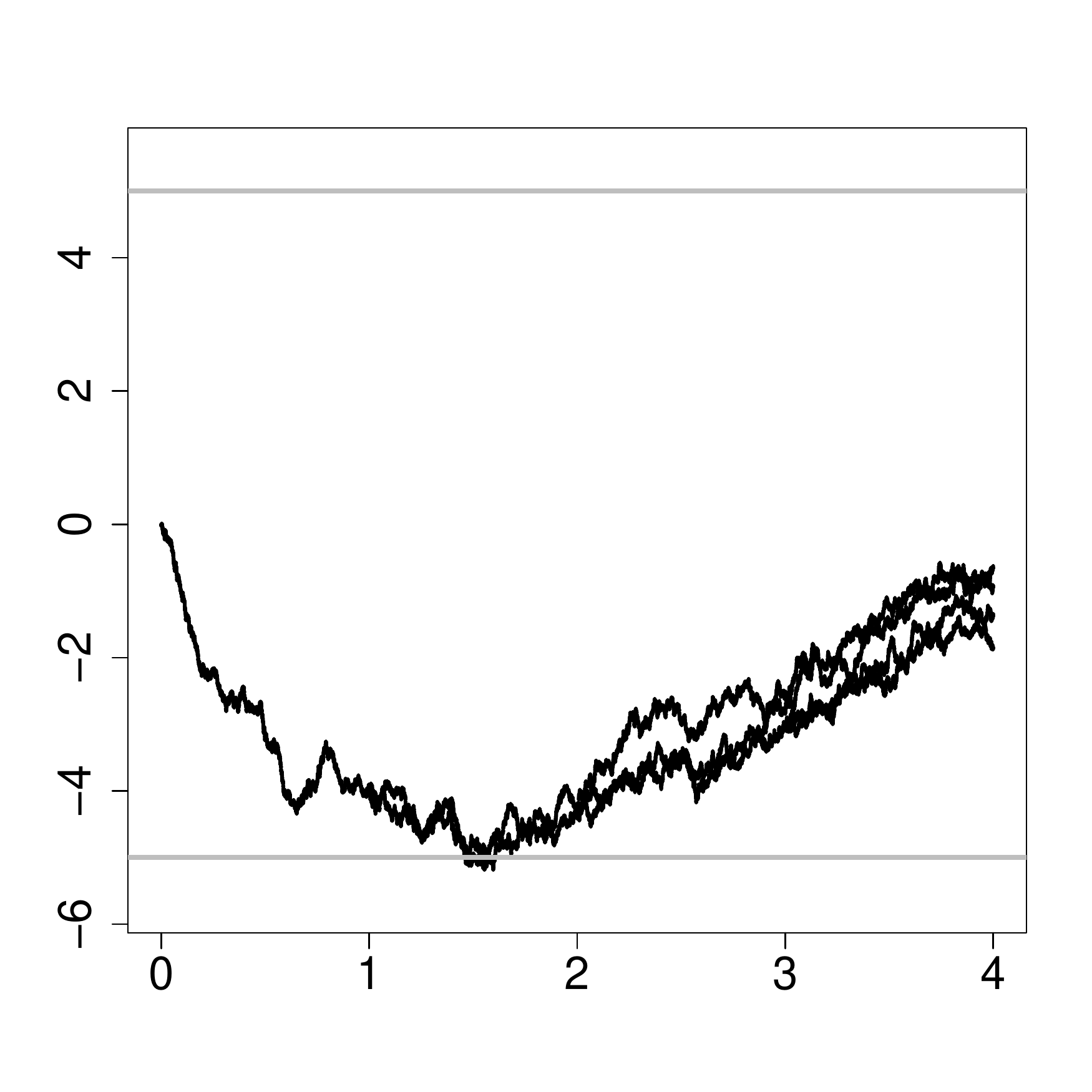}

\caption{Simulated trajectories on the phylogeny depicted at the top.
First column: $m=0$, second column: $m=0.5$, third column $m=10$,
first row: all species have $\theta=5$, second row:
half of species have $\theta=-5$, other half $\theta=5$.
The solid line branches have $\theta=-5$ as the optimum while the dashed have $\theta=5$.
In all presented simulations $\alpha=1$, $X_{0}=0$, $\sigma^{2}=1$.
We can clearly observe that migration has an averaging out effect
on the ``optimum'' value.
\label{figtrajsim}}
\end{center}
\end{figure}

We now present the most illustrative histograms of the effects of
migration, while all the simulation results can be found in the
supplementary material.
In the simulations we can immediately see that, as expected, the
migration parameter $m$ mixes the individuals. When all have the same
optimum value then increasing $m$ brings all individuals closer around
the mean, by decreasing the observed trait variance. This effect is
thus similar to increasing selection so that neglecting migration
could affect selection estimates. In fact this increase of the
selection effect is clearly visible in the simulated trajectories
presented in Fig. \ref{figtrajsim}.  In the contrasting situation of
different optima then an increase of the migration parameter $m$
pushes individuals away from their optimum and towards an average of
these optima. This is again visible in the simulated trajectories of
Fig. \ref{figtrajsim}. When migration is weak the trait values still
manage to tend towards their optima. However in the case of strong
migration ($m=10$, $\alpha=1$) the traits move away from their optima.

\begin{figure}[!ht]
\begin{center}
\includegraphics[width=0.3\textwidth]{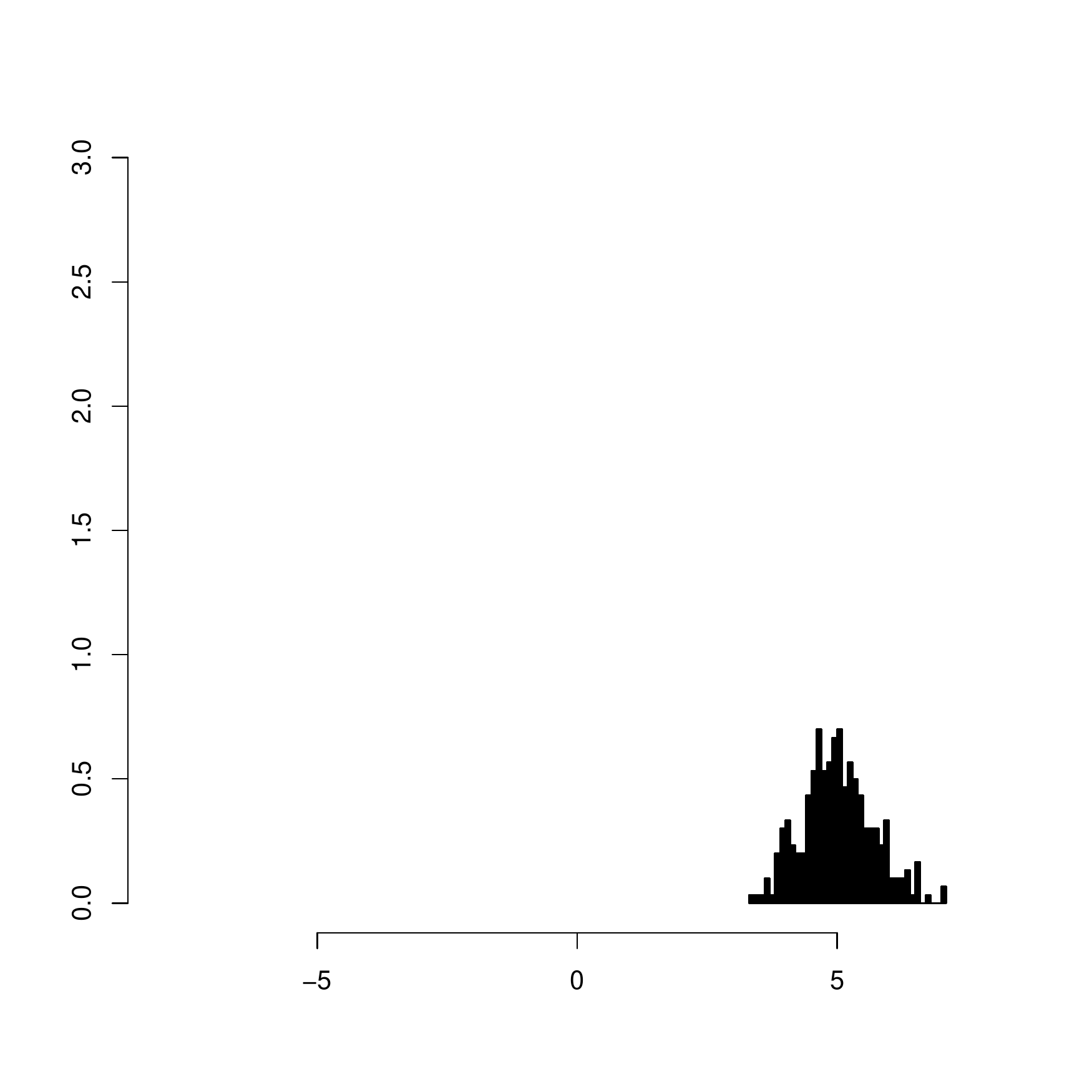}
\includegraphics[width=0.3\textwidth]{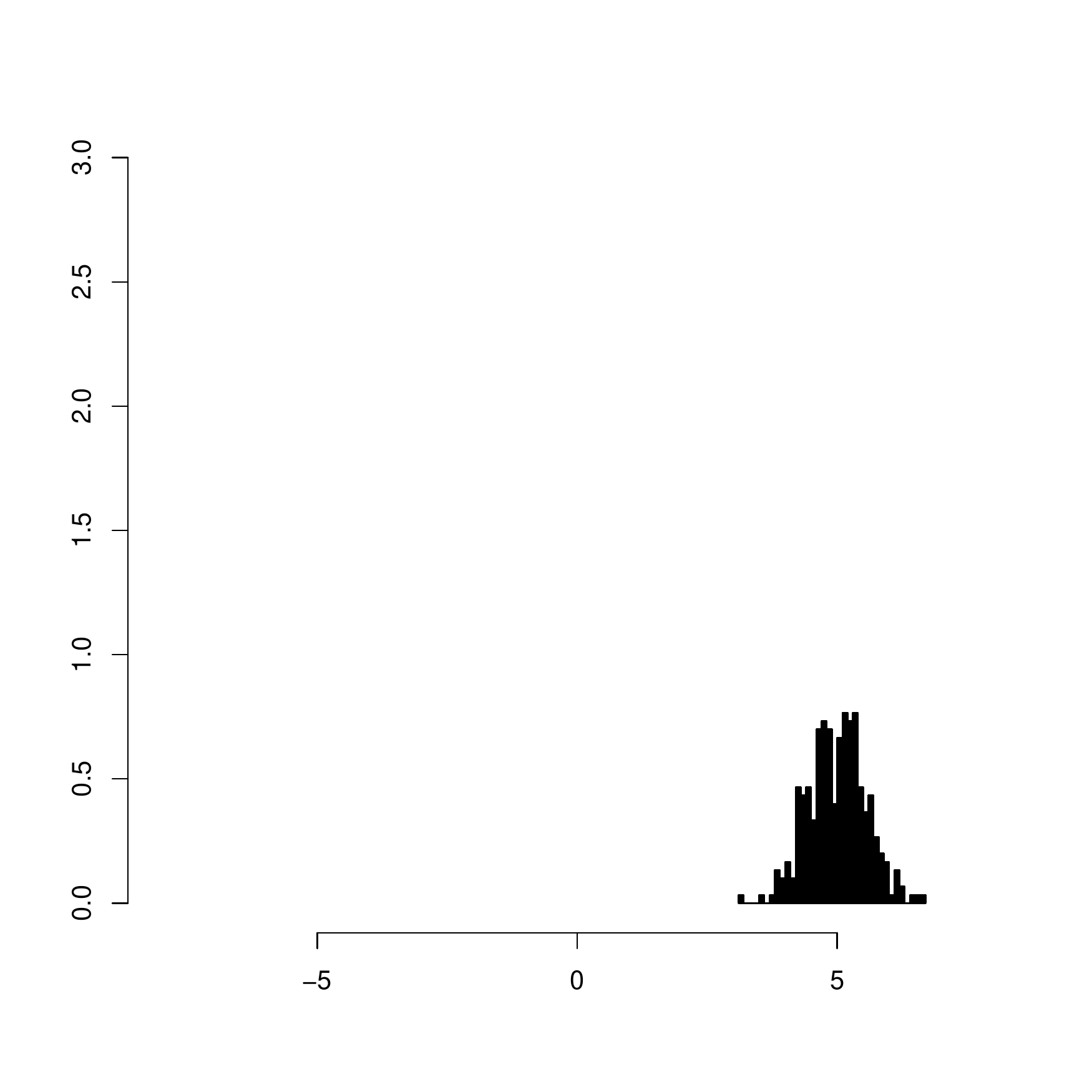}
\includegraphics[width=0.3\textwidth]{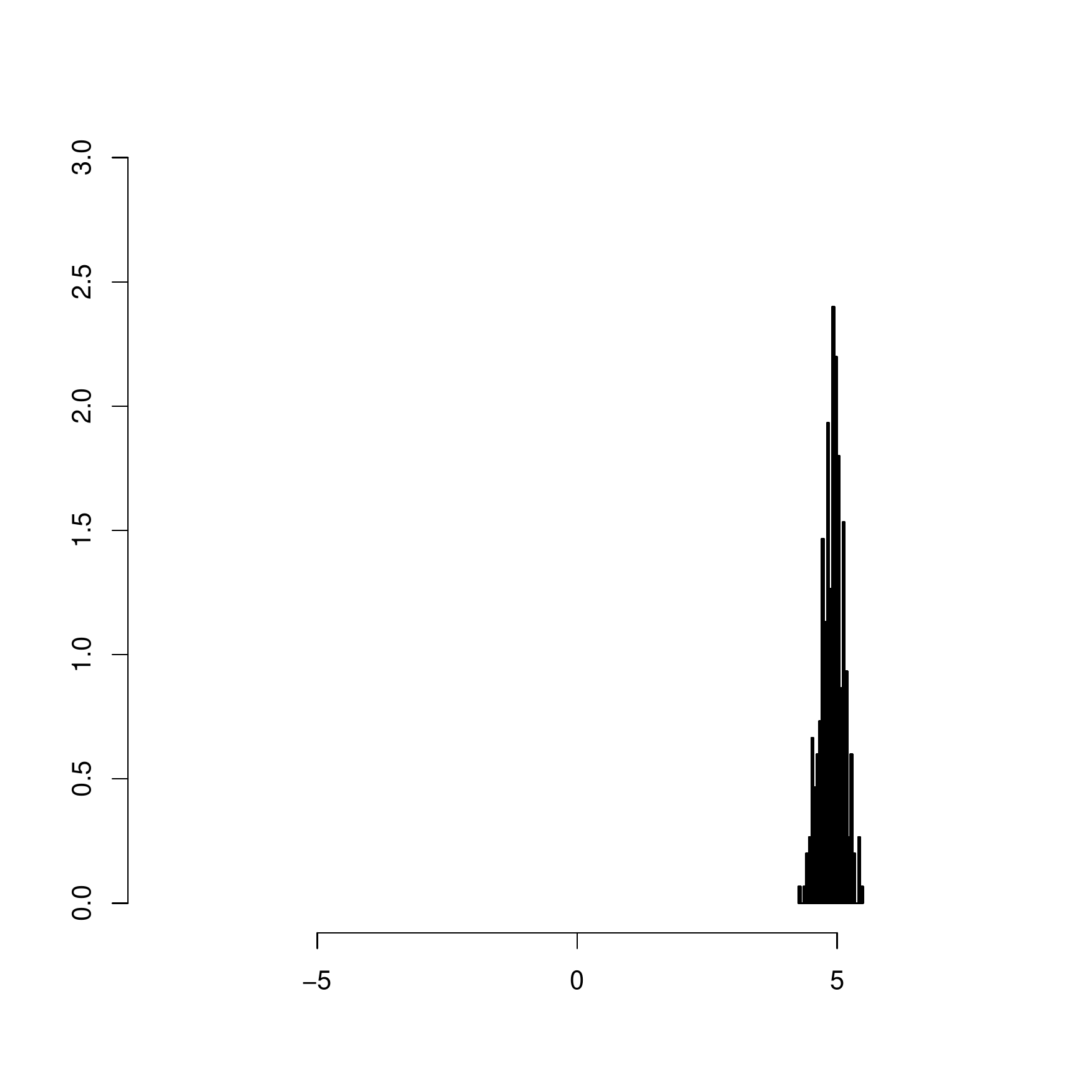} \\
\includegraphics[width=0.3\textwidth]{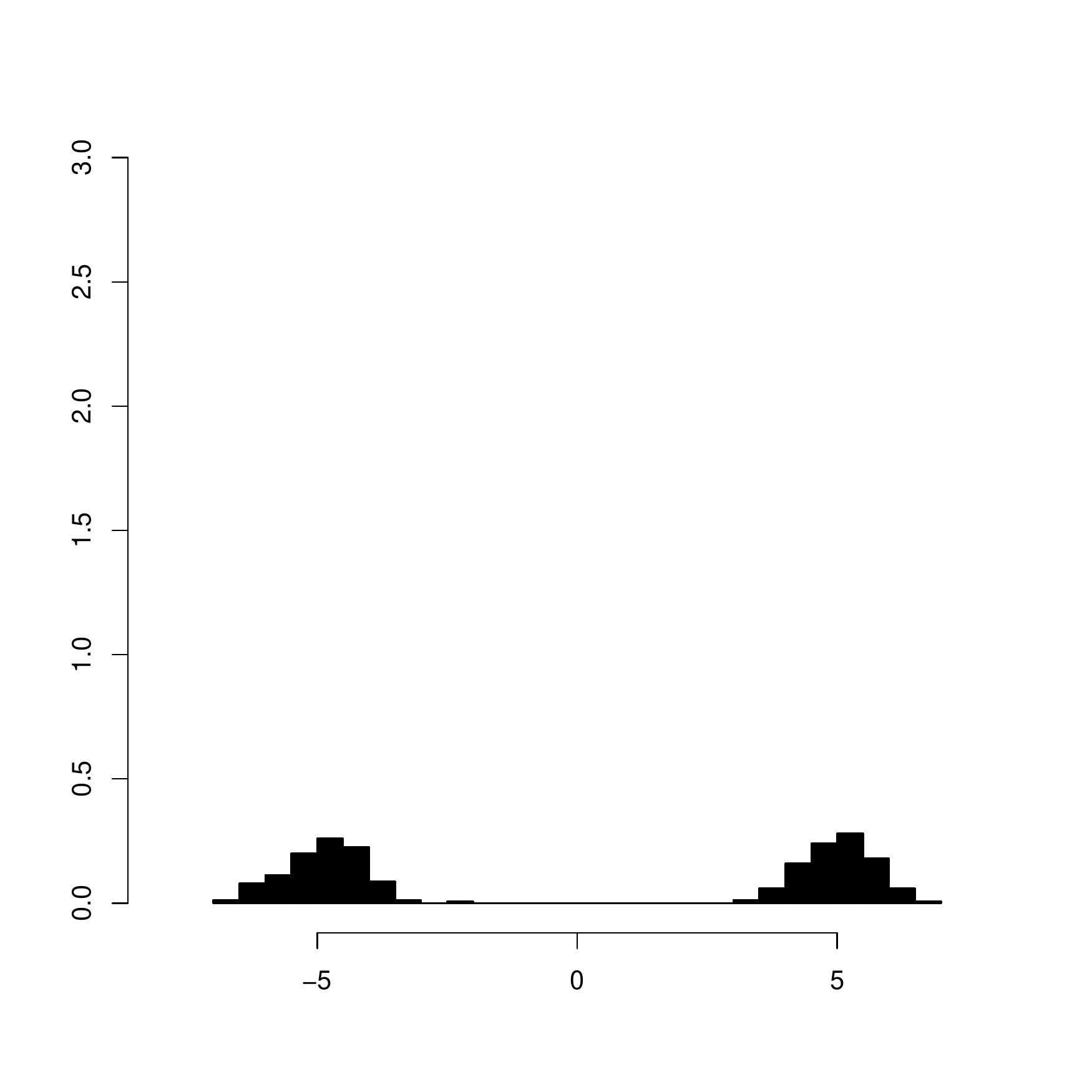}
\includegraphics[width=0.3\textwidth]{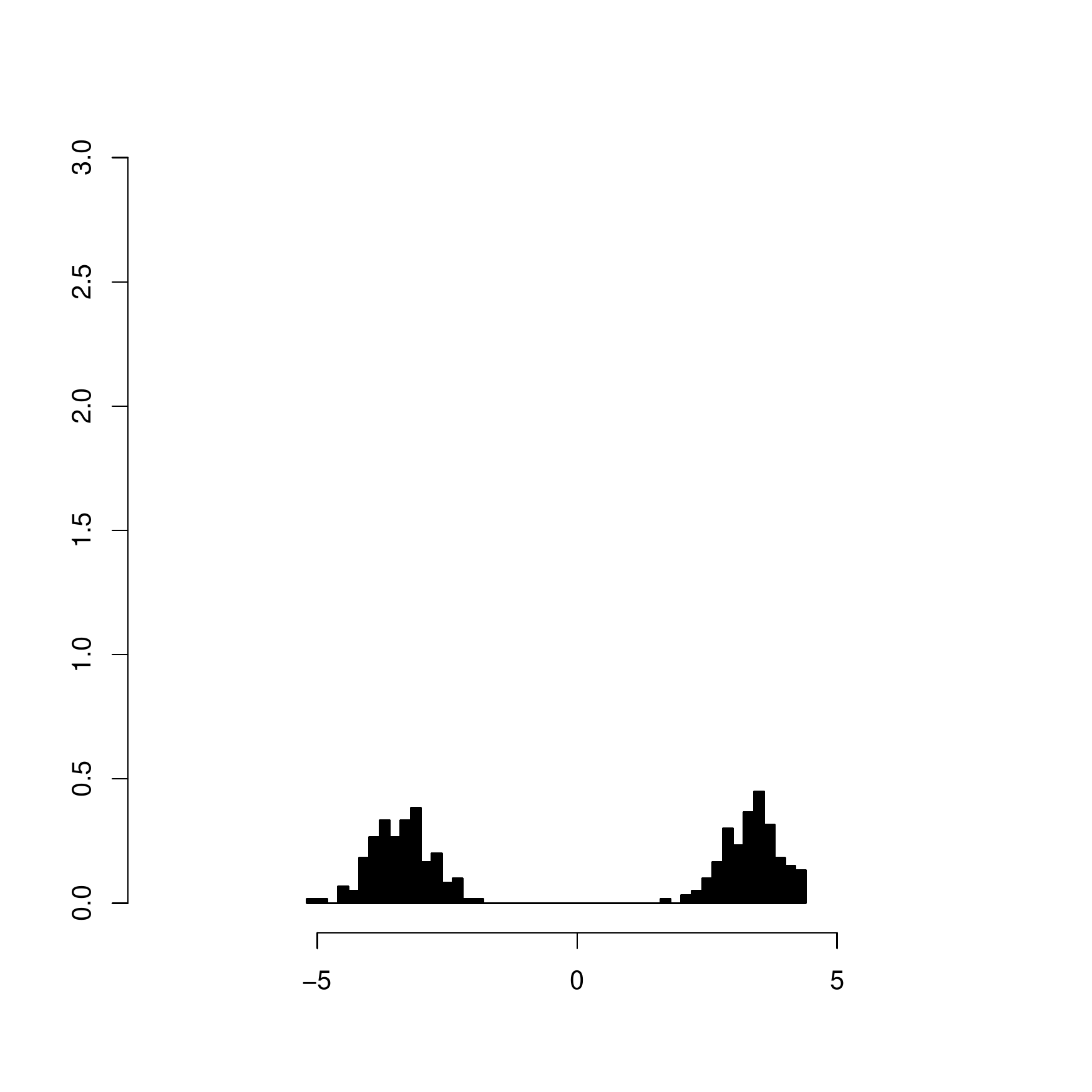}
\includegraphics[width=0.3\textwidth]{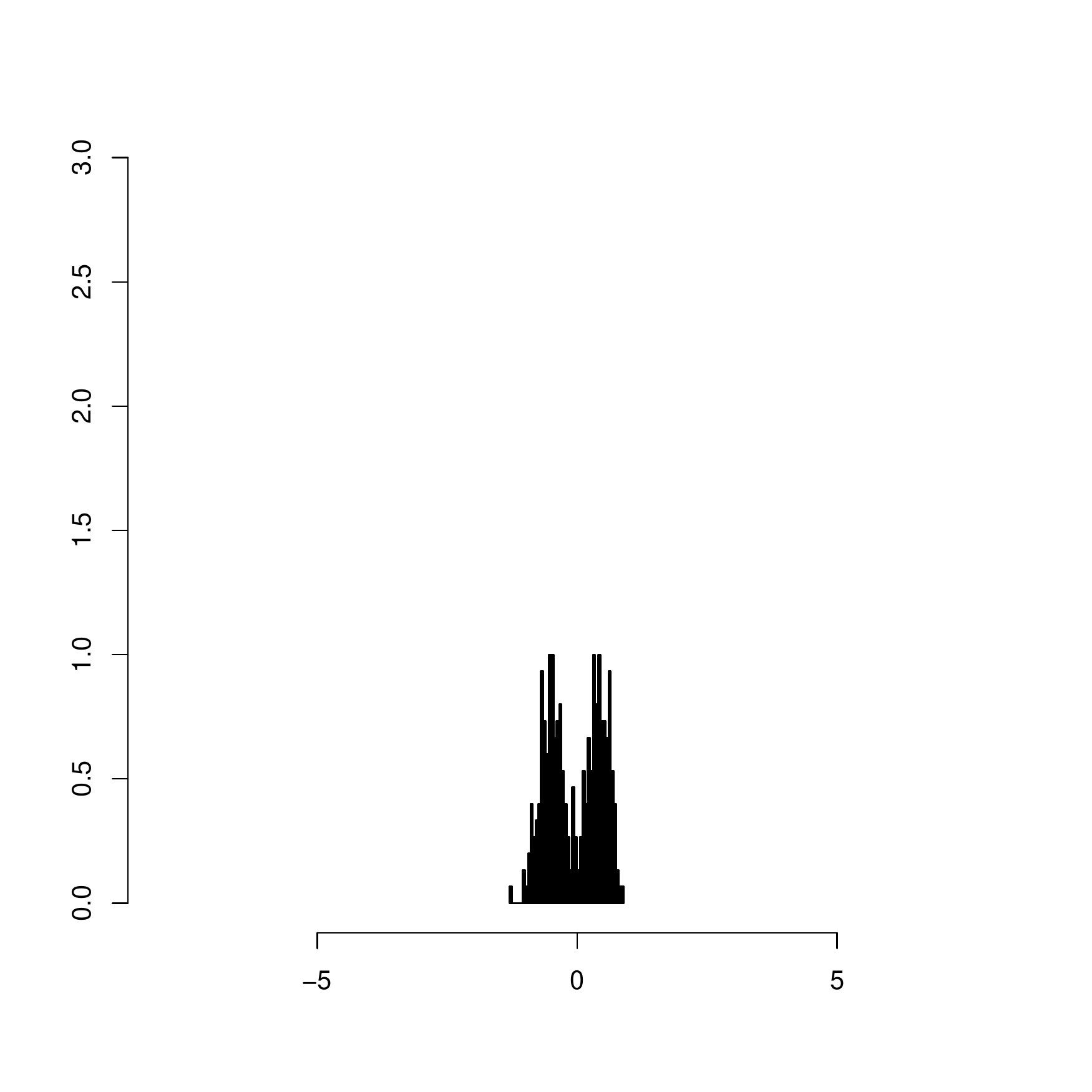}
\caption{Histograms of estimates of $\theta$ values. 
First column: $m=0$, second column: $m=0.5$, third column $m=10$,
first row: all individuals have $\theta=5$, second row:
half of individuals have $\theta=-5$, other half $\theta=5$.
In all presented simulations $\alpha=1$, $X_{0}=0$, $\sigma^{2}=1$.
\label{figpopsim}}
\end{center}
\end{figure}

\begin{figure}[!ht]
\begin{center}
\includegraphics[width=0.3\textwidth]{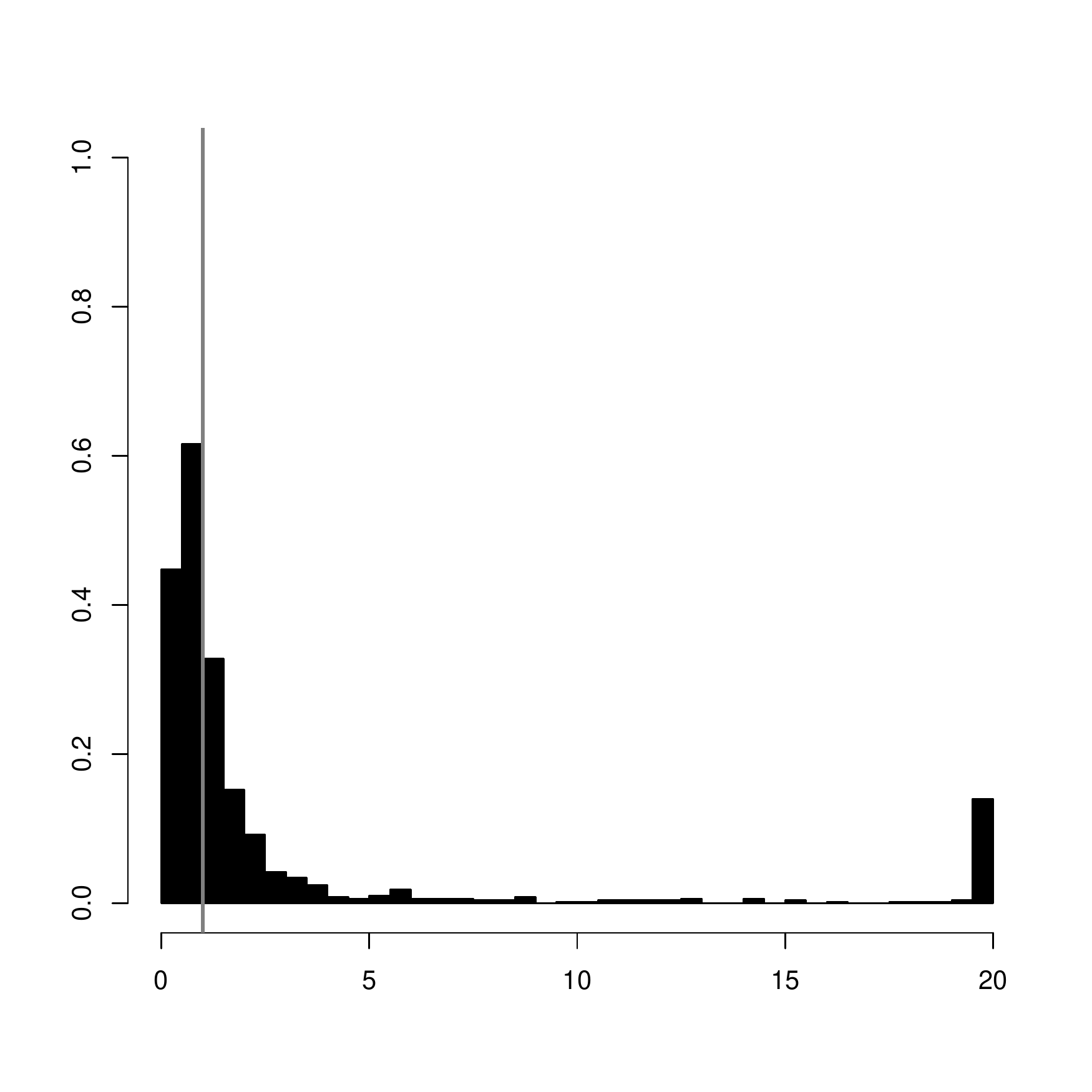}
\includegraphics[width=0.3\textwidth]{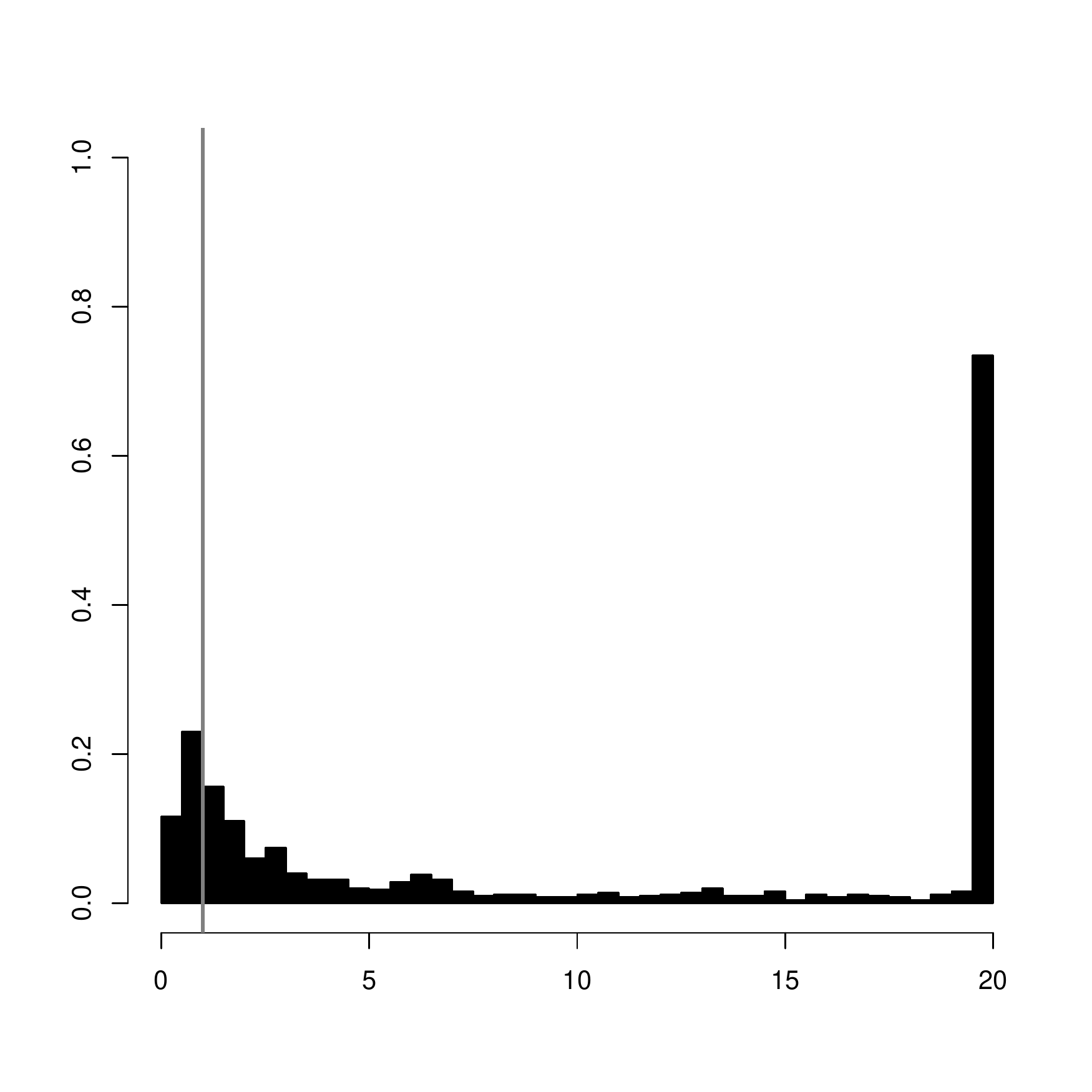}
\includegraphics[width=0.3\textwidth]{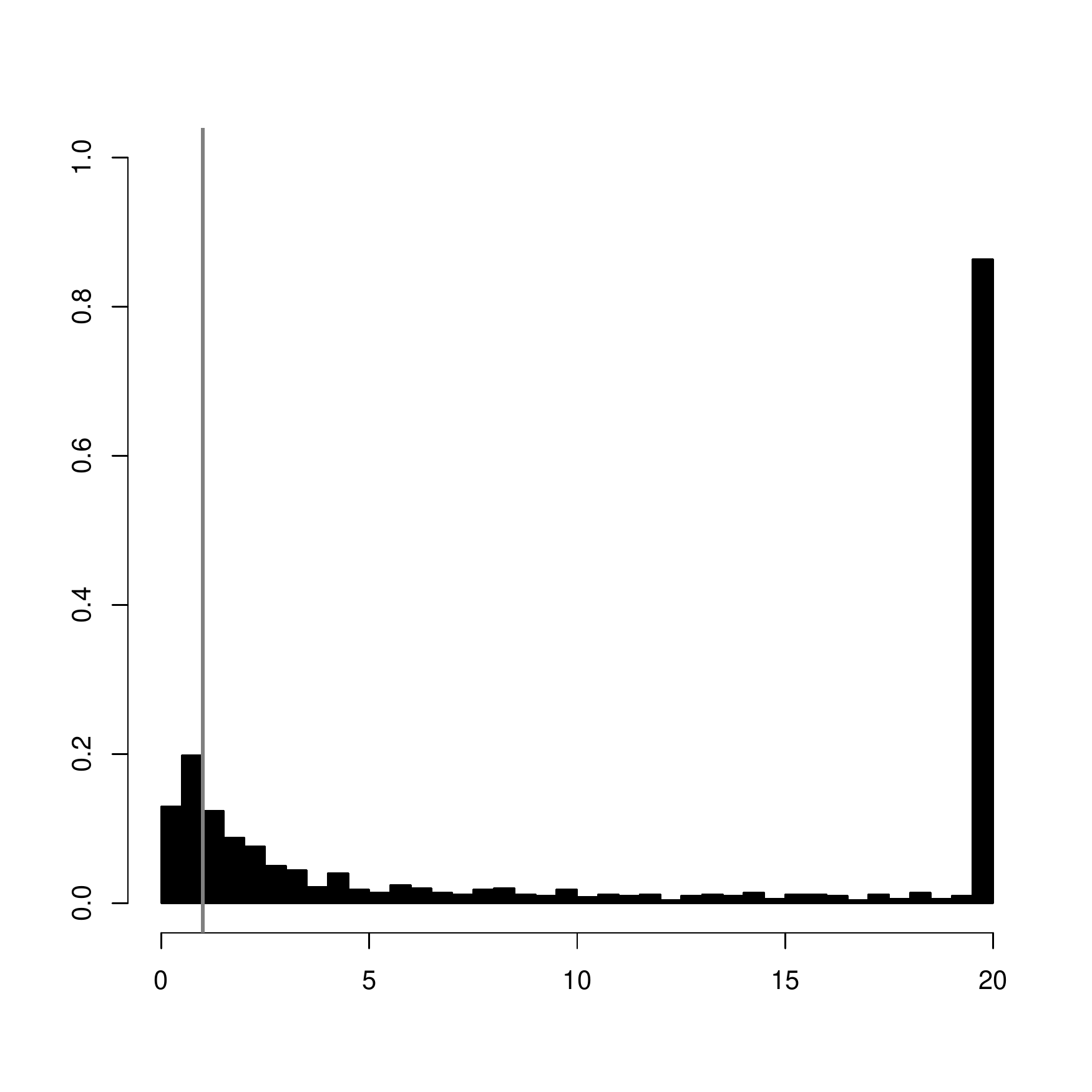} \\
\includegraphics[width=0.3\textwidth]{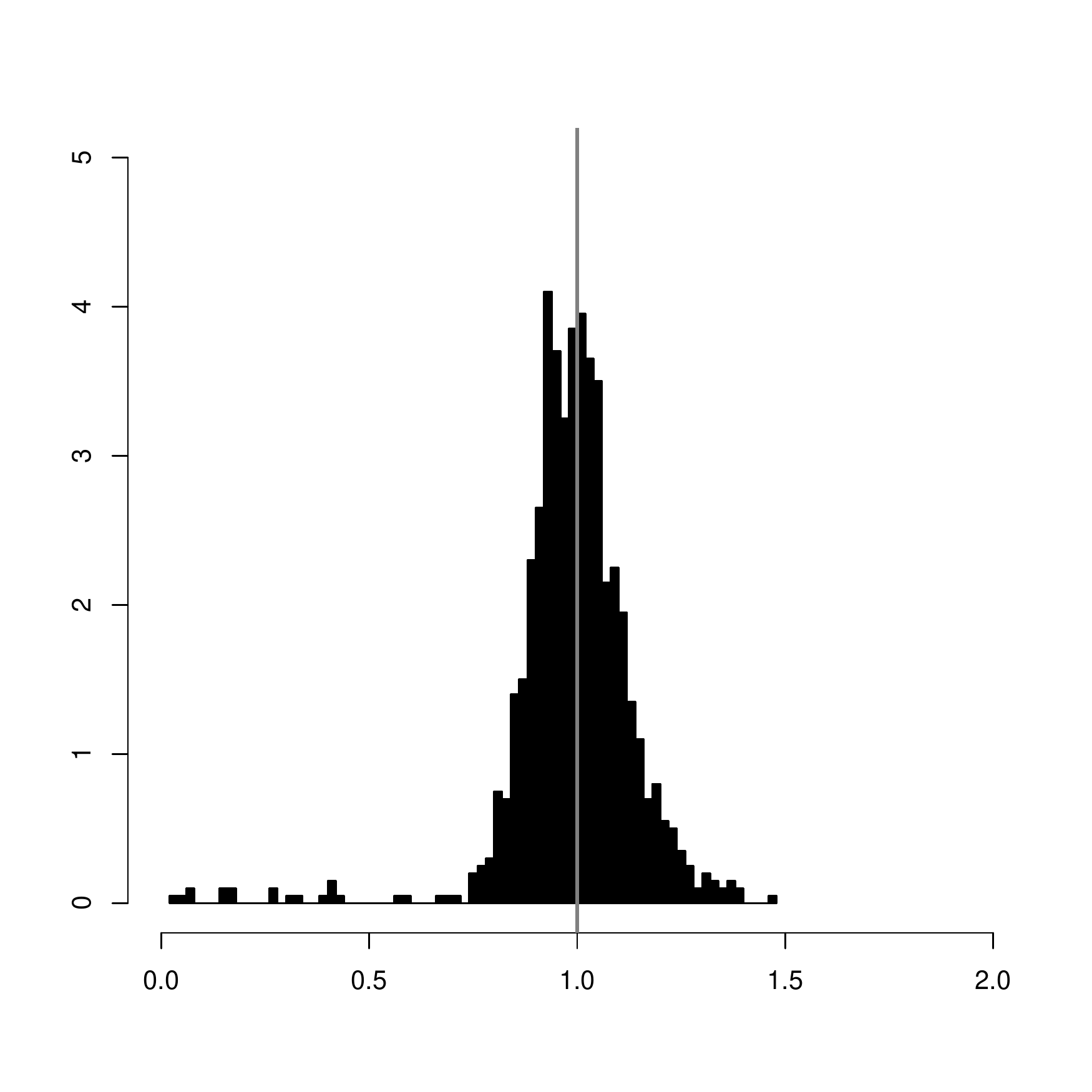}
\includegraphics[width=0.3\textwidth]{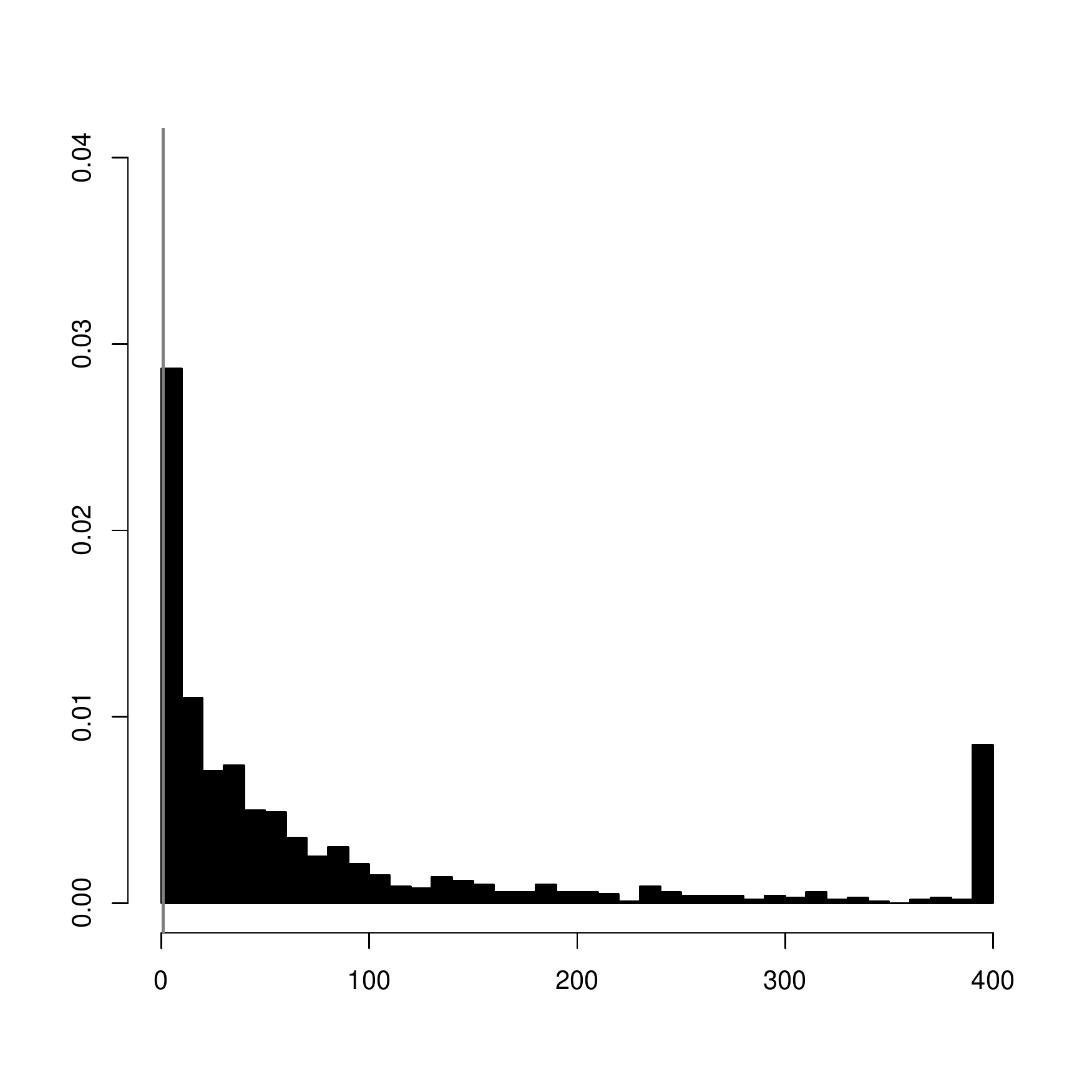}
\includegraphics[width=0.3\textwidth]{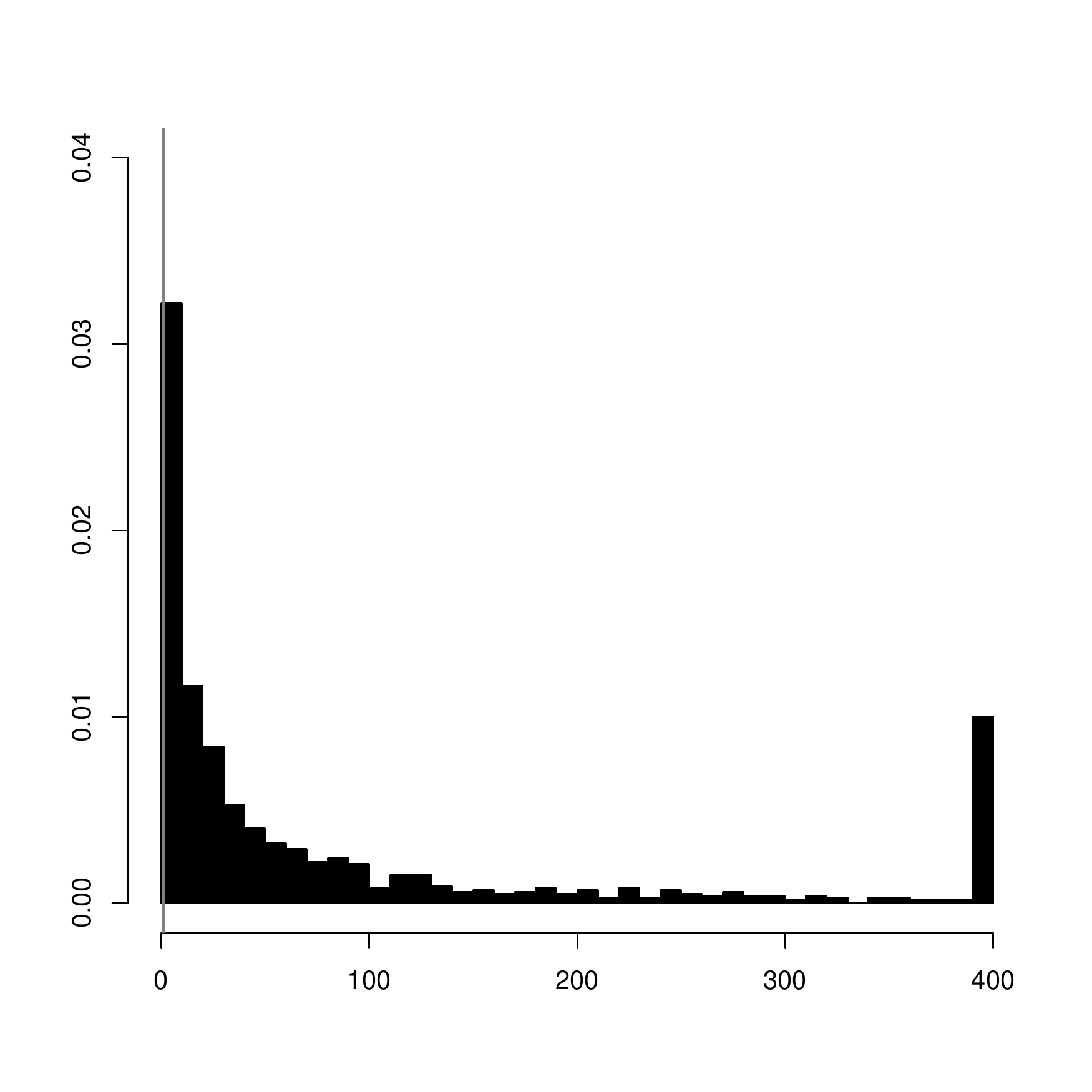}
\caption{Histograms of estimates of $\alpha$.
First column: $m=0$, second column: $m=0.5$, third column $m=10$,
first row: all individuals have $\theta=5$, second row:
half of individuals have $\theta=-5$, other half $\theta=5$.
In all presented histograms $\alpha=1$, $X_{0}=0$, $\sigma^{2}=1$.
The vertical line represents the true value. 
Note that the histograms are not all on the same scale.
The left and right--most bars are values equal
or more extreme than the appropriate $x$--value.
\label{fighistalpha}}
\end{center}
\end{figure}

\begin{figure}[!ht]
\begin{center}
\includegraphics[width=0.3\textwidth]{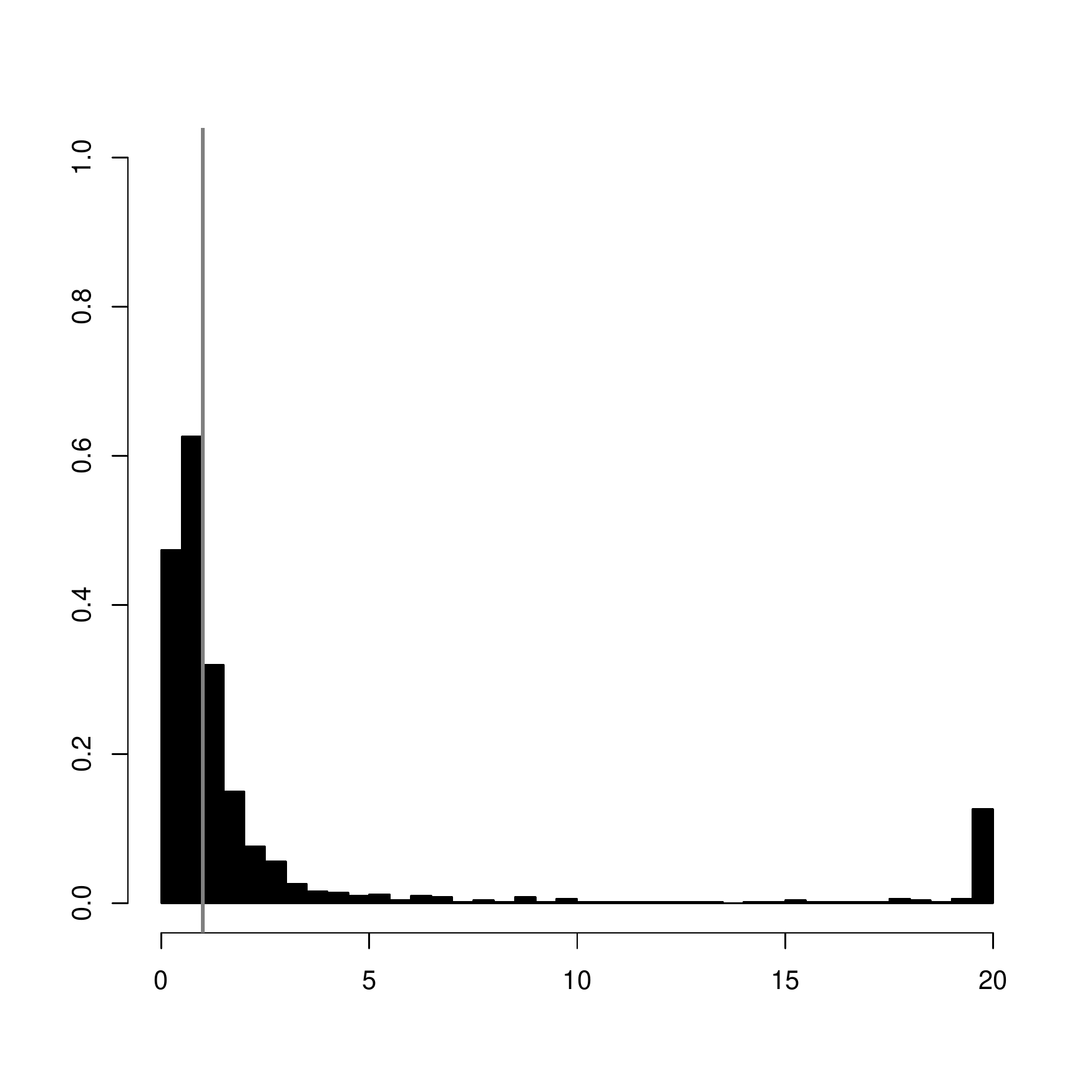}
\includegraphics[width=0.3\textwidth]{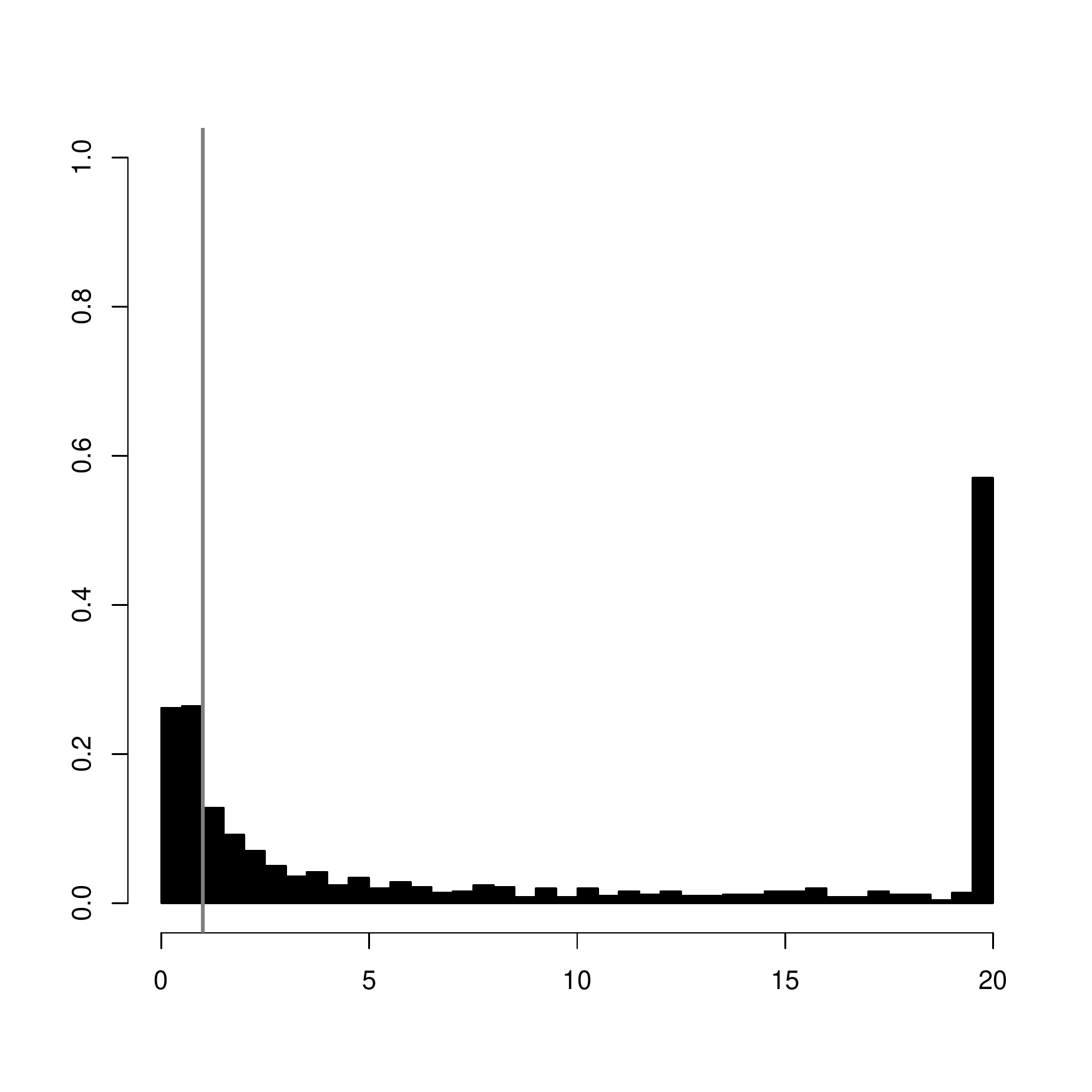}
\includegraphics[width=0.3\textwidth]{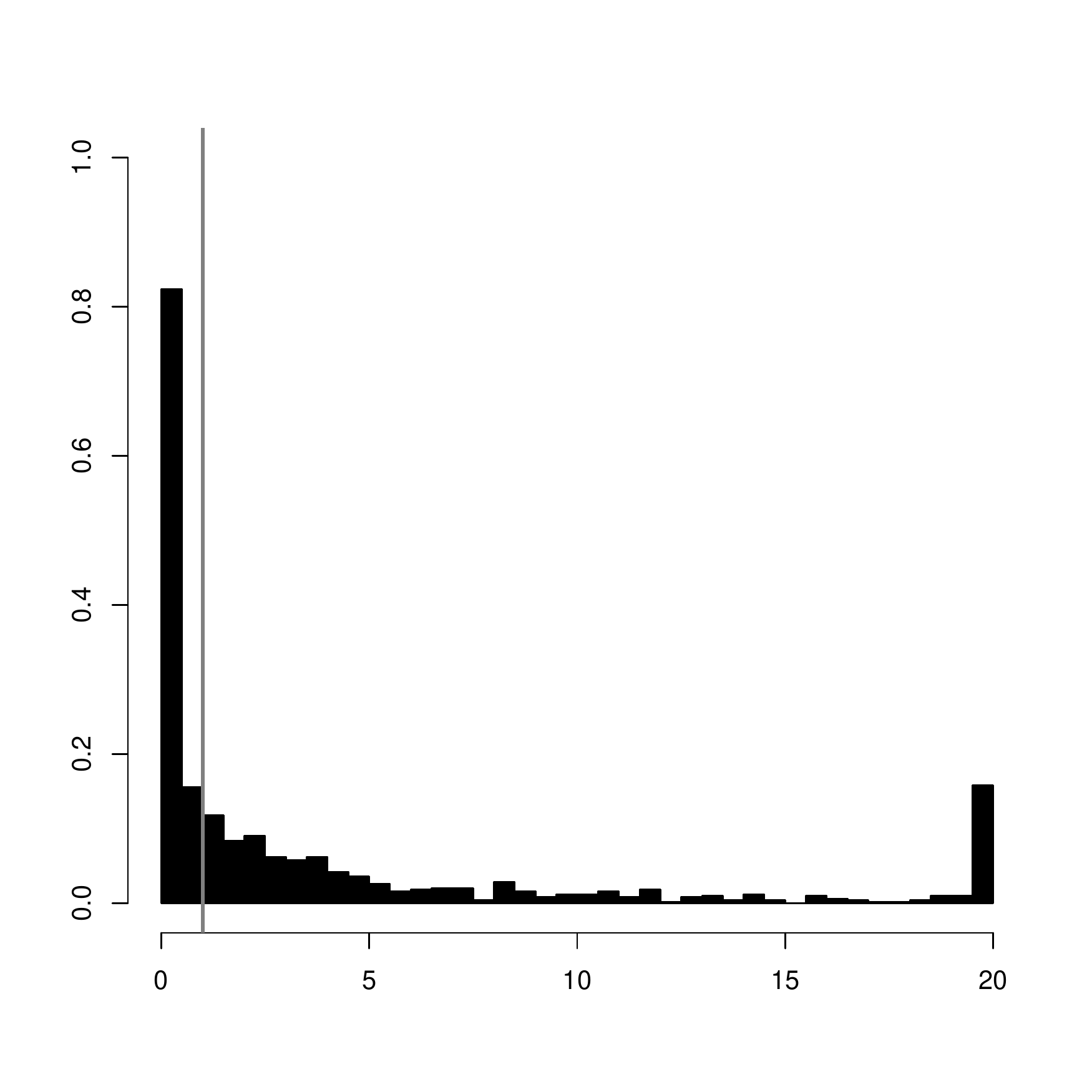} \\
\includegraphics[width=0.3\textwidth]{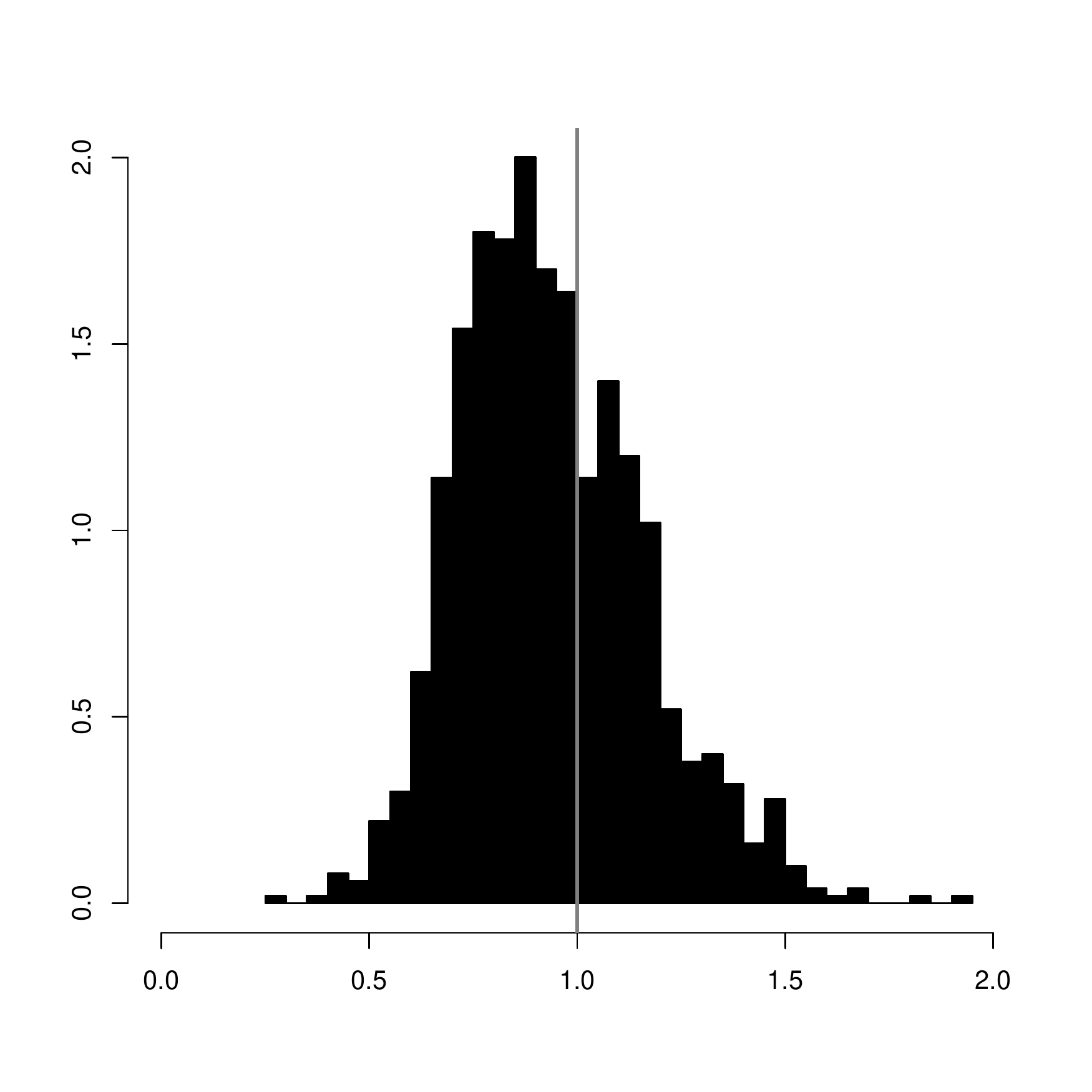}
\includegraphics[width=0.3\textwidth]{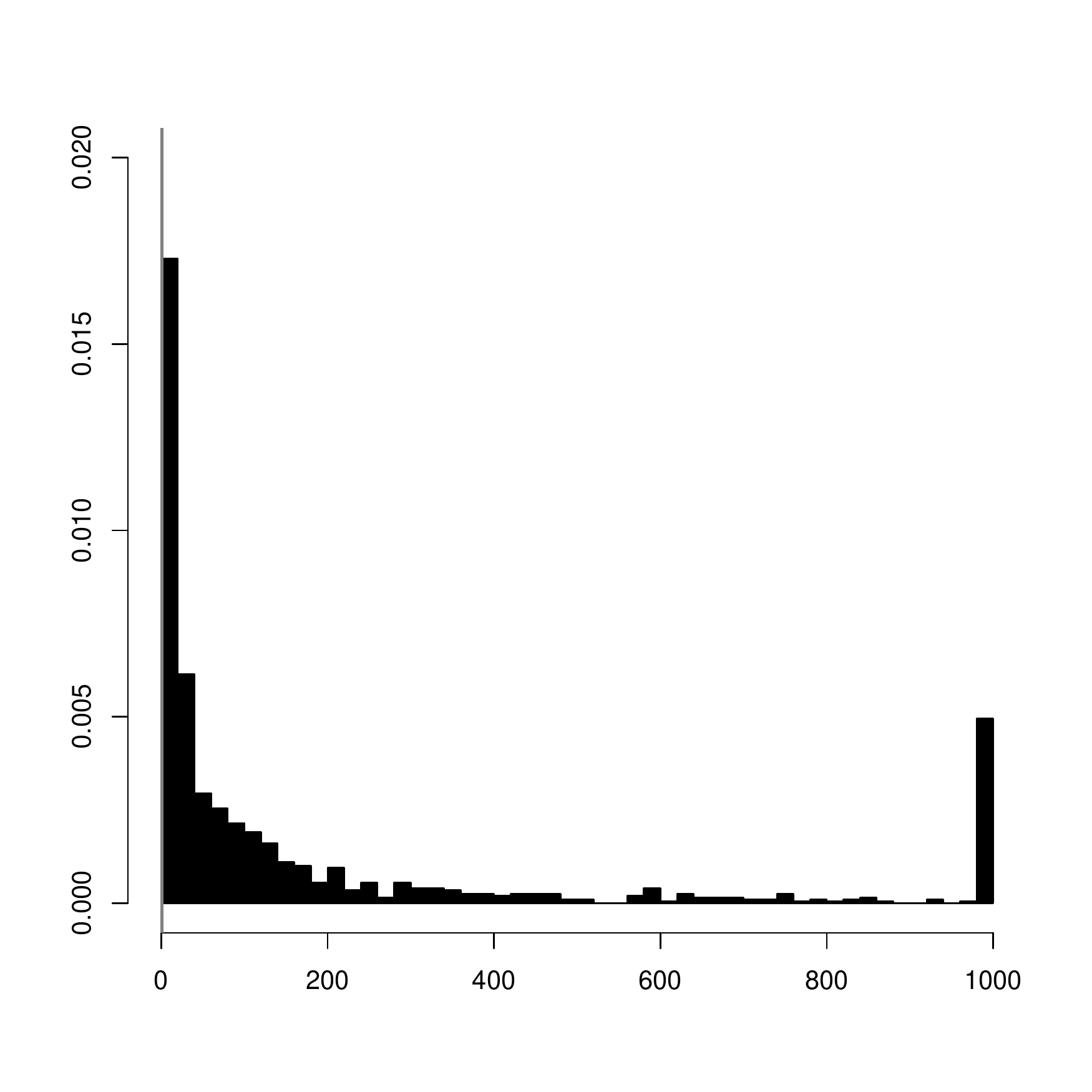}
\includegraphics[width=0.3\textwidth]{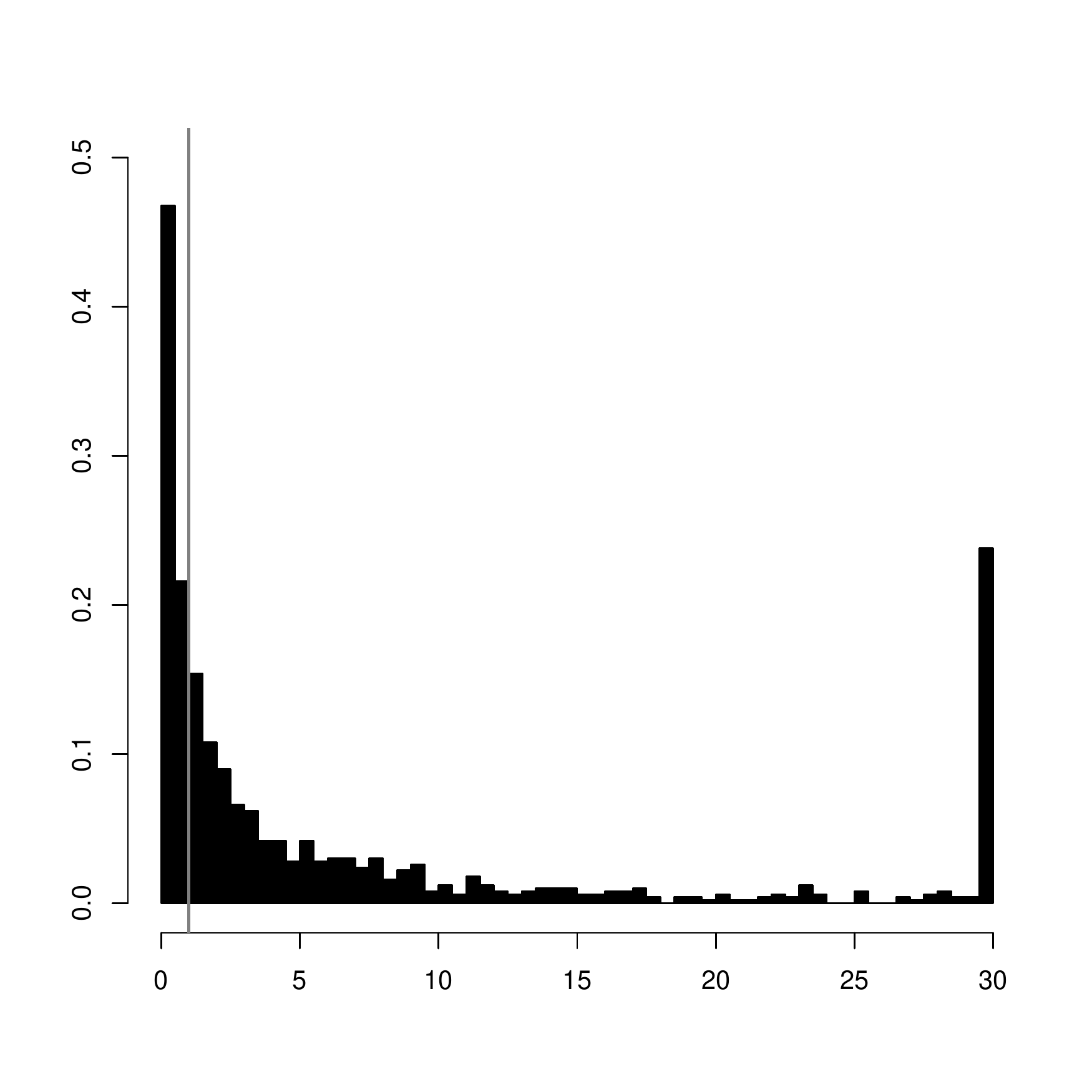}
\caption{Histograms of estimates of 
$\sigma^{2}$.
First column: $m=0$, second column: $m=0.5$, third column $m=10$,
first row: all individuals have $\theta=5$, second row:
half of individuals have $\theta=-5$, other half $\theta=5$.
In all presented histograms $\alpha=1$, $X_{0}=0$, $\sigma^{2}=1$.
The vertical line represents the true value.
Note that the histograms are not all on the same scale.
The left and right--most bars are values equal
or more extreme than the appropriate $x$--value.
\label{fighistsigma2a}}
\end{center}
\end{figure}

\begin{figure}[!ht]
\begin{center}
\includegraphics[width=0.3\textwidth]{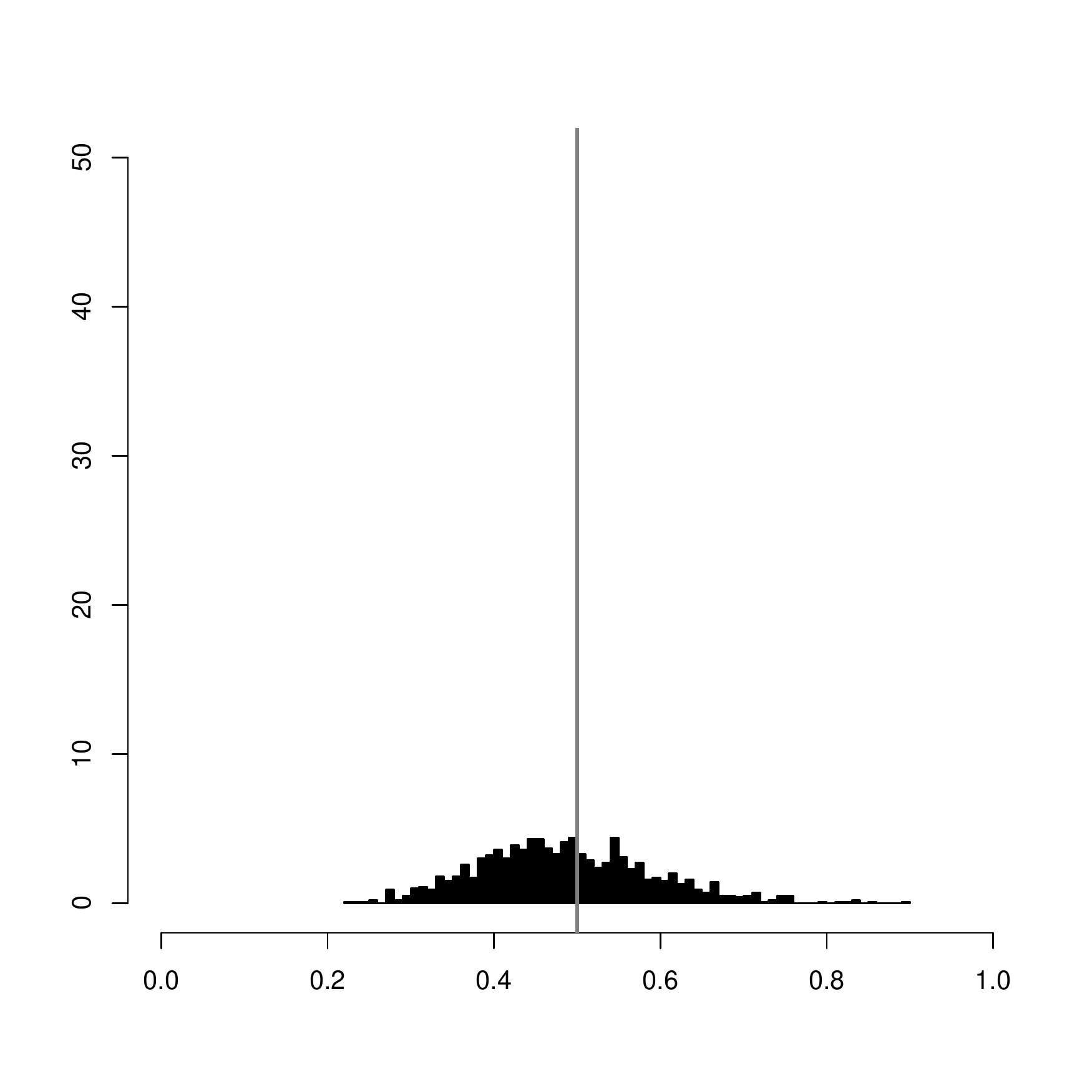}
\includegraphics[width=0.3\textwidth]{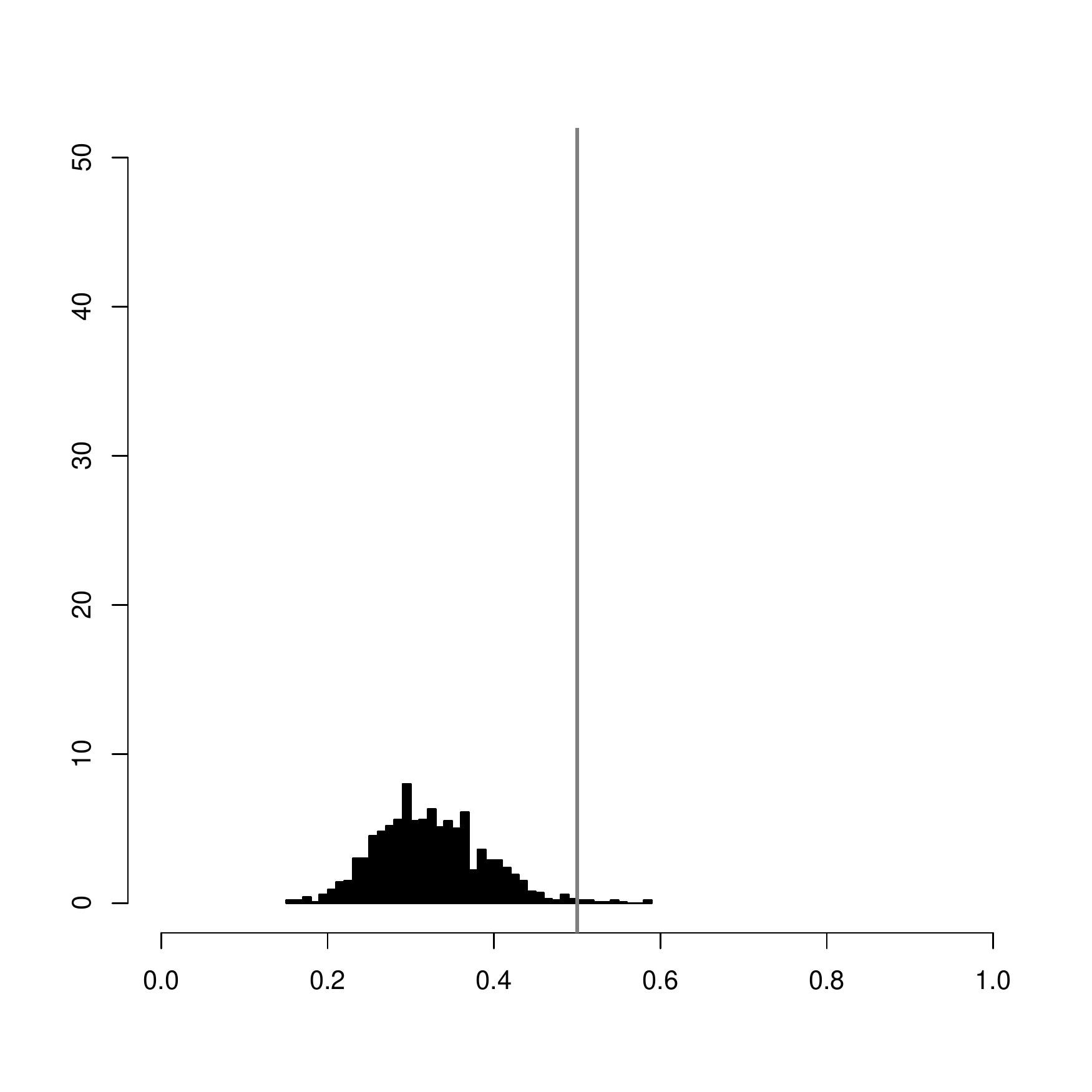}
\includegraphics[width=0.3\textwidth]{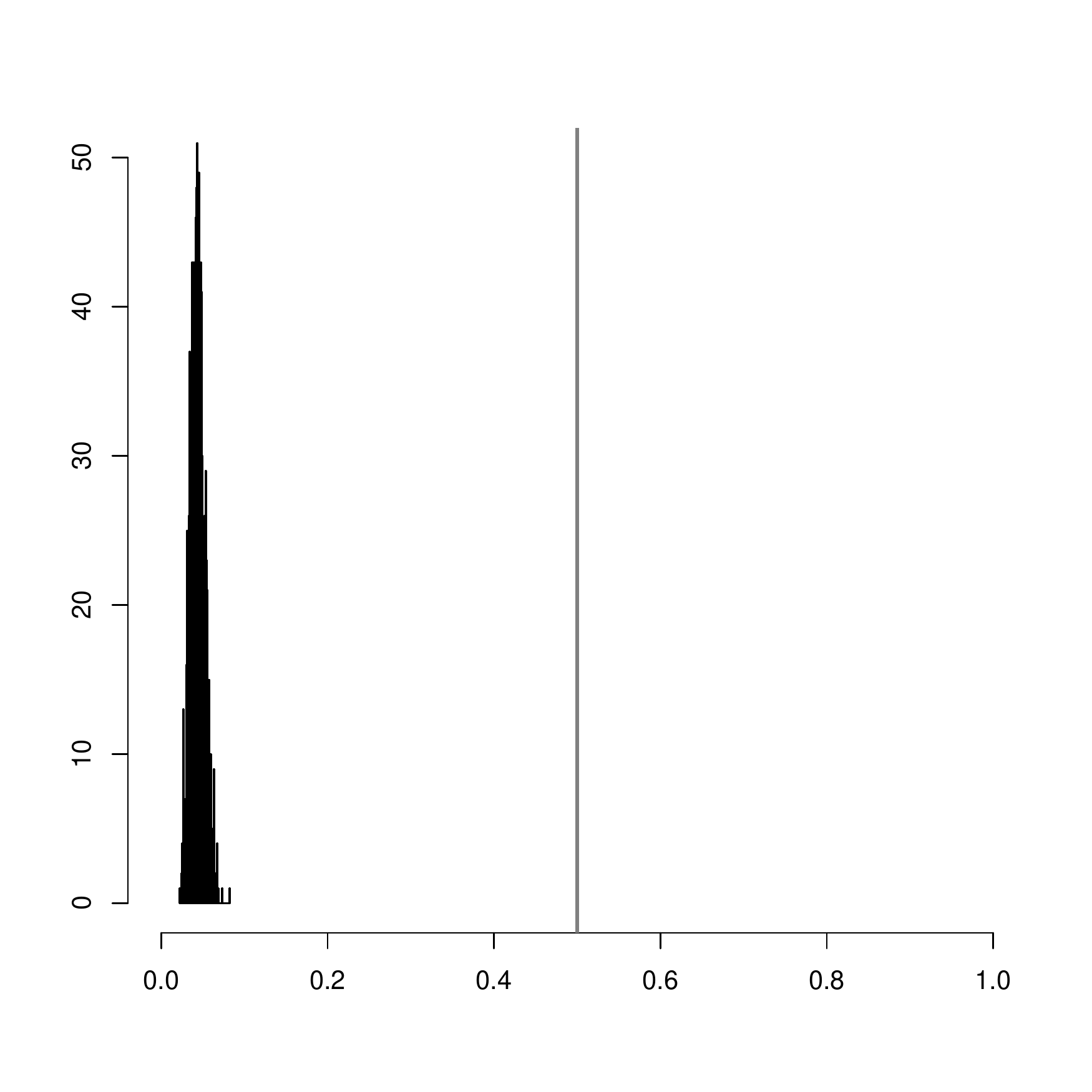} \\
\includegraphics[width=0.3\textwidth]{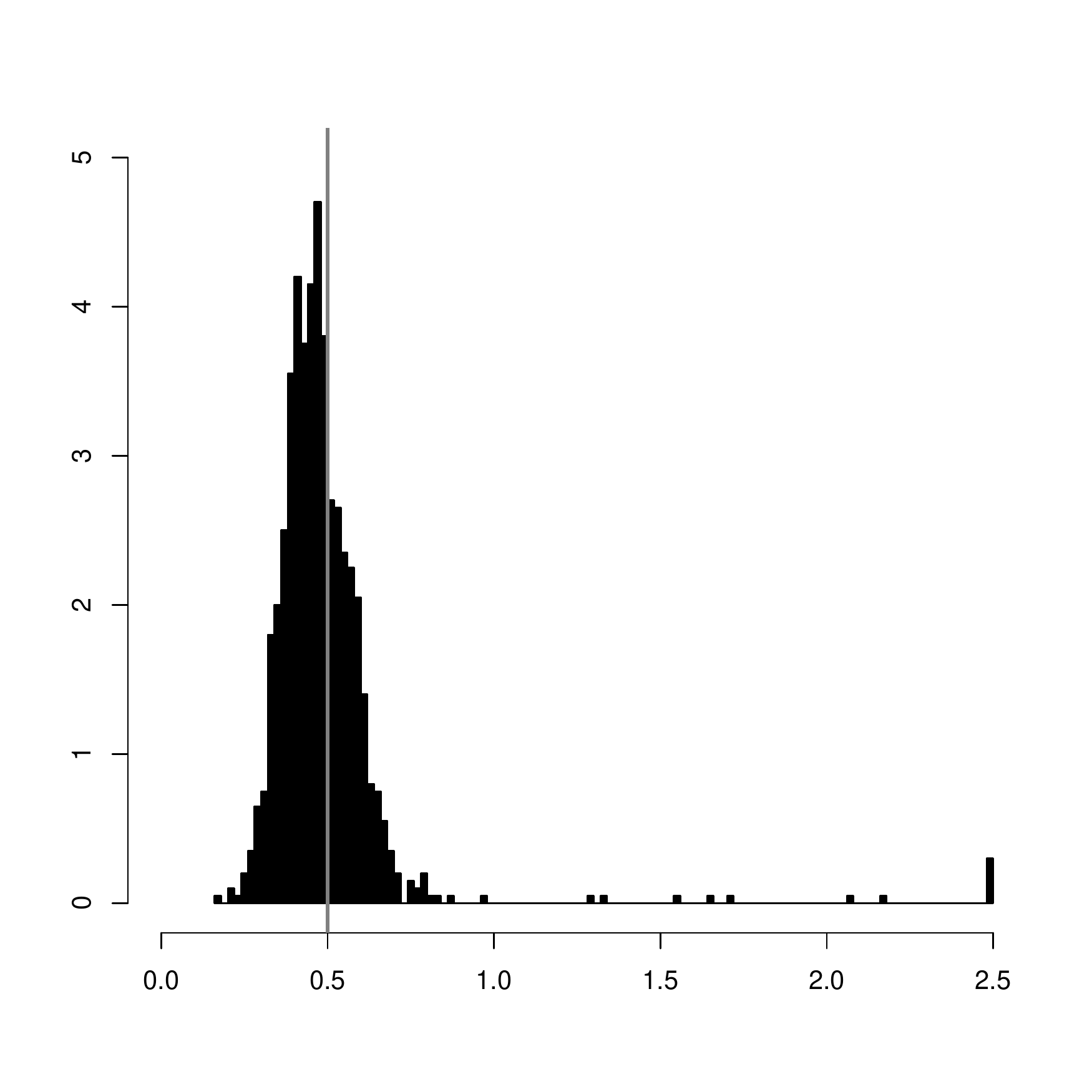}
\includegraphics[width=0.3\textwidth]{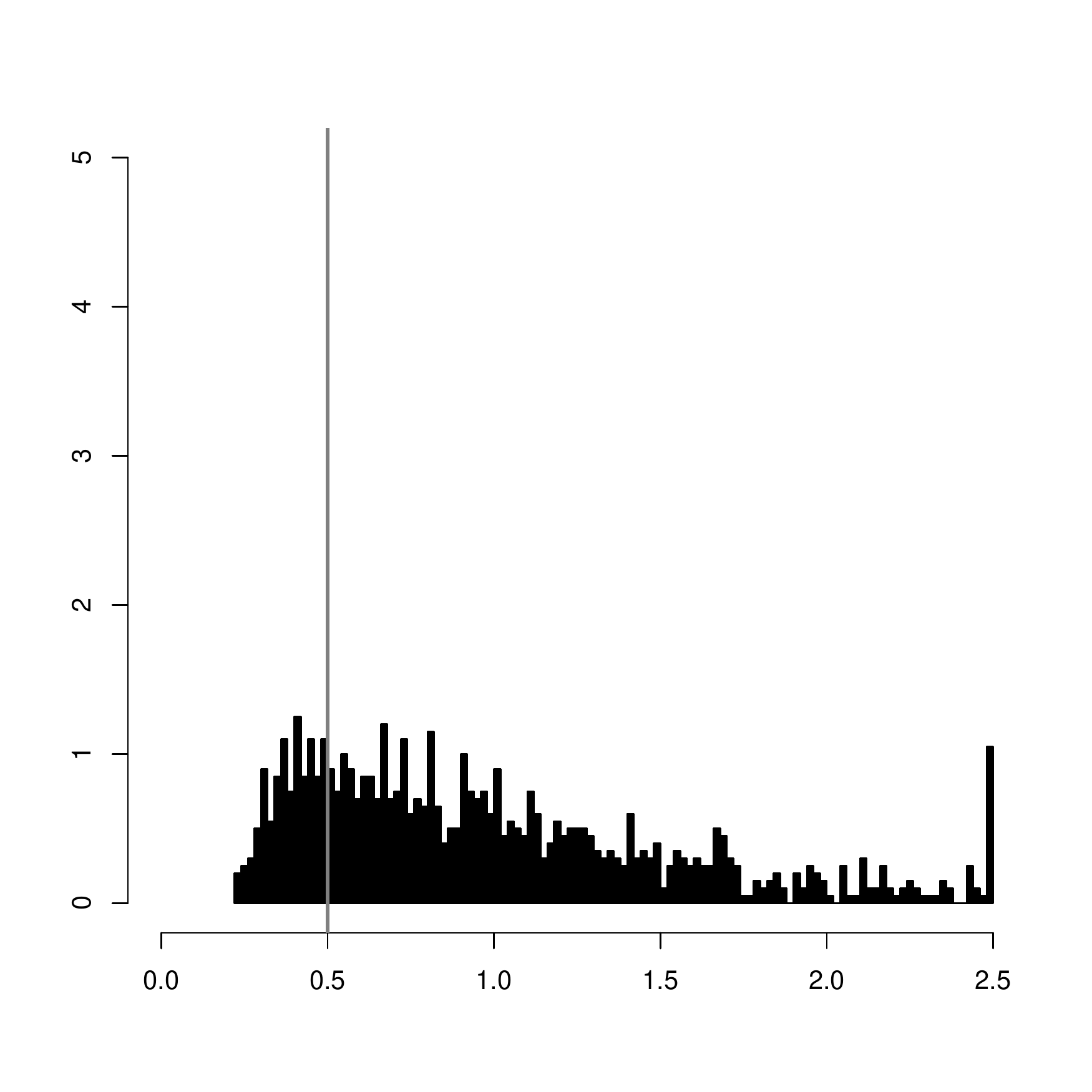}
\includegraphics[width=0.3\textwidth]{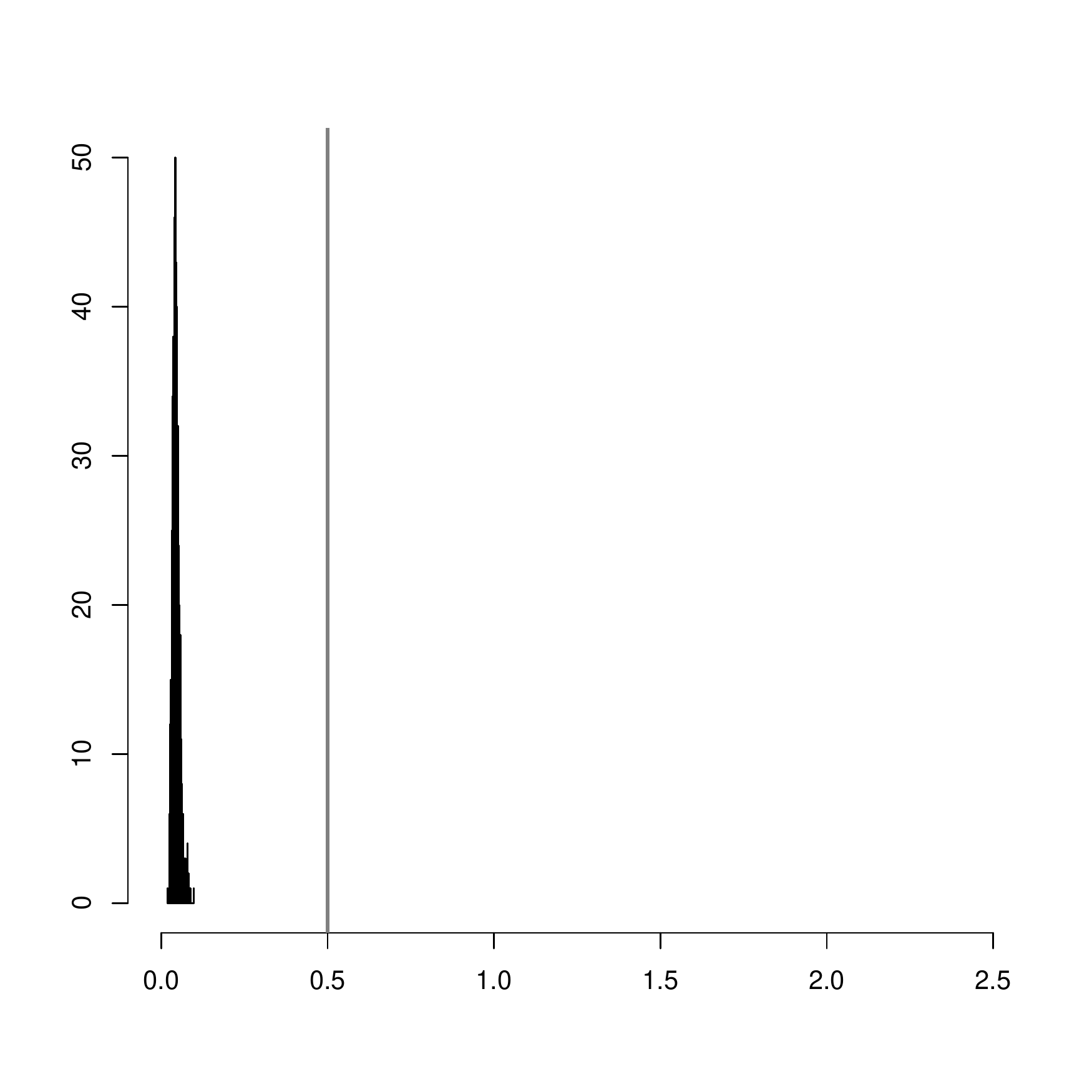}
\caption{Histograms of estimates of the stationary variance 
$\sigma^{2}/(2\alpha)$ of the OU model. 
First column: $m=0$, second column: $m=0.5$, third column $m=10$,
first row: all individuals have $\theta=5$, second row:
half of individuals have $\theta=-5$, other half $\theta=5$.
In all presented histograms $\alpha=1$, $X_{0}=0$, $\sigma^{2}=1$.
Hence the stationary variance is $\sigma^{2}/(2\alpha)=0.5$.
The vertical line represents the true value.
Note that the histograms are not all on the same scale.
The left and right--most bars are values equal
or more extreme than the appropriate $x$--value.
\label{fighistvy}}
\end{center}
\end{figure}

\begin{figure}[!ht]
\begin{center}
\includegraphics[width=0.3\textwidth]{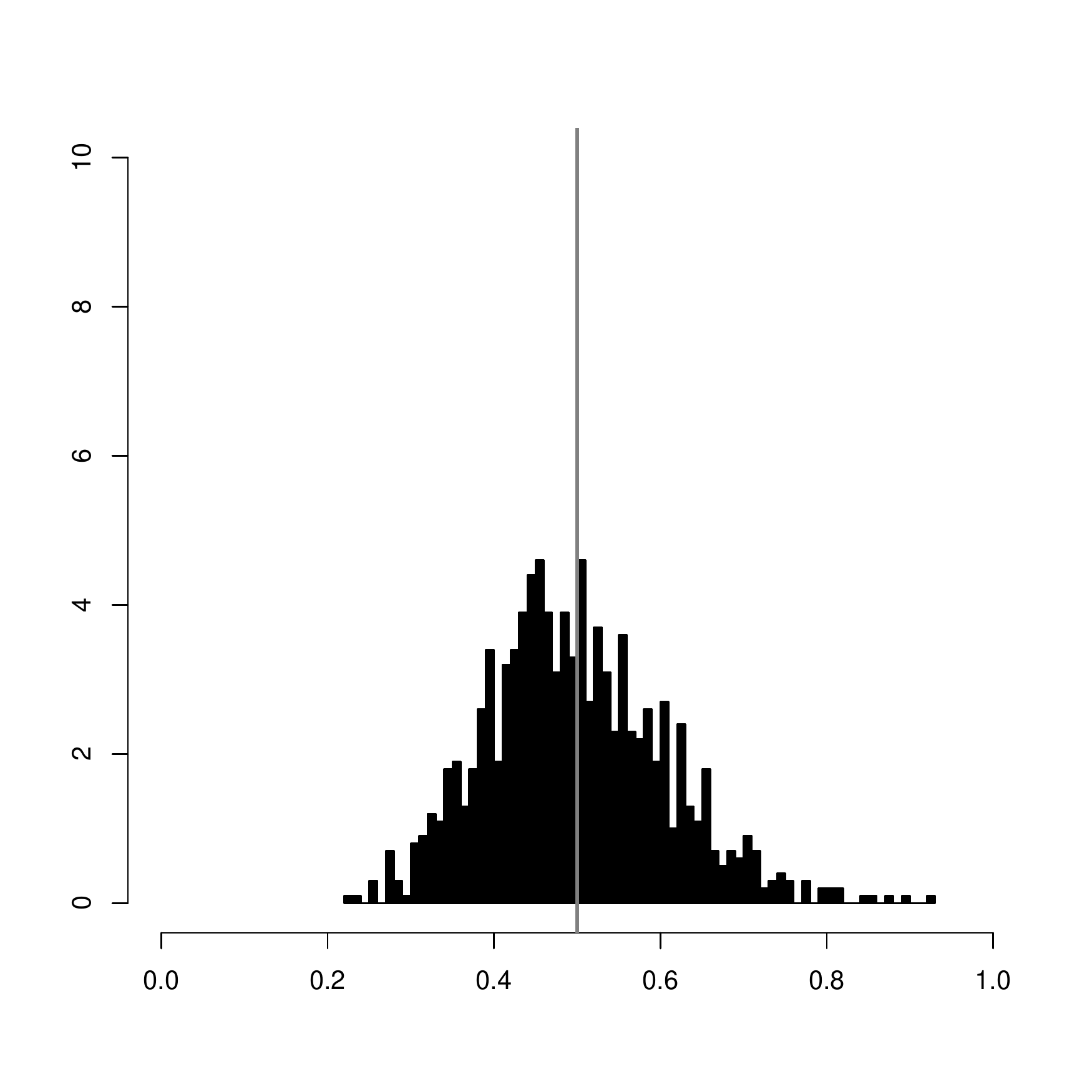}
\includegraphics[width=0.3\textwidth]{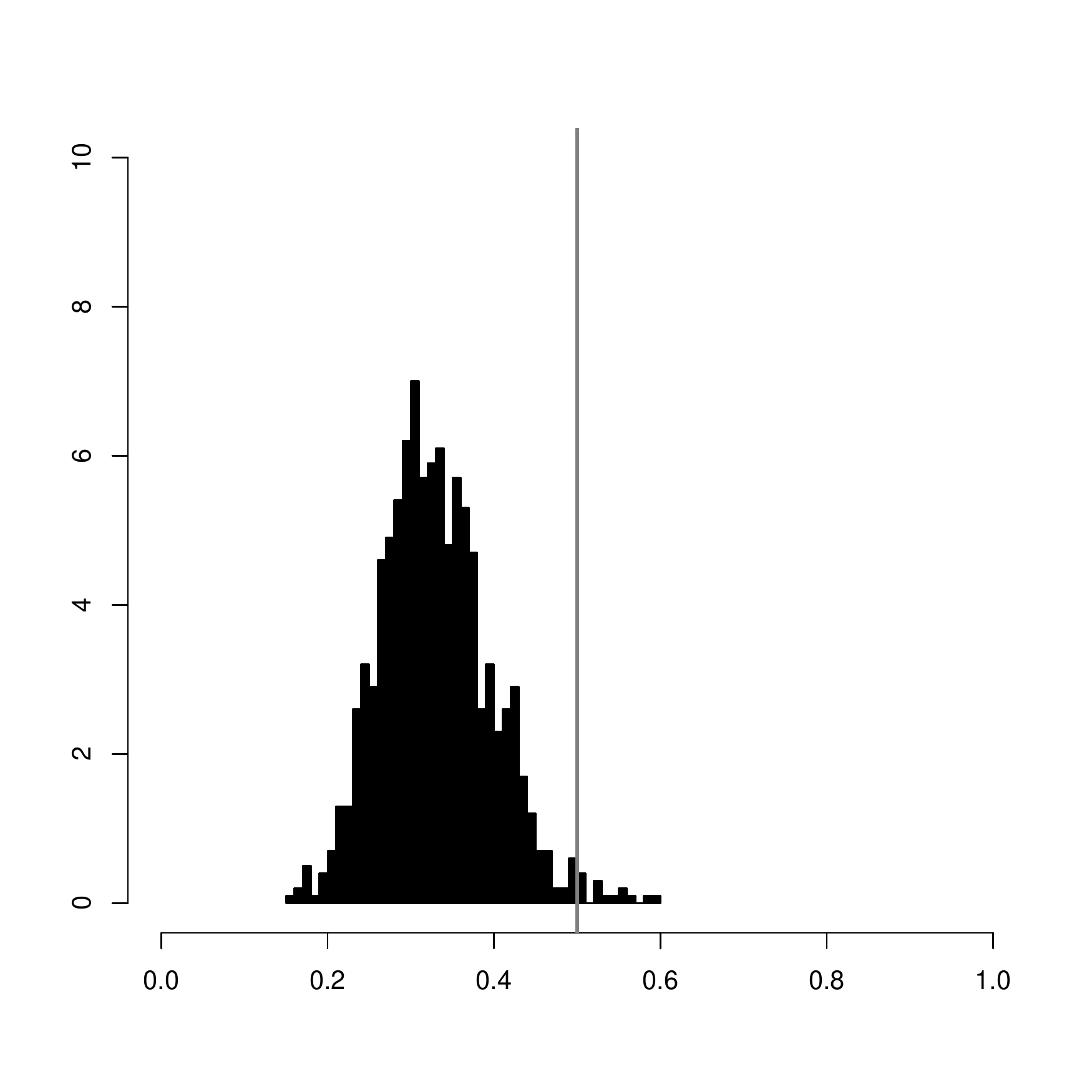}
\includegraphics[width=0.3\textwidth]{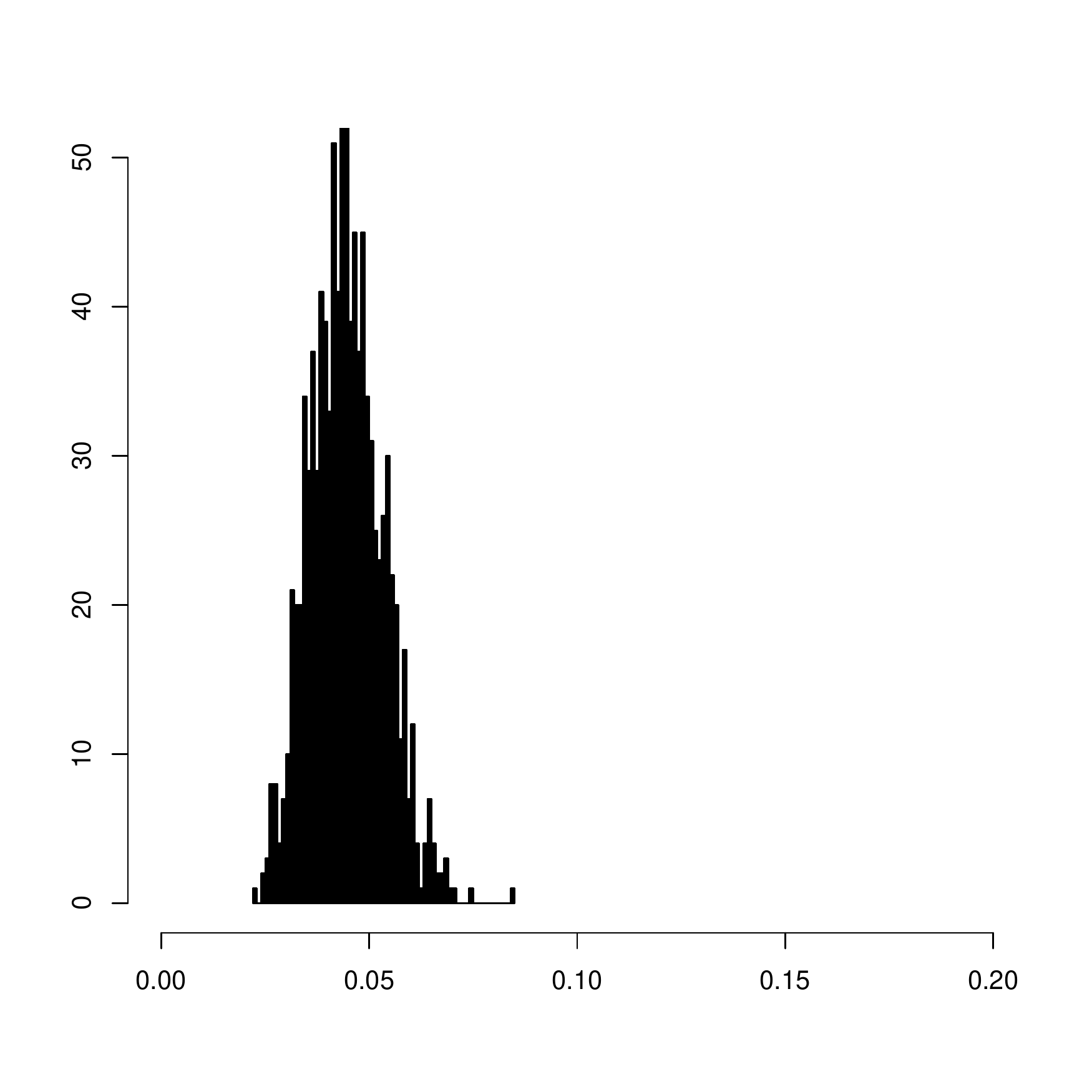} \\
\includegraphics[width=0.3\textwidth]{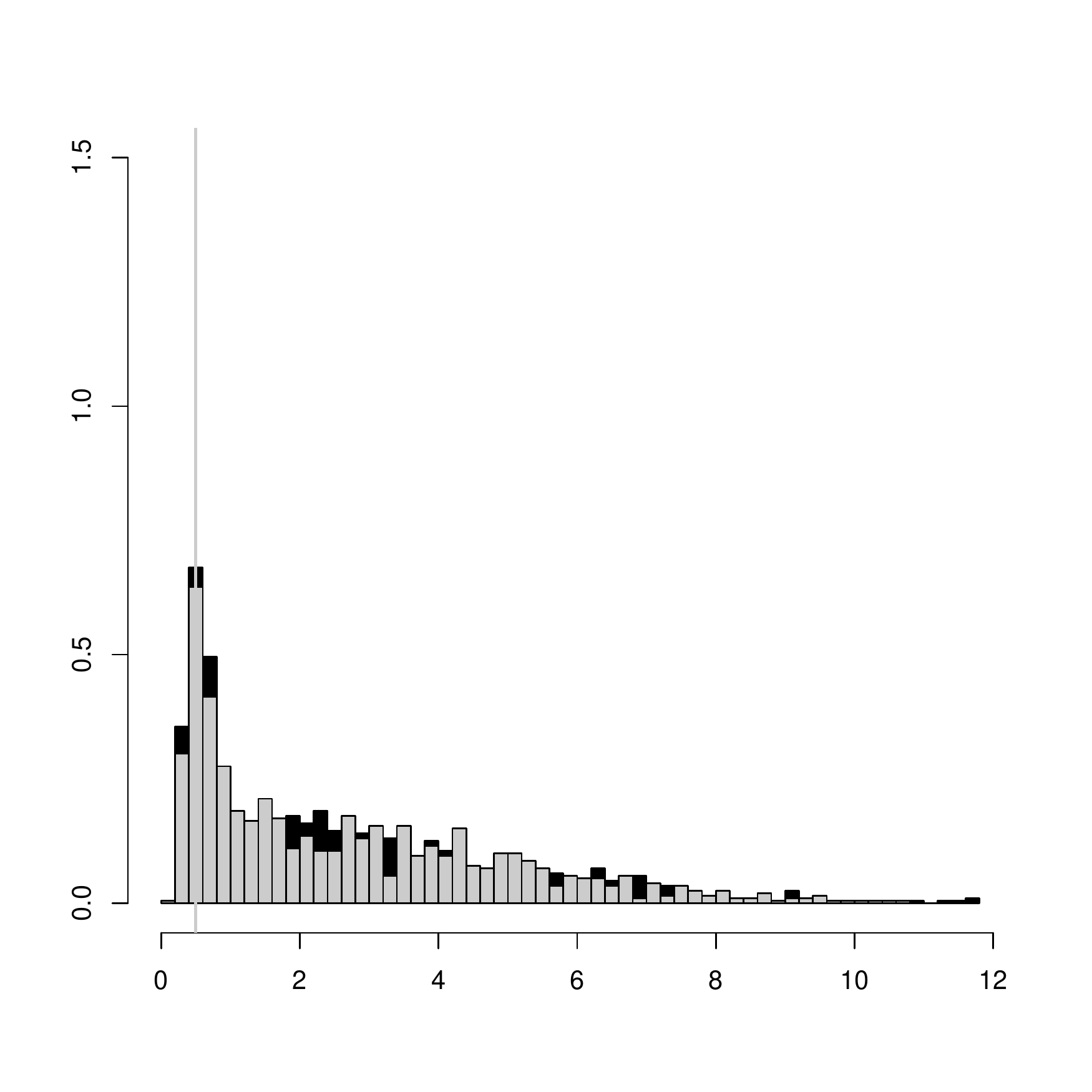}
\includegraphics[width=0.3\textwidth]{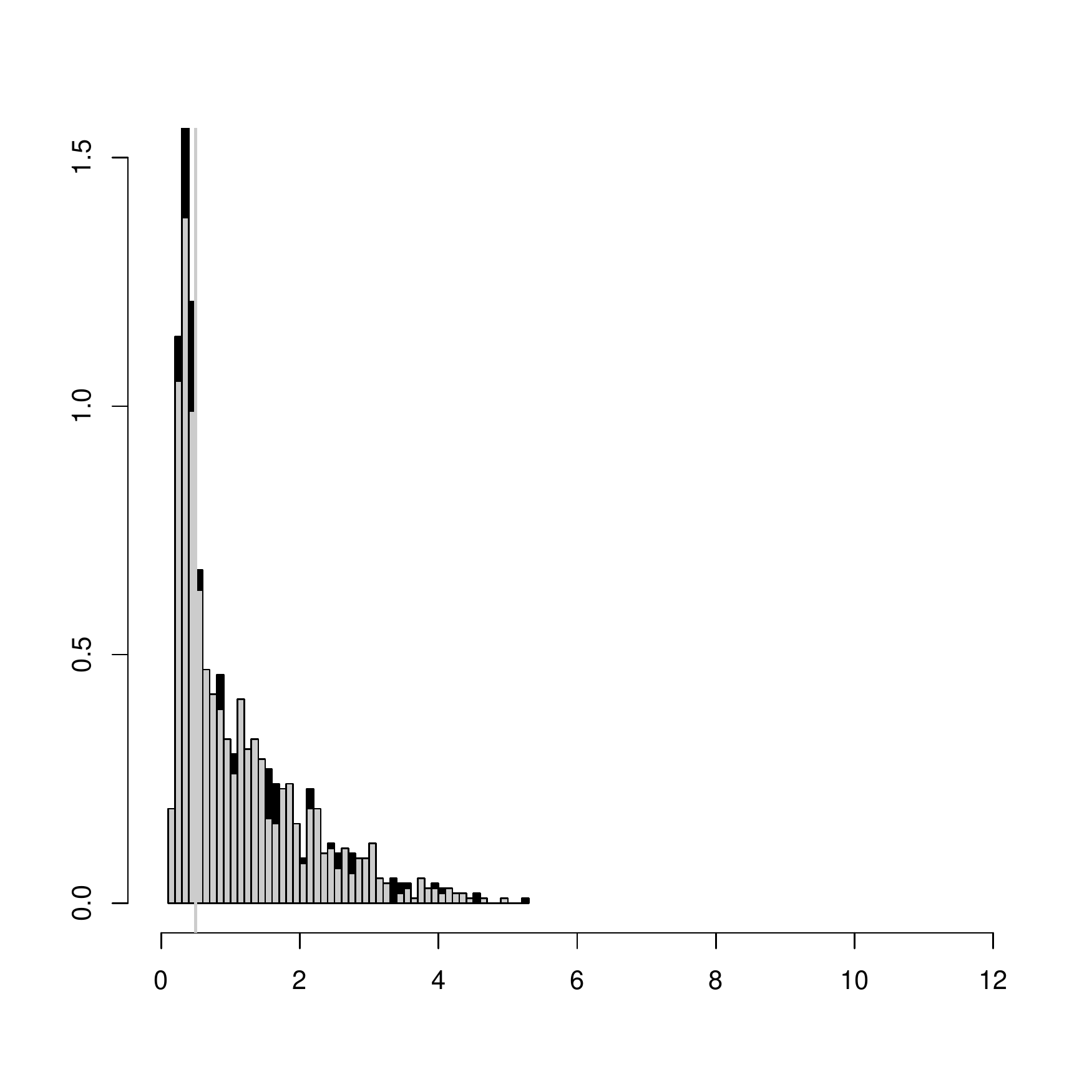}
\includegraphics[width=0.3\textwidth]{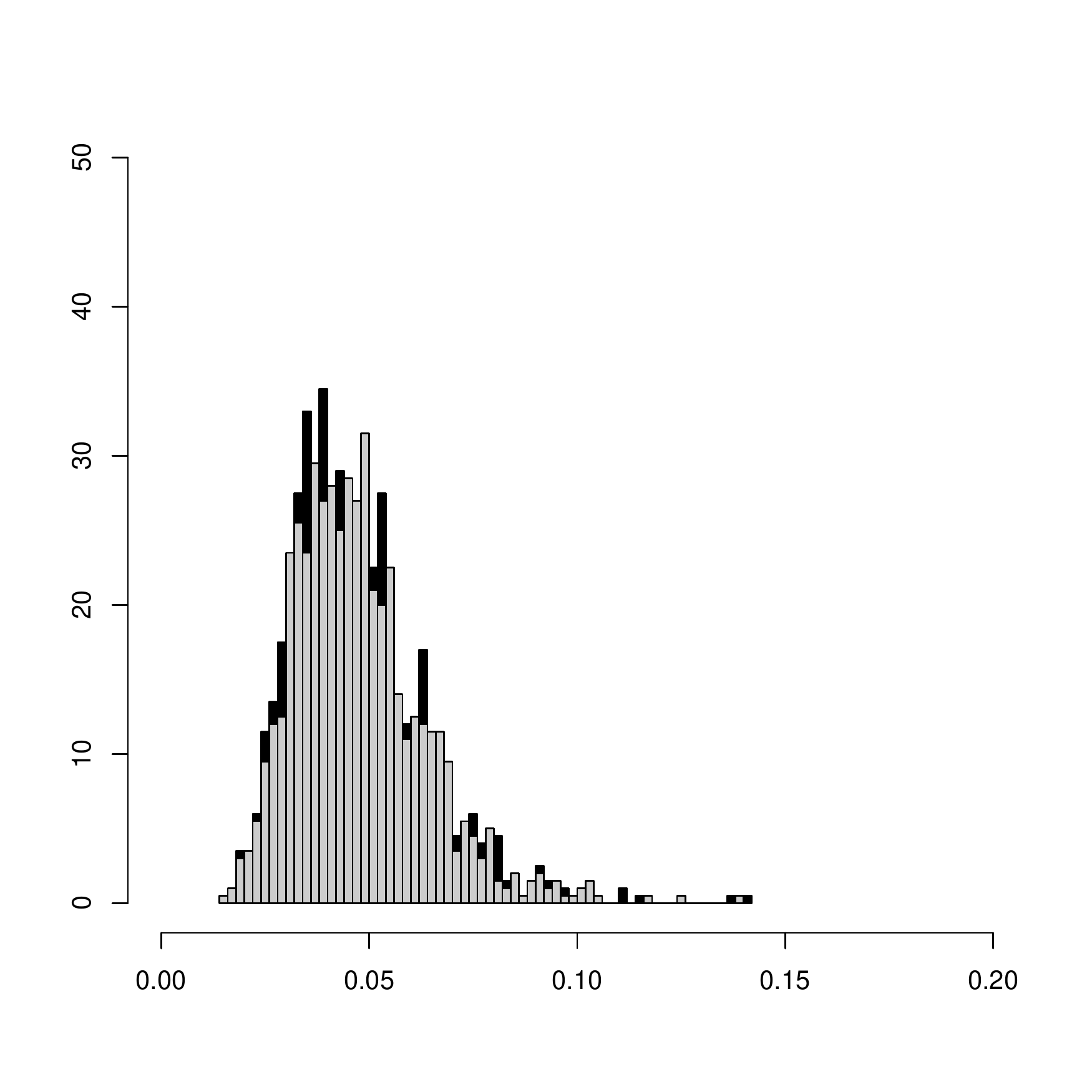}
\caption{Histograms of sample variances of data simulated under
the migration model.
First column: $m=0$, second column: $m=0.5$, third column $m=10$,
first row: all individuals have $\theta=5$, second row:
half of individuals have $\theta=-5$, other half $\theta=5$.
In all presented histograms $\alpha=1$, $X_{0}=0$, $\sigma^{2}=1$.
In the second row we plot separately the sample variances
of $\theta=-5$ individuals (black) and $\theta=5$ individuals (gray).
The vertical line is the OU process stationary variance
$\sigma^{2}/(2\alpha)=0.5$.
Note that the histograms are not all on the same scale.
The left and right--most bars are values equal
or more extreme than the appropriate $x$--value.
\label{fighistdatavar}}
\end{center}
\end{figure}

\begin{figure}[!ht]
\begin{center}
\includegraphics[width=0.3\textwidth]{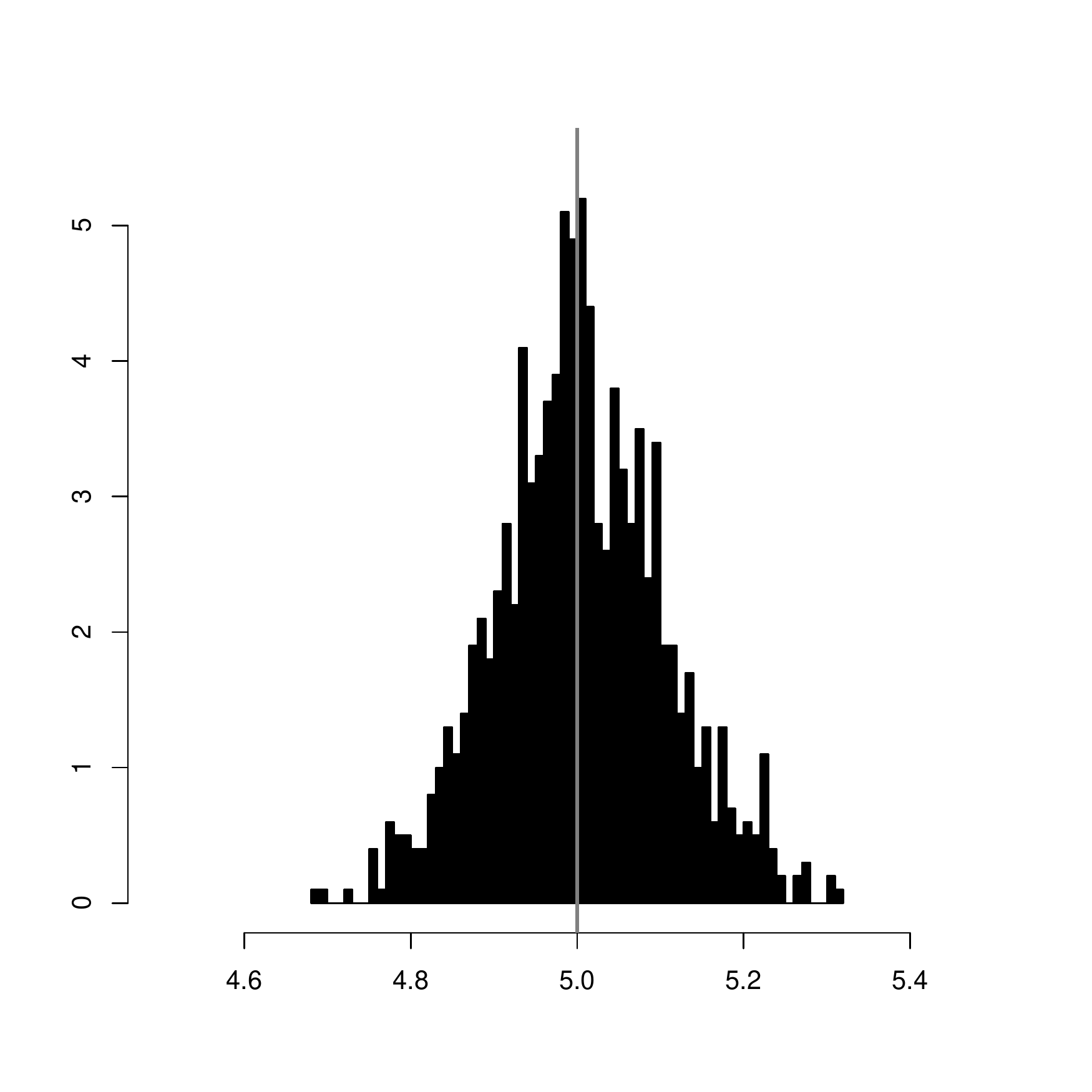}
\includegraphics[width=0.3\textwidth]{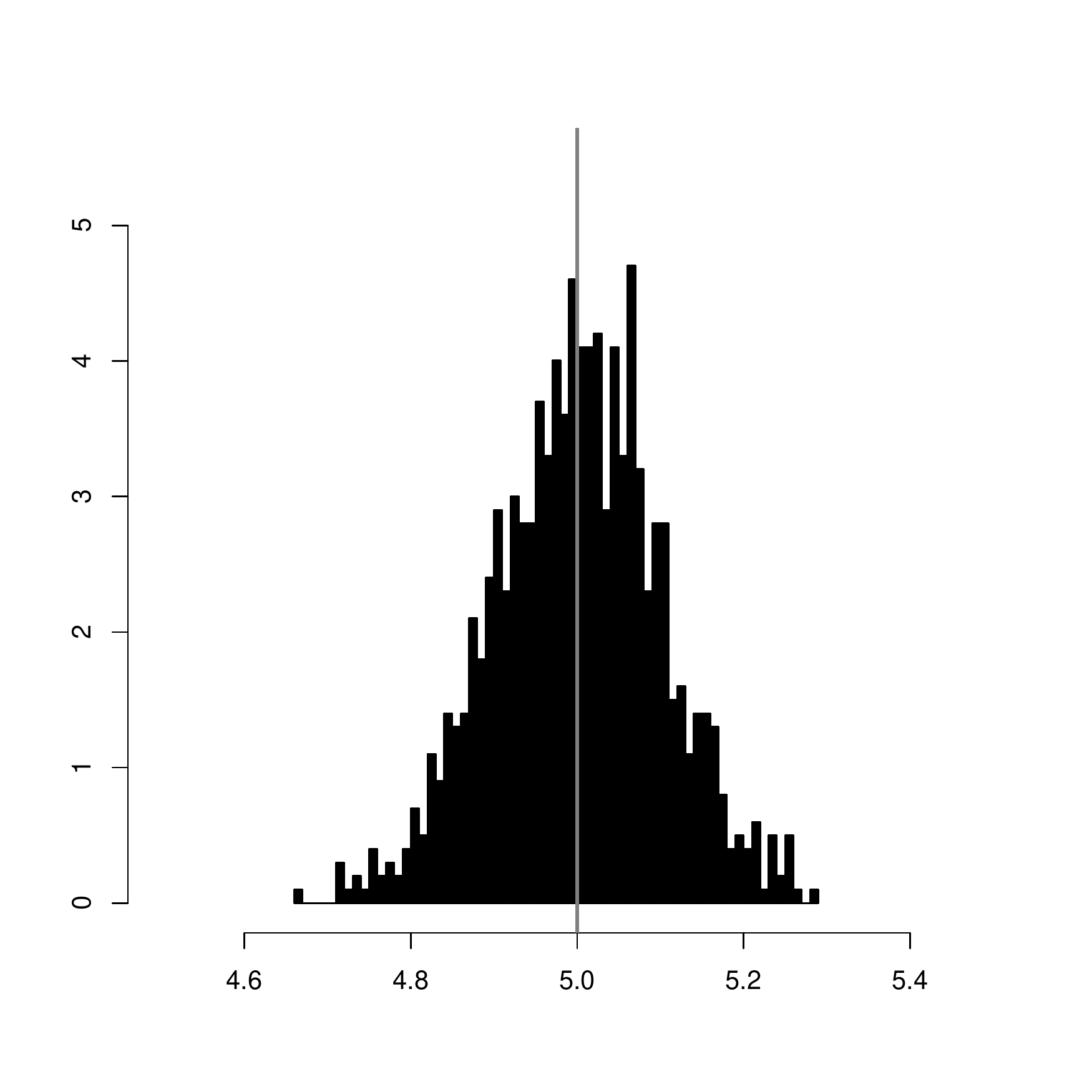}
\includegraphics[width=0.3\textwidth]{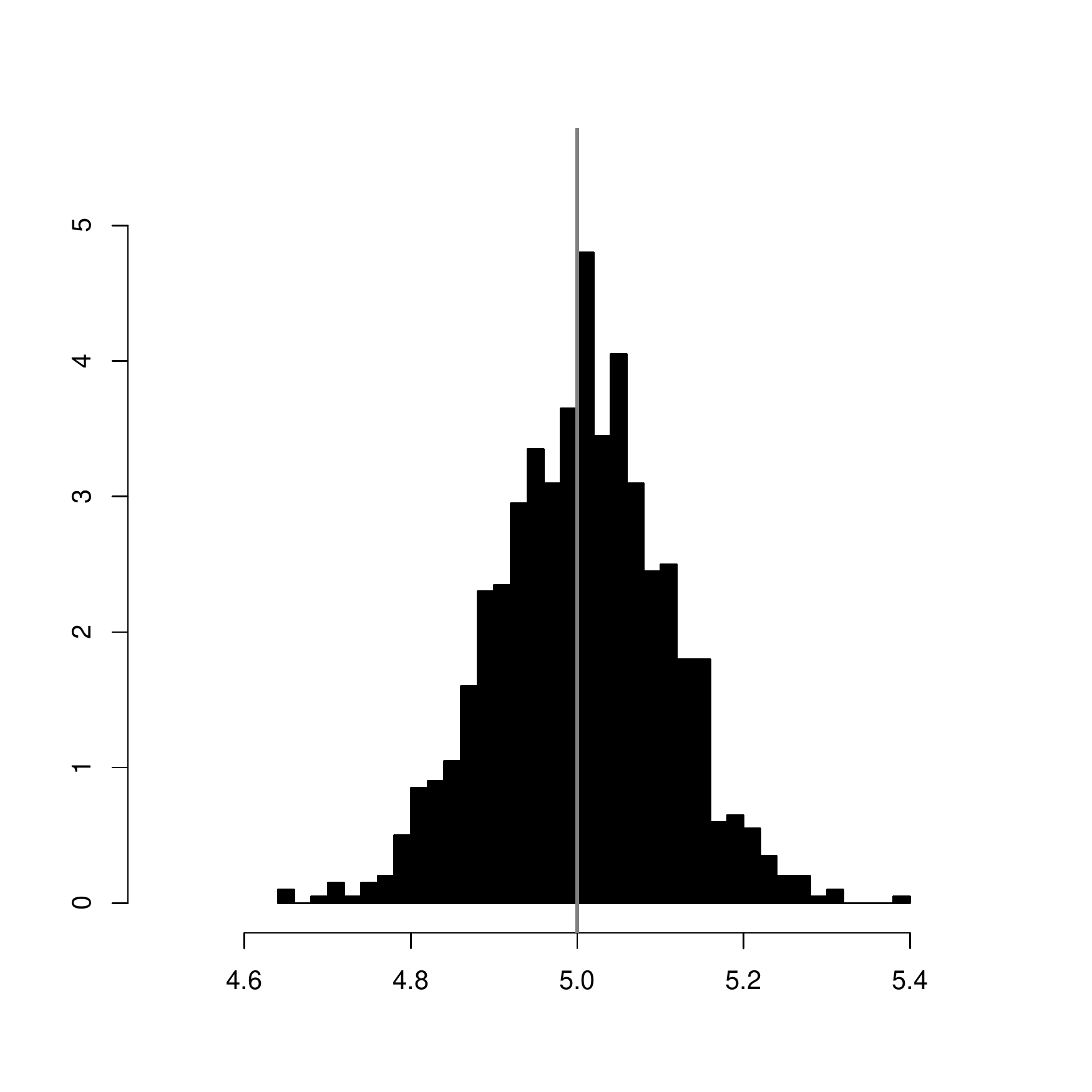} \\
\includegraphics[width=0.3\textwidth]{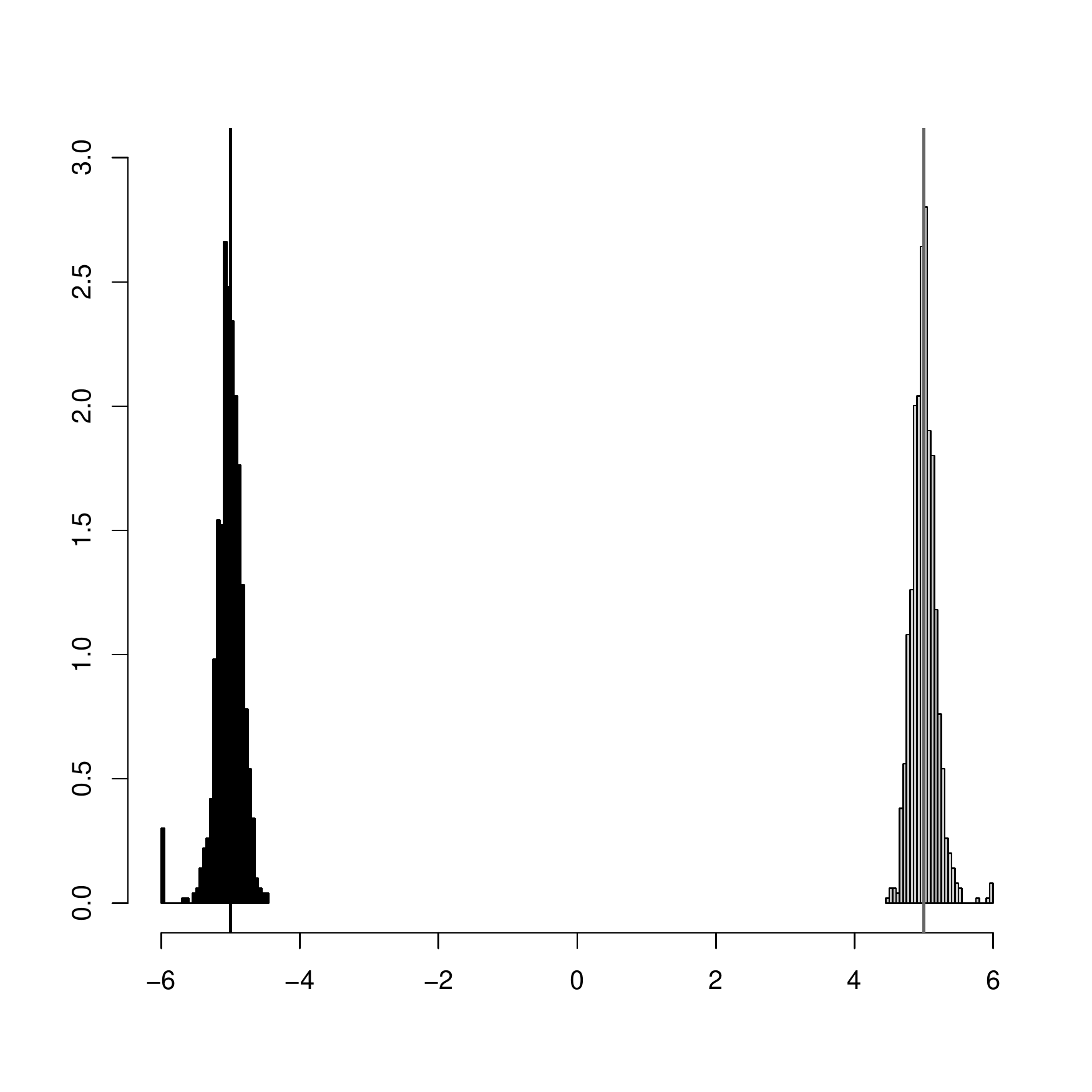}
\includegraphics[width=0.3\textwidth]{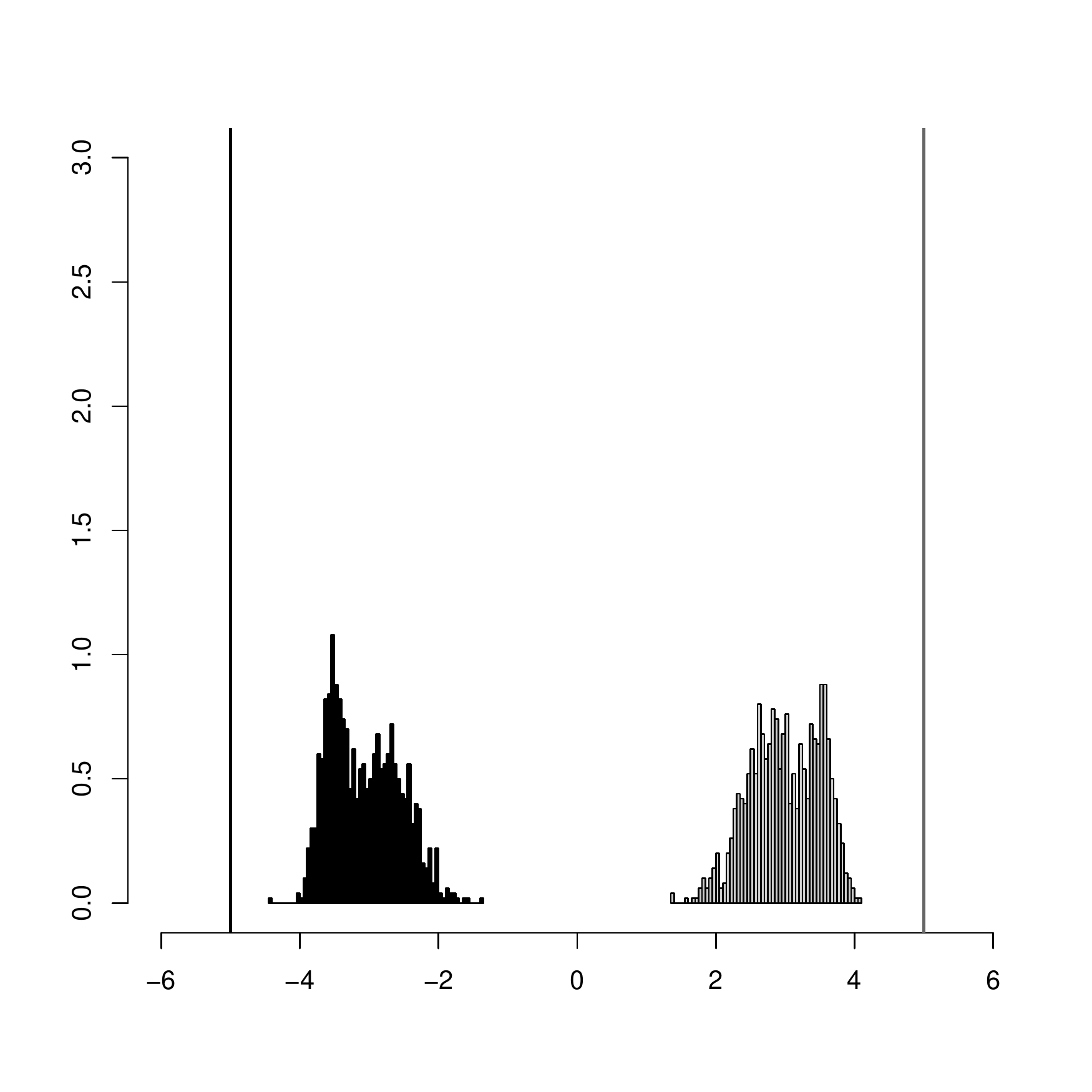}
\includegraphics[width=0.3\textwidth]{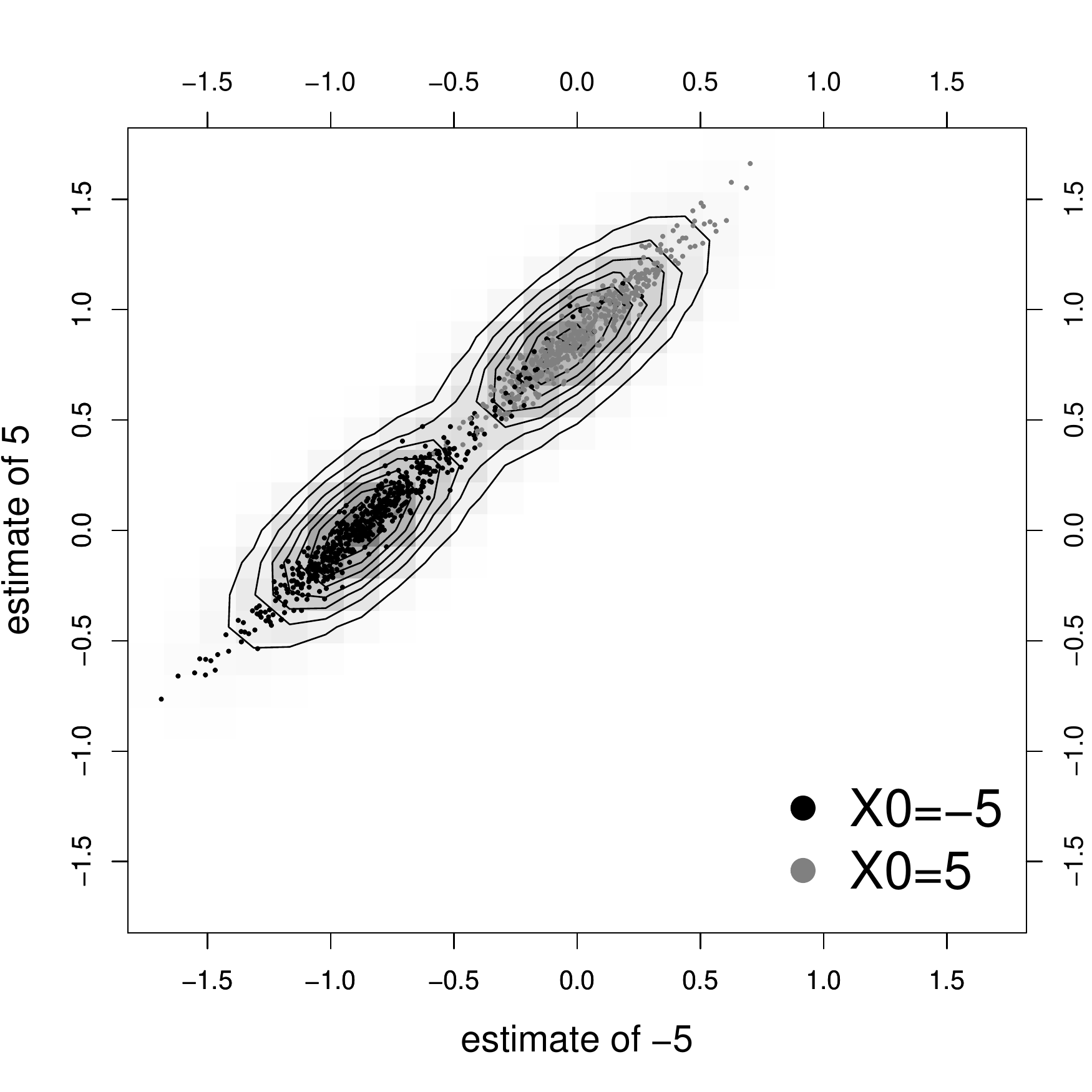}
\caption{Histograms of estimates of $\theta$.
First column: $m=0$, second column: $m=0.5$, third column $m=10$,
first row: all individuals have $\theta=5$, second row:
half of individuals have $\theta=-5$ (black estimates), other half $\theta=5$ (gray estimates).
In all presented histograms $\alpha=1$, $X_{0}=0$, $\sigma^{2}=1$.
In the second row we plot separately the
$\theta=-5$ individuals (black) and $\theta=5$ individuals (gray).
The vertical lines represent the true values.
In the case of strong migration (bottom right figure) we do not present the 
histogram, but plot the actually estimated values in each of the runs. 
Note that the histograms in the two rows are not on the same scale.
The left and right--most bars are values equal
or more extreme than the appropriate $x$--value.
\label{fighisttheta}}
\end{center}
\end{figure}

The simulations presented in Figs. \ref{figtrajsim} and
\ref{figpopsim} suggest that not taking into consideration migration
can seriously bias parameter estimation.  We did a simulation study to
test this and the resulting histograms are presented in
Figs. \ref{fighistvy}, \ref{fighistdatavar}, \ref{fighisttheta},
\ref{fighistsigma2a}, \ref{fighistalpha} and Tab. \ref{tabOUrest}. For
each parameter combination we simulated $1000$ $50$--tip Yule trees
with speciation rate $0.1$ (actually with speciation rate $1$ and
rescaled all branch lengths by $10$) using TreeSim.  On each tree we
simulated a trait process under the law defined by the parameter
combination. If $m=0$ (i.e. usual OU process) we used the mvSLOUCH
package and when $m\neq 0$ R code that drew trait values from the law
described in this paper. Then, using mvSLOUCH we estimated parameters
of the OU process based on the simulated data. The mvSLOUCH package
does not take migration into account, hence we can observe the
consequence of ignoring migration effects.  When there were two
$\theta$ values, half $(25)$ the tip species had one value and half
the other. We tried to cluster this as far as possible (the trees are
random). Nodes with indices $1$--$25$ (in phylo format) had one value
and those with indices $26$--$50$ the other. We drew (with equal
probability) which half had which value of $\theta$. Afterwards the
regimes were painted back on the tree using the Fitch algorithm
\citep{MFit1971,DSwoWMad1987} implemented in the fitch.mvsl() function
in mvSLOUCH.

The results of the simulation study show that the consequences of
ignoring migration in a phylogenetic comparative study are as those
indicated in Fig. \ref{figpopsim} (populations evolving under an OU
with migration). Estimates of $\alpha$ (Fig. \ref{fighistalpha}), the
selection parameter are very heavily inflated. This is as the positive
$m$ parameter brings the observations closer together (compare
Figs. \ref{figpopsim}, \ref{fighistdatavar}) --- indicating a stronger
selection effect, but to a different optimum when there are multiple
optima.  We can see that with a single optimum it is difficult to
estimate $\alpha$.  This is not surprising, \citet{CCreMButAKin2015}
observed the same issue.  However, we can see that $\alpha$ is much
easier to estimate when there are multiple optima
(Fig. \ref{fighistalpha}, bottom left). The intuition is that when
there is a switching of optima, especially near the tips, one can
observe how quickly the trait moves from one optimum to the other ---
giving information on $\alpha$. A similar situation is for estimation
of $\sigma^{2}$ (Fig. \ref{fighistsigma2a}).  The presence of
migration makes it more difficult to estimate and multiple optima make
it easier.

In OU phylogenetic models it is often easier to estimate the
stationary variance of the process, $v_{y}:=\sigma^{2}/(2\alpha)$. As
with strong selection the variance of the contemporary sample should
be essentially given by this formula \citep[cf.][]{KBarSSag2015}.
With no migration (Fig. \ref{fighistvy}, left column) the parameter
$v_{y}$ is well estimated. However when we increase migration the
estimate of this value can differ significantly. The maximum
likelihood estimation procedure tries to fit such
$(\hat{\alpha},\hat{\sigma^{2}})$ values so that $\hat{v}_{y}$ will
correspond to the variance of the contemporary sample. As migration
significantly decreases the variability around the mean (see
Fig. \ref{fighistdatavar}) the estimation procedure (unaware of
migration) will fixate in a parameter region where $v_{y}$ is small.

Estimating the value of the optimum parameter, $\theta$, was one the
original motivations behind phylogenetic comparative methods. This
parameter (or vector of parameters if there are multiple optima) is
usually the easiest to estimate
\citep[][]{KBaretal2012,KBarSSag2015,CCreMButAKin2015}.  In fact when
there is a single optimum, migration just makes it easier to estimate
it: migration decreases sample variance and increases the speed of
convergence of the mean.  However, when there are multiple optima
(Fig. \ref{fighisttheta}, bottom row) ignoring migration makes it
difficult to identify the different optima.  The averages of the
subsamples (i.e. those under the different regimes) are much closer
together, as migration mixes the individuals of the different
populations.  With very strong migration (Fig.\ref{fighisttheta},
bottom right) we can actually observe a very interesting effect: the
ancestral and selective regime effects interact with each other.  Each
dot on the scatter plot is the estimated
$(\hat{\theta}_{1},\hat{\theta}_{2})$ pair. When the ancestor was at
$\theta_{i}$ then the value of $\theta_{i'}$, $i'\neq i$ was estimated
around $0$ while in $\hat{\theta}_{i}$ we may observe a lagging
ancestral effect. If there were no ancestral effects we would expect
$\theta_{i}\approx \theta_{i'}\approx 0$ (Fig. \ref{figpopsim}) as
migration is symmetric between all species. However, when there are
ancestral effects they will show up in both estimates. One will have
an interplay between the force towards the average of
$(\theta_{i}+\theta_{i'})/2$, ancestral effects and weak (compared to
migration) selection effects.

In summary, ignoring migration can have a large impact on the
identification of the strength of selection and the magnitude of
disruptive effects.  If populations are mixing and are otherwise in
separate selective regimes it is impossible to estimate the optima of
these regimes correctly.  However, if migration is only between
individuals inside the same selective regime, then identification of
its optimum value is significantly easier.

\begin{table}
\centering
{\small
\begin{tabular}{c|cccccc}
Setup & Parameter & Mean & Variance & MSE & rbias & rMSE \\
\hline
1 & $\alpha=1$ & $5.154$ & $557.168$ & $ 574.425$ & $ 4.154$ & $ 574.425$\\
& $\sigma^{2}=1$ & $4.421$ & $ 260.531$ & $ 272.235$ & $ 3.421$ & $ 272.235$ \\
& $\sigma^{2}/(2\alpha)=0.5$ & $0.486$ & $ 0.011$ & $ 0.011$ & $ -0.028$ & $ 0.022$\\
& $\theta=5$ & $5.004$ & $ 0.01$ & $ 0.01$ & $ 0.001$ & $ 0.002$\\
& $\hat{\sigma^{2}}$ & $0.497$ & $0.012$ & --- & --- & ---\\
\hline
2 & $\alpha=1$ & $101.076$ & $ 258167.8$ & $ 268182.9$ & $ 100.076$ & $ 268182.9$\\
& $\sigma^{2}=1$ & $64.541$ & $ 103060.5$ & $ 107097.9$ & $ 63.541$ & $ 107097.9$\\
& $\sigma^{2}/(2\alpha)=0.5$ & $0.324$ & $ 0.004$ & $ 0.035$ & $ -0.352$ & $ 0.071$ \\
& $\theta=5$ & $5.001$ & $ 0.01$ & $ 0.01$ & $ 0$ & $ 0.002$\\
& $\hat{\sigma^{2}}$ & $0.33$ & $ 0.004$ & --- & --- & ---\\
\hline
3 & $\alpha=1$ & $141.391$ & $ 841122.1$ & $ 860831.8$ & $ 140.391$ & $ 860831.8$\\
& $\sigma^{2}=1$ & $11.674$ & $ 4906.333$ & $ 5020.262$ & $ 10.674$ & $ 5020.262$\\
& $\sigma^{2}/(2\alpha)=0.5$ & $0.044$ & $ 0 0.208$ & $ -0.912$ & $ 0.416$\\
& $\theta=5$ & $5.004$ & $ 0.011$ & $ 0.011$ & $ 0.001$ & $ 0.002$\\
& $\hat{\sigma^{2}}$ & $0.045$ & $ 0$ & --- & --- & ---\\
\hline
4 & $\alpha=1$ & $0.99$ & $ 0.022$ & $ 0.022$ & $ -0.01$ & $ 0.022$\\
& $\sigma^{2}=1$ & $0.939$ & $ 0.049$ & $ 0.053$ & $ -0.061$ & $ 0.053$\\
& $\sigma^{2}/(2\alpha)=0.5$ & $0.523$ & $ 0.452$ & $ 0.452$ & $ 0.046$ & $ 0.905$ \\
& $\theta_{1}=-5$ & $-5.22$ & $ 7.223$ & $ 7.271$ & $ 0.044$ & $ -1.454$ \\
& $\theta_{2}=5$ & $5.02$ & $ 0.214$ & $ 0.215$ & $ 0.004$ & $ 0.043$ \\
& $\hat{\sigma^{2}_{1}}$ & $2.622$ & $ 4.934$ & --- & --- & ---\\
& $\hat{\sigma^{2}_{2}}$ & $2.674$ & $ 4.928$ & --- & --- & ---\\
\hline
5 & $\alpha=1$ & $998.951$ & $ 5.915\cdot 10^{8}$ & $ 5.925 \cdot 10^{8}$ & $ 997.951$ & $ 5.925\cdot 10^{8}$\\
& $\sigma^{2}=1$ & $2167.767$ & $ 2.199\cdot 10^{9}$ & $ 2.203\cdot 10^{9}$ & $2166.767$ & $ 2.203\cdot 10^{9}$\\
& $\sigma^{2}/(2\alpha)=0.5$ & $0.994 $ & $0.338$ & $ 0.582$ & $ 0.988$ & $ 1.164$ \\
& $\theta_{1}=-5$ & $-3.061$ & $ 0.253$ & $ 4.011$ & $ -0.388$ & $ -0.802$ \\
& $\theta_{2}=5$ & $2.996$ & $ 0.257$ & $ 4.271$ & $ -0.401$ & $ 0.854$  \\
& $\hat{\sigma^{2}_{1}}$ & $1.036$ & $ 0.82$ & --- & --- & ---\\
& $\hat{\sigma^{2}_{2}}$ & $1.099$ & $ 0.831$ & --- & --- & ---\\ 
\hline
6 & $\alpha=1$ & $62750.66$ & $ 3.244\cdot 10^{12}$ & $ 3.248\cdot 10^{12}$ & $ 62749.66$ & $ 3.248\cdot 10^{12}$\\
& $\sigma^{2}=1$ & $6703.932$ & $ 3.814\cdot 10^{10}$ & $ 3.819\cdot 10^{10}$ & $ 6702.932$ & $ 3.819 \cdot 10^{10}$ \\
& $\sigma^{2}/(2\alpha)=0.5$ & $0.044$ & $ 0$ & $ 0.208$ & $ -0.911$ & $ 0.415$\\
& $\theta_{1}=-5$ & $-0.459$ & $ 0.248$ & $ 20.873$ & $ -0.908$ & $ -4.175$ \\
& $\theta_{2}=5$ & $0.43$ & $ 0.248$ & $ 21.137$ & $ -0.914$ & $ 4.227$ \\
& $\hat{\sigma^{2}_{1}}$ & $0.047$ & $ 0$ & --- & --- & ---\\
& $\hat{\sigma^{2}_{2}}$ & $0.047$ & $ 0$ & --- & --- & ---
\end{tabular}
}
\caption{Summary of simulation results. For each setup and parameter we report the 
mean, variance, mean square error (MSE), relative bias (rbias) and relative MSE (rMSE)
of the estimates. The means are visualized on histograms.
The parameter $\hat{\sigma^{2}}$ is the sample variance,
$\hat{\sigma^{2}_{1}}$ refers to the sample variance of $\theta=-5$ individuals,
$\hat{\sigma^{2}_{2}}$ to $\theta=5$ individuals. 
Migration is set to $m=0$ (setup 1 and 4), $m=0.5$ (setup 2 and 5), or $m=10$ (setup 3 and 6).
\label{tabOUrest}}
\end{table}

It is important to notice that in the modelling setup considered here
the layout of the regimes is known. Only the values of the optima are
considered unknown.  An interesting model would of course be where the
layout is not known but is estimated from the data. However this is a
very complex problem, with possible identifiability issues \citep[see
e.g.][]{TIngDMah2013,MKhaRKriKRohCAne2016}.  Allowing for modelling
the regime layout is a further development step.

\section{Comparison with ecological interaction between species}

So far we built our model by considering migration between species or
populations.  However, the structure of the model allows for a wider
interpretation as that of interaction between species mediated
through their phenotypes, especially ecological interactions. Our
model applies as far as these interactions are described by linear
combinations of the phenotype of the different species, which then
correspond to a particular matrix $\mathbf{H}$. 

For instance, \citet{JDruJClaMManHMor2016} used a special submodel of
Eq. \eqref{eqOUMsde} to describe competition between evolving species
and study \textit{Anoles} morphology.  Their model is the special case
$\alpha=0$, with their $S\equiv m$.  However, as they focused on the
effect of competition, $S<0$ in their analysis. This essentially is a
Brownian motion with migration model (without selection). We can see
that the process does not have a stationary distribution as one of the
eigenvalues of $\mathbf{H}=\alpha=0$. In fact its variance tends to
infinity.  A slight difference between the model of
Eq. \eqref{eqmeanvarOUM} and \citet{JDruJClaMManHMor2016}'s is that
they did not assume that all species interact with each other. Instead
they considered subsets of species interacting with each other. This
can be seen in their Fig. $1$, where different \textit{Anoles} species
interact if they live on the same island. However this setup means
that $\mathbf{M}$ and $\mathbf{H}$ are block--diagonal with different
``$n$'' in different blocks, but same $\alpha$ and $m$. Hence the
eigen decomposition is the same as of ``our'' $\mathbf{H}$ except that
it is done independently for each block.

If we assume on the contrary that $S>0$, the model in
\citet{JDruJClaMManHMor2016} can be interpreted as a mimicry model
where selection favour a common phenotype among different species,
used for example as a warning message against predators
\citep{JMalMJor1999}.  For each species $i$, the equivalent ODE can be
written as:

\be\label{eqMimicry}
\ud X_{i}(t) = -S \left(X_{i}(t) - \frac{1}{n}\sum_{j=1}^{n} X_{j}(t) \right) \ud t
\ee
which corresponds to our equation \eqref{eqOUMode} with $\alpha=0$ and
an $\mathbf{M}$ matrix with elements $m_{i,i}=S(1-\frac{1}{n})$ and
$m_{i,j}=-\frac{S}{n}$. We remark that this can be generalized to a
weighted mean of phenotypes, hence species-dependent values of $m_{i,j}$, if
the different species do not contribute equally to the definition of
the optimum (for example as a function of their abundance).

As for competition, this mimicry process does not have a stationary
distribution as one of the eigenvalues is zero. However, contrary to
the situation in \citet{JDruJClaMManHMor2016}, we should note that
as the remaining eigenvalues are positive in this case, there is a
stationary component in the mimicry process.  If we condition on one
species then the conditional distribution of the remaining ones will
tend to stationarity \citep[compare with the mvSLOUCH
model][]{KBaretal2012}. This can be biologically interpreted that if
one \emph{arbitrarily} chooses one species then the remaining ones
will ``shadow'' it with stable fluctuations. This one species defines
a randomly changing primary optimum \citep{THanJPieSOrz2008} for the
others.  It is important to notice that this does not imply that the
remaining species will be in any way more similar to each other.
Rather it implies that the deviations of these species from the
\emph{arbitrarily} chosen focal species will be exhibiting the same
stochastic behaviour.  The mimicry trait can randomly diverge but the
variation among species for this trait will reach a stationary
distribution. Biologically what is selected for is not the absolute
value of the trait but the fact that all species are sufficiently
similar.

Our model generalizes these special cases (where $\alpha=0$) to more
realistic conditions where there is stabilizing selection around fixed
optima in addition to selection (stabilizing or disruptive) due to
species interactions. Moreover, both migration and ecological
interaction effects could be encoded in the $\mathbf{H}$ matrix,
although the two components would not be necessarily identifiable
without additional biological information. Our model thus provides a
general framework to study the evolution of traits under selection,
drift and species interactions \textit{sensu lato}.

Finally, we also note that \citet{JDruJClaMManHMor2016} numerically
solve a system of ODEs, their Eqs. ($3$a) and ($3$b), in order to
obtain the mean and covariance structure of the system and in effect
are able to calculate the likelihood. However the solution of their
ODE system is exactly our Eq. \eqref{eqmeanvarOUM}. Hence an
analytical solution is possible instead of a numerical one.  As long
as $\mathbf{M}$ has a nice enough structure (e.g. is block diagonal)
the solution can be written in closed form.  For a general
$\mathbf{M}$ the solution will be in closed form up to the eigen
decomposition of $\mathbf{H}$ (for which there are numerous numerical
libraries).

\section{Discussion}
The Ornstein--Uhlenbeck process has been the model of choice to assess
the relative importance of genetic drift and selection in phenotypic
trait evolution across species.  The availability of large phenotypic
datasets -- in particular, genomewide transcriptomic datasets, that
inherently are equivalent to a myriad of phenotypic traits -- has led
to a renewed interest in the OU process and its application to
phenotypic traits evolution
\citep[e.g.][]{RRohPHarRNie2014,Rohlfs:2015bd, Nourmohammad:2015}.

Until recently species were assumed to evolve independently of each
other in OU models.  However, this is not always realistic and new
models have started to appear where interactions between species are
permitted \citep{SNuiLHar2015, JDruJClaMManHMor2016}.  These studies
were motivated by the presence of ecological interactions between
species, while our motivation in the present work stemmed from the
occurrence of demographic interactions, mainly the exchange of
migrants between species.  Hybridization and introgression are now
well documented in both plants and animals \citep{RAbbetal2013} and
have played an important role in speciation and in the resulting
species trees. As pointed out recently by \citet{SolisLemus:2016kp}
gene trees discordances have often been solely attributed to
incomplete lineage sorting but gene flow is another possibility that
has received less attention. However, ignoring gene flow can lead to
the reconstruction of the wrong evolutionary history. This, of course,
will also apply when reconstruction phenotypic trait evolution and it
is therefore important to integrate gene flow in addition to 
ecological interactions in OU models.  

An interesting finding of our study is that the linking of species through ecological or 
demographic interactions 
lead to the same modified  OU process 
compared to the OU model with species interaction. 
In addition to modeling migration, our model also extend previous models of species interaction. 
Moreover, it also has implications 
for our ability to estimate the importance of stabilizing selection. 
Our simulations showed that ignoring migration when it is actually present can lead to  
severely overestimate selection. Below we discuss some consequences and possible extensions 
of our results.

The analytical derivations and the simulation study presented here show the 
(intuitively obvious) effect of gene flow. The gene flow between species clearly leads them 
to become more similar. The means of the different species tend to a weighted average of the 
optima of individual species and the variance of the species decays very quickly. 
Furthermore it is intrinsically difficult to distinguish between the effect of the pull 
towards the environmental optimum due to selection for a given species from the push effect 
of the interactions with different species. Hence, analyzing data where gene flow does occur 
between species with a OU process without migration estimation procedure could naturally lead 
to the conclusion that adaptation is rapidly moving to an optimum that is common for all species 
and the inference of evidence of convergent evolution. For instance, the program SURFACE that 
explicitly aims at identifying cases of convergent evolution by fitting OU models does not 
incorporate gene flow in the model \citep{Ingram:2013cm}. Even other approaches to assess 
convergence in phenotypic evolution do not explicitly incorporate gene flow or refer to 
it as a possible contributing factor \citep{Moen:2016gm}. Although, the observation that 
convergence, for instance in \textit{Anolis} is spectacular at a regional scale but is not 
observed at a larger scale would fit well with an effect of gene flow on convergence pattern 
\citep[references in][]{Moen:2016gm}. In many cases the assumption that gene flow did not occur 
between species will be reasonable, but the evidence in favor of the existence of gene exchange 
with congeners in may species, especially in plants, suggest that care is required when drawing 
inferences based on a OU model that does not consider gene flow. 
It should also be noted that the trees used in the phylogenetic reconstruction are 
often based on the assumption that gene flow is absent and on a limited number of markers, 
often on solely cytoplasmic markers, and that they may therefore simply miss it. 
The availability of genomic data and of multiple individuals within species together with 
an increasing awareness of gene exchange among species and of the limits of the OU model 
\citep{cooper:2016} should, however, alleviate this issue. 

One of the original motivations for introducing phylogenetic
comparative methods was to ``correct'' for the correlations
between observations induced by the evolutionary tree. 
It is important to keep in mind that these correlations are in fact more than a ``nuisance'' 
factor since the dependency structure
allows one to identify the phylogenetic half--lives
\citep{THanJPieSOrz2008}, something which would have
been impossible from an i.i.d. sample. However, if
the correlations are stronger, due to interaction effects,
then the ``tree correction'' is not sufficient. 
Identity through similarity is mistaken for
identity through convergence, resulting in over--estimation
of selection. The rule of thumb is that if very fast
adaptation is observed to very few regimes then one 
should consider if this might not be the result
of past/present migration or interactions effects.

Adaptive migration models should be widely applicable
in evolutionary biology. One immediate application is to include
population structure in phylogenetic comparative studies. The tips 
of the tree are usually species but inside them there may be 
distinct subpopulations that can have diverged significantly if local adaptation is strong 
\citep{Savolainen:2013df}. 
Neither treating them as one \citep[including intra--species variance][]{THanKBar2012}
or separate entities is fully satisfactory. In the first case one ignores 
the sub--population structure and assumes common adaptive regimes for all.
In the second approach the problem is taken care of, but one ignores
the fact that one is looking at a single species. The different sub--populations
may mix, reproduce and generally interact. The migration model allows for correct
treatment of such situations, with the size of $m$ controlling how much flow there is 
between the different sub--populations. 
More generally, the migration matrix could be parameterized using additional information on 
gene flow among diverging populations, typically using molecular markers. 
The OU model with migration is also a generalization of the two-population model proposed by 
\citet{AHenTDayETay2001} to analyze local adaptation. 
First, it generalizes it to any number of populations. 
Second, it allows to include a branching structure and not only a single 
stationary migration matrix structure. It thus provides an interesting basis for further 
local adaptation models.

Beyond the effect of migration on trait evolution, we have showed the equivalence with 
species interaction models and how it can be applied to competition or mimicry. 
We can also consider the use of OU with migration model
to study e.g. host--parasite or plant--pollinator systems
\citep[combining interactions with phylogenies in such systems has been recently considered, e.g.][]{
HMinCHuiJTerSKosKSch2014,TPoiDSto2016}.
With the current phylogenetic comparative methods one would need to treat
the two clades separately, e.g. modelling the evolution of the host and parasite
trait on either of the trees. In a migration setup the two phylogenies
can be included separately, with a very distant time of coalescence. 
Then the appropriate host--parasite pairs interact through the $\mathbf{M}$ matrix.

Currently evolutionary models assume that speciation is an instantaneous
event. This might be the case on long--time scales but
with recent clades the time during which species split could be comparable
with the tree height. Furthermore, phylogenetic tree estimation
software, can result in non--binary splitting. Such multiple 
radiations are commonly interpreted not as true multi--furcations 
but rather as resulting from a lack of information. In both cases the migration framework 
allows one to use the tree in a phylogenetic comparative analysis. 
Around the point of multi--furcation/species splitting then one would model that all involved
lineages interact with each other. 

All of the above setups require the availability of software to estimate
parameters of migration models. Such are most certainly under development.
However, their inference power remains to be assessed.
With a contemporary sample only it is already difficult to make
inference on the value of $\alpha$ in the absence of migration. 
One would not expect the addition of migration to make estimation any easier.
It is a question whether without fossil measurements it will
be possible to separately estimate the effects of $\alpha$ and $m$
or only a combination of them. This immediately leads to the question
whether different regimes optima are estimable or whether we can only estimate their
linear combinations (species averages). There might be some hope 
however, as our simulation study indicated that (in the situation of $m=0$)
with multiple regimes $\alpha$ is well identifiable. Also the linear
transformation of the optima vector $(\vec{\theta})$ depends
only on $\alpha$ and $m$ --- hence if one is able to identify them,
or rather appropriate ratios of them (see Appendix) then adaptive
regimes should be identifiable. All of this, however, requires 
a careful study alongside inference software development.

\section*{Acknowledgments}
KB was supported by the Knut and Alice Wallenbergs Foundation. 
SG was supported by the French CNRS and the Marie Curie IEF Grant ‘SELFADAPT’ 623486. 
ML acknowledges support from the Swedish Research Council: VR grants 2012--04999 and 2015--03797 
to which this work is related. 

\appendix

\section{Worked 
$3$ species example}

\begin{figure}[!ht]
\begin{center}
\includegraphics[width=0.6\textwidth]{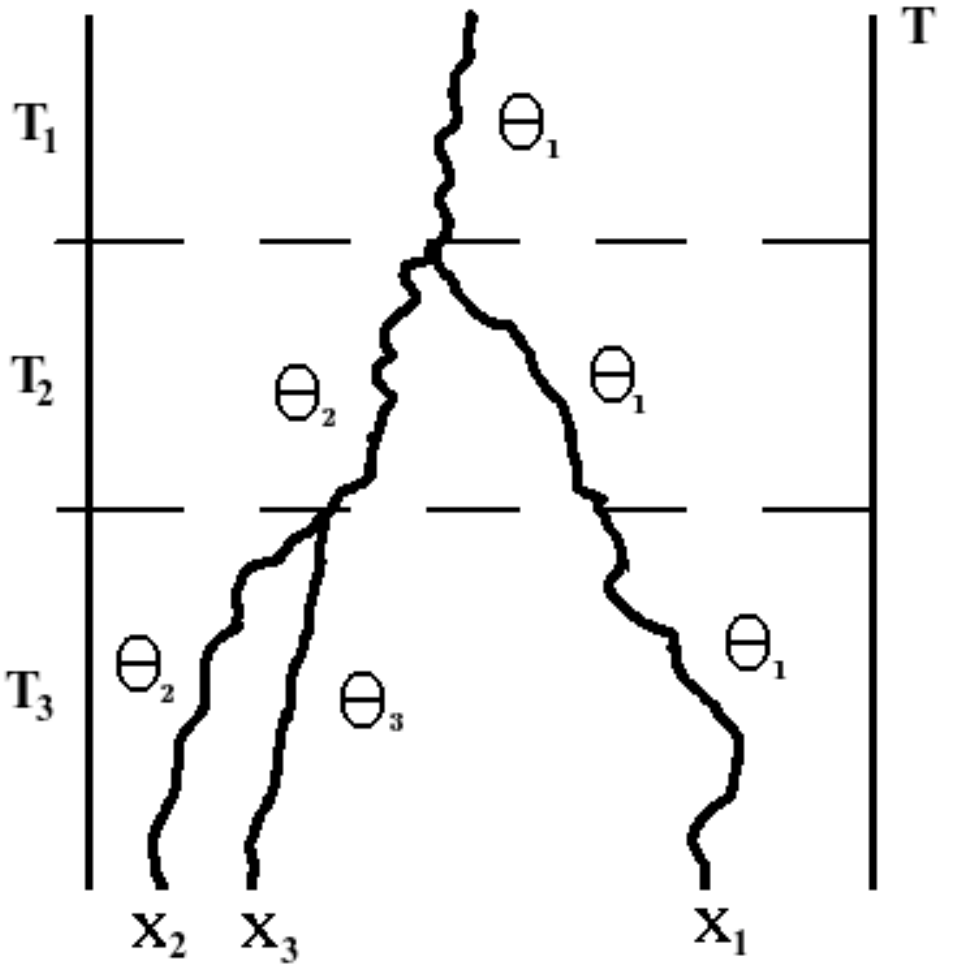}
\caption{A model with constant migration between the populations. It is modelled as a three--dimensional
stochastic process on the time interval $[0,T]$. On the time interval $T_{1}$ all three
coordinates evolve equally, on the time interval $T_{2}$ a split occurs two coordinates
evolve equally while the third evolves in a correlated fashion. Then at the start of interval 
$T_{3}$ the two identical processes split into two populations and gene flow is allowed between all three populations. 
\label{figCMig3}}
\end{center}
\end{figure}

Let us consider the simple model as in Fig. \ref{figCMig3}. 
We will go through all the calculations for this toy $3$ species model in detail. 
Let us write that the length of time interval $i$ is $T_{i}$. Assume that the gene flow rates are the same everywhere
and also let us assume that on each branch we have a constant $\theta$ (see notation in Fig. \ref{figCMig3}). 
We consider the process time interval by time interval. 
\begin{itemize}
\item On time interval $T_{1}$ we have only one species so no migration can take place.
Therefore the system is described by the SDE

$$
\ud X^{1}_{t} = -\alpha(X^{1}_{t} - \theta_{1})\ud t + \sigma \ud B^{1}_{t}
$$
and then at time $T_{1}$ 

$$
X^{1}_{T_{1}} = e^{-\alpha T_{1}}X(0)  +(1-e^{-\alpha T_{1}})\theta_{1} 
+  \sigma e^{-\alpha T_{1}}\int\limits_{0}^{T_{1}} e^{\alpha s}\ud B^{1}_{s}.
$$
\item On the time interval $T_{2}$ we have two lineages between which there is gene flow.
Therefore the system has to be described by a bivariate process

$$
\ud \left[ \begin{array}{c}X^{1}_{t} \\ X^{2}_{t} \end{array} \right]
=
\left(
-\left[
\begin{array}{cc} \alpha + m & -m \\ -m & \alpha +m \end{array}
\right]
\left[ \begin{array}{c}X^{1}_{t} \\ X^{2}_{t} \end{array} \right]
+
\left[ \begin{array}{c}\alpha \theta_{1} \\ \alpha \theta_{2} \end{array} \right]
\right) \ud t
+
\sigma \ud \left[ \begin{array}{c}B^{1}_{t} \\ B^{2}_{t} \end{array} \right].
$$
To solve it we recall the following eigen decomposition of the ``deterministic part'' matrix
$$
\left[
\begin{array}{cc} \alpha + m & -m \\ -m & \alpha +m \end{array}
\right]
=
\left[
\begin{array}{cc} 1 & -1 \\ 1 & 1 \end{array}
\right]
\left[
\begin{array}{cc} \alpha  & 0 \\ 0 & \alpha +2m \end{array}
\right]
\left[
\begin{array}{cc} 1/2 & 1/2 \\ -1/2 & 1/2 \end{array}
\right].
$$
Then we have at the end of the interval
$$
\begin{array}{l}
\left[ \begin{array}{c}X^{1}_{T_{2}} \\ X^{2}_{T_{2}} \end{array} \right]  = 
\left[
\begin{array}{cc} 1 & -1 \\ 1 & 1 \end{array}
\right]
\left(
\left[
\begin{array}{cc} e^{-\alpha T_{2}}  & 0 \\ 0 & e^{-(\alpha +2m)T_{2}} \end{array}
\right]
\left[
\begin{array}{cc} 1/2 & 1/2\\ -1/2 & 1/2 \end{array}
\right]
\left[ \begin{array}{c}X^{1}_{T_{1}} \\ X^{1}_{T_{1}} \end{array} \right]
\right. \\~\\  \left.
+ 
\left[
\begin{array}{cc} 1-e^{-\alpha T_{2}}  & 0 \\ 0 & \frac{\alpha}{\alpha +2m}(1-e^{-(\alpha +2m)T_{2}}) \end{array}
\right]
\left[
\begin{array}{cc} 1/2 & 1/2 \\ -1/2 & 1/2 \end{array}
\right]
\left[ \begin{array}{c}\theta_{1} \\ \theta_{2} \end{array} \right]
\right. \\~\\  \left.
+
\sigma
\left[
\begin{array}{cc} e^{-\alpha T_{2}}  & 0 \\ 0 & e^{-(\alpha +2m)T_{2}} \end{array}
\right]
\int\limits_{0}^{T_{2}}
\left[
\begin{array}{cc} e^{\alpha s}  & 0 \\ 0 & e^{(\alpha +2m)s} \end{array}
\right]
\left[
\begin{array}{cc} 1/2 & 1/2 \\ -1/2 & 1/2 \end{array}
\right]
\ud\left[ \begin{array}{c}B^{1}_{s} \\ B^{2}_{s} \end{array} \right]
\right)
\\~\\  = 

\left[
\begin{array}{cc} 1 & -1 \\ 1 & 1 \end{array}
\right]
\left(
\left[
\begin{array}{cc} e^{-\alpha T_{2}}  & 0 \\ 0 & e^{-(\alpha +2m)T_{2}} \end{array}
\right]
\left[ \begin{array}{c}X^{1}_{T_{1}} \\ 0 \end{array} \right]
\right. \\  \left.
+ \frac{1}{2}
\left[
\begin{array}{cc} 1-e^{-\alpha T_{2}}  & 0 \\ 0 & \frac{\alpha}{\alpha +2m}(1-e^{-(\alpha +2m)T_{2}}) \end{array}
\right]
\left[ \begin{array}{c}\theta_{1} + \theta_{2} \\ \theta_{2} - \theta_{1} \end{array} \right]
\right. \\~\\  \left.
+
\frac{\sigma}{2}
\left[
\begin{array}{cc} e^{-\alpha T_{2}}  & 0 \\ 0 & e^{-(\alpha +2m)T_{2}} \end{array}
\right]
\int\limits_{0}^{T_{2}}
\left[
\begin{array}{cc} e^{\alpha s}  & 0 \\ 0 & e^{(\alpha +2m)s} \end{array}
\right]
\ud\left[ \begin{array}{c}B^{1}_{s} + B^{2}_{s} \\ B^{2}_{s} - B^{1}_{s} \end{array} \right]
\right)
\end{array}
$$

$$
\begin{array}{l}
= 
\left[
\begin{array}{cc} 1 & -1 \\ 1 & 1 \end{array}
\right]
\left(
\left[ \begin{array}{c}e^{-\alpha T_{2}}X^{1}_{T_{1}} \\ 0 \end{array} \right]
+ 
\frac{1}{2}
\left[ \begin{array}{c}(1-e^{-\alpha T_{2}})(\theta_{1} + \theta_{2}) \\ 
\frac{\alpha}{\alpha +2m}(1-e^{-(\alpha +2m)T_{2}}(
\theta_{2} - \theta_{1}) \end{array} \right]
\right. \\~\\  \left.
+
\frac{\sigma}{2}
\left[
\begin{array}{cc} e^{-\alpha T_{2}}  & 0 \\ 0 & e^{-(\alpha +2m)T_{2}} \end{array}
\right]
\left[ \begin{array}{c}\int\limits_{0}^{T_{2}}e^{\alpha s}\ud B^{1}_{s} 
+ \int\limits_{0}^{T_{2}}e^{\alpha s}\ud B^{2}_{s} \\ 
\int\limits_{0}^{T_{2}}e^{(\alpha +2m)s}\ud B^{2}_{s} - 
\int\limits_{0}^{T_{2}}e^{(\alpha +2m)s}\ud B^{1}_{s} \end{array} \right]
\right)
\\~\\
=
\left[ \begin{array}{c}e^{-\alpha T_{2}}X^{1}_{T_{1}} \\ e^{-\alpha T_{2}}X^{1}_{T_{1}} \end{array} \right]
 \\  
+ 
\frac{1}{2}
\left[ \begin{array}{c}
(1-e^{-\alpha T_{2}} + \frac{\alpha}{\alpha +2m}(1-e^{-(\alpha +2m)T_{2}}))
\theta_{1} 
+
(1-e^{-\alpha T_{2}} - \frac{\alpha}{\alpha +2m}(1-e^{-(\alpha +2m)T_{2}}))
\theta_{2}
\\ 
(1-e^{-\alpha T_{2}} - \frac{\alpha}{\alpha +2m}(1-e^{-(\alpha +2m)T_{2}}))\theta_{1} 
+ 
(1-e^{-\alpha T_{2}} + \frac{\alpha}{\alpha +2m}(1-e^{-(\alpha +2m)T_{2}}))\theta_{2}
\end{array} \right]
 \\~\\  
+
\frac{\sigma}{2}
\left[ \begin{array}{c}
\int\limits_{0}^{T_{2}} e^{\alpha (s-T_{2})} + e^{(\alpha +2m)(s-T_{2})} \ud B^{1}_{s} 
+ 
\int\limits_{0}^{T_{2}}e^{\alpha (s-T_{2})} - e^{(\alpha +2m)(s-T_{2})}\ud B^{2}_{s} 
\\ 
\int\limits_{0}^{T_{2}}e^{\alpha (s-T_{2})} - e^{(\alpha +2m)(s-T_{2})}\ud B^{1}_{s} 
+ 
\int\limits_{0}^{T_{2}}e^{\alpha (s-T_{2})} + e^{(\alpha +2m)(s-T_{2})}\ud B^{2}_{s} 
\end{array} \right]
\end{array}
$$

$$
\begin{array}{l}
=
\left[ \begin{array}{c} e^{-(\alpha T_{1}+\alpha T_{2})}X(0) \\  e^{-(\alpha T_{1}+\alpha T_{2})}X(0) \end{array} \right]
+ 
\left[ \begin{array}{c}
(e^{-\alpha T_{2}}-e^{-\alpha (T_{1}+T_{2})})\theta_{1}\\ (e^{-\alpha T_{2}}-e^{-\alpha (T_{1}+T_{2})})\theta_{1}
\end{array} \right]
\\~\\
+
\frac{1}{2}
\left[ \begin{array}{c}
(1-e^{-\alpha T_{2}} + \frac{\alpha}{\alpha +2m}(1-e^{-(\alpha +2m)T_{2}}))\theta_{1} 
+
(1-e^{-\alpha T_{2}} - \frac{\alpha}{\alpha +2m}(1-e^{-(\alpha +2m)T_{2}}))\theta_{2}
\\~\\ 
(1-e^{-\alpha T_{2}} - \frac{\alpha}{\alpha +2m}(1-e^{-(\alpha +2m)T_{2}}))\theta_{1} 
+ 
(1-e^{-\alpha T_{2}} + \frac{\alpha}{\alpha +2m}(1-e^{-(\alpha +2m)T_{2}}))\theta_{2}
\end{array} \right]
 \\~\\  
+
\sigma 
\left[ \begin{array}{c}
e^{-\alpha (T_{1}+T_{2})}\int\limits_{0}^{T_{1}} e^{\alpha s}\ud B^{1}_{s}\\
e^{-\alpha (T_{1}+T_{2})}\int\limits_{0}^{T_{1}} e^{\alpha s}\ud B^{1}_{s}
\end{array} \right]
\\~\\
+
\frac{\sigma}{2}
\left[ \begin{array}{c}
\int\limits_{0}^{T_{2}} e^{\alpha (s-T_{2})} + e^{(\alpha +2m)(s-T_{2})} \ud B^{1}_{s} 
+ 
\int\limits_{0}^{T_{2}}e^{\alpha (s-T_{2})} - e^{(\alpha +2m)(s-T_{2})}\ud B^{2}_{s} 
\\ 
\int\limits_{0}^{T_{2}}e^{\alpha (s-T_{2})} - e^{(\alpha +2m)(s-T_{2})}\ud B^{1}_{s} 
+ 
\int\limits_{0}^{T_{2}}e^{\alpha (s-T_{2})} + e^{(\alpha +2m)(s-T_{2})}\ud B^{2}_{s} 
\end{array} \right]
\end{array}
$$

\item On the time interval $T_{3}$ all three lineages are interacting hence the system
has to be described by a trivariate process

$$
\ud \left[ \begin{array}{c}X^{1}_{t} \\ X^{2}_{t} \\ X^{3}_{t} \end{array} \right]
=
\left(
-\left[
\begin{array}{ccc} \alpha + m & -m/2 & -m/2  \\ -m/2 & \alpha +m & -m/2 \\ -m/2 & -m/2 & \alpha +m   \end{array}
\right]
\left[ \begin{array}{c}X^{1}_{t} \\ X^{2}_{t}\\ X^{3}_{t} \end{array} \right]
+
\left[ \begin{array}{c}\alpha \theta_{1} \\ \alpha \theta_{2} \\ \alpha \theta_{3} \end{array} \right]
\right) \ud t
+
\sigma \ud \left[ \begin{array}{c}B^{1}_{t} \\ B^{2}_{t}\\ B^{3}_{t} \end{array} \right].
$$
To solve it we recall the following eigen decomposition of the ``deterministic part'' matrix

$$
\begin{array}{rcl}
\left[
\begin{array}{ccc} \alpha + m & -m/2& -m/2 \\ -m/2 & \alpha +m & -m/2 \\ -m/2 & -m/2& \alpha +m \end{array}
\right]
& = &
\left[
\begin{array}{ccc} 1 & -1 & -1 \\ 1 & 0 & 1 \\ 1 & 1 & 0 \end{array}
\right]
\left[
\begin{array}{ccc} \alpha  & 0 & 0 \\ 0 & (2\alpha +3m)/2 & 0 \\ 0 & 0 & (2\alpha +3m)/2  \end{array}
\right]
\\~\\&&
\left[
\begin{array}{ccc} 1/3 & 1/3 & 1/3 \\ -1/3 & -1/3 & 2/3 \\ -1/3 & 2/3 & -1/3 \end{array}
\right]
\end{array}
$$
and the solution at the end of the interval is $(\beta:=\alpha + 3m/2)$

$$
\begin{array}{rcl}
\left[ \begin{array}{c}X^{1}_{T_{3}} \\ X^{2}_{T_{3}} \\ X^{3}_{T_{3}}  \end{array} \right]
& = &
\left[
\begin{array}{ccc} 1 & -1 & -1 \\ 1 & 0 & 1 \\ 1 & 1 & 0 \end{array}
\right]
\left(
\left[
\begin{array}{ccc} e^{-\alpha T_{3}}  & 0 & 0\\ 0 & e^{-\beta T_{3}} & 0 \\ 0 &0 & e^{-\beta T_{3}}\end{array}
\right]
\left[
\begin{array}{ccc} \frac{1}{3} & \frac{1}{3} & \frac{1}{3} \\ -\frac{1}{3} & -\frac{1}{3} & \frac{2}{3} \\ -\frac{1}{3} & \frac{2}{3} & -\frac{1}{3} \end{array}
\right]
\left[ \begin{array}{c}X^{1}_{T_{2}} \\ X^{2}_{T_{2}} \\ X^{2}_{T_{2}} \end{array} \right]
\right. \\~\\ && \left.
+ 
\left[
\begin{array}{ccc} 1-e^{-\alpha T_{3}}  & 0& 0 \\ 0 & \frac{\alpha}{\beta}(1-e^{-\beta T_{3}}) & 0 \\ 0 &0 & \frac{\alpha}{\beta}(1-e^{-\beta T_{3}}) \end{array}
\right]
\left[
\begin{array}{ccc} \frac{1}{3} & \frac{1}{3} & \frac{1}{3} \\ -\frac{1}{3} & -\frac{1}{3} & \frac{2}{3} \\ -\frac{1}{3} & \frac{2}{3} & -\frac{1}{3} \end{array}
\right]
\left[ \begin{array}{c}\theta_{1} \\ \theta_{2} \\ \theta_{3} \end{array} \right]
\right. \\~\\ && \left.
+
\sigma
\int\limits_{0}^{T_{3}}
\left[
\begin{array}{ccc} e^{-\alpha(T_{3}- s)}  & 0  & 0\\ 0 & e^{-\beta(T_{3} -s)} & 0 \\ 0& 0 & e^{-\beta(T_{3} -s)} \end{array}
\right]
\left[
\begin{array}{ccc} \frac{1}{3} & \frac{1}{3} & \frac{1}{3} \\ -\frac{1}{3} & -\frac{1}{3} & \frac{2}{3} \\ -\frac{1}{3} & \frac{2}{3} & -\frac{1}{3} \end{array}
\right]
\ud\left[ \begin{array}{c}B^{1}_{s} \\ B^{2}_{s} \\ B^{3}_{s} \end{array} \right]
\right)
\end{array}
$$

$$
\begin{array}{l}
=
\left[
\begin{array}{ccc} 1 & -1 & -1 \\ 1 & 0 & 1 \\ 1 & 1 & 0 \end{array}
\right]
\left(
\frac{1}{3}
\left[ \begin{array}{c}
e^{-\alpha T_{3}}\left( 
X^{1}_{T_{2}} + 2X^{2}_{T_{2}} \right) \\
e^{-\beta T_{3}}\left(
 X^{2}_{T_{2}} - X^{1}_{T_{2}}  \right) \\
e^{-\beta T_{3}} \left( X^{2}_{T_{3}} - X^{1}_{T_{2}} \right)
\end{array} \right]
+ 
\frac{1}{3}
\left[ \begin{array}{c}
(1-e^{-\alpha T_{3}})\left(\theta_{1} +  \theta_{2} + \theta_{3}  \right)\\
\frac{\alpha}{\beta}(1-e^{-\beta T_{3}})\left(2\theta_{3} - \theta_{1} - \theta_{2}\right)   \\
\frac{\alpha}{\beta}(1-e^{-\beta T_{3}})\left(2\theta_{2}-\theta_{1}   - \theta_{3} \right)
\end{array} \right]
\right. \\~\\  \left.
+
\frac{\sigma}{3}
\left[ \begin{array}{c}
\int\limits_{0}^{T_{3}}
e^{-\alpha(T_{3}- s)}
\ud B^{1}_{s} + 
\int\limits_{0}^{T_{3}}
e^{-\alpha(T_{3}- s)}
\ud B^{2}_{s} + 
\int\limits_{0}^{T_{3}}
e^{-\alpha(T_{3}- s)}
\ud B^{3}_{s} \\
2\int\limits_{0}^{T_{3}}
e^{-\beta(T_{3} -s)}
\ud B^{3}_{s} - 
\int\limits_{0}^{T_{3}}
e^{-\beta(T_{3} -s)}
\ud B^{1}_{s} - 
\int\limits_{0}^{T_{3}}
e^{-\beta(T_{3} -s)}
\ud B^{2}_{s} \\  
2\int\limits_{0}^{T_{3}}
e^{-\beta(T_{3} -s)}
\ud B^{2}_{s} -
\int\limits_{0}^{T_{3}}
e^{-\beta(T_{3} -s)}
\ud B^{1}_{s} - 
\int\limits_{0}^{T_{3}}
e^{-\beta(T_{3} -s)}
\ud B^{3}_{s} 
\end{array} \right]
\right)
\end{array}
$$

$$
\begin{array}{l}
=
{\tiny
\frac{1}{3} }
\left(
{\tiny
\left[ \begin{array}{c}
(e^{-\alpha T_{3}}+2e^{-\beta T_{3}})X^{1}_{T_{2}} +
2(e^{-\alpha T_{3}}-e^{-\beta T_{3}})X^{2}_{T_{2}} 
\\
(e^{-\alpha T_{3}}-e^{-\beta T_{3}})X^{1}_{T_{2}} +
(2e^{-\alpha T_{3}}+e^{-\beta T_{3}})X^{2}_{T_{2}} 
\\
(e^{-\alpha T_{3}}-e^{-\beta T_{3}})X^{1}_{T_{2}} +
(2e^{-\alpha T_{3}}+e^{-\beta T_{3}})X^{2}_{T_{2}} 
\end{array} \right]}
\right. \\~\\  \left.
+ 
{\tiny
\left[ \begin{array}{c}
(1-e^{-\alpha T_{3}}+2\frac{\alpha}{\beta}(1-e^{-\beta T_{3}}))\theta_{1} +  
(1-e^{-\alpha T_{3}}-\frac{\alpha}{\beta}(1-e^{-\beta T_{3}}))\theta_{2} + 
(1-e^{-\alpha T_{3}}-\frac{\alpha}{\beta}(1-e^{-\beta T_{3}}))\theta_{3}
\\
(1-e^{-\alpha T_{3}}-\frac{\alpha}{\beta}(1-e^{-\beta T_{3}}))\theta_{1} +  
(1-e^{-\alpha T_{3}}+2\frac{\alpha}{\beta}(1-e^{-\beta T_{3}}))\theta_{2} + 
(1-e^{-\alpha T_{3}}-\frac{\alpha}{\beta}(1-e^{-\beta T_{3}}))\theta_{3}
\\
(1-e^{-\alpha T_{3}}-\frac{\alpha}{\beta}(1-e^{-\beta T_{3}}))\theta_{1} +  
(1-e^{-\alpha T_{3}}-\frac{\alpha}{\beta}(1-e^{-\beta T_{3}}))\theta_{2} + 
(1-e^{-\alpha T_{3}}+2\frac{\alpha}{\beta}(1-e^{-\beta T_{3}}))\theta_{3}
\end{array} \right]
}
\right. \\~\\  \left.
+
{\tiny
\sigma
\left[ \begin{array}{c}
\int\limits_{0}^{T_{3}} e^{-\alpha(T_{3}- s)}+2e^{-\beta(T_{3} -s)}  \ud B^{1}_{s} + 
\int\limits_{0}^{T_{3}} 
e^{-\alpha(T_{3}- s)}-e^{-\beta(T_{3} -s)} \ud B^{2}_{s} + 
\int\limits_{0}^{T_{3}} 
e^{-\alpha(T_{3}- s)}-e^{-\beta(T_{3} -s)} \ud B^{3}_{s} \\
\int\limits_{0}^{T_{3}} e^{-\alpha(T_{3}- s)}-e^{-\beta(T_{3} -s)}  \ud B^{1}_{s} + 
\int\limits_{0}^{T_{3}} 
e^{-\alpha(T_{3}- s)}+2e^{-\beta(T_{3} -s)} \ud B^{2}_{s} + 
\int\limits_{0}^{T_{3}} 
e^{-\alpha(T_{3}- s)}-e^{-\beta(T_{3} -s)} \ud B^{3}_{s} \\
\int\limits_{0}^{T_{3}} e^{-\alpha(T_{3}- s)}-e^{-\beta(T_{3} -s)}  \ud B^{1}_{s} + 
\int\limits_{0}^{T_{3}} 
e^{-\alpha(T_{3}- s)}-e^{-\beta(T_{3} -s)} \ud B^{2}_{s} + 
\int\limits_{0}^{T_{3}} 
e^{-\alpha(T_{3}- s)}+2e^{-\beta(T_{3} -s)} \ud B^{3}_{s} 
\end{array} \right]
}
\right)
\end{array}
$$
\end{itemize}
We can see that in the same fashion as above we can plug in the ancestral values to obtain 
the solution at the tips. We have not done this here as the formula would be too long to be readable.
After this obtaining the mean, variance and covariance will be immediate as the Brownian motions 
are independent on disjoint intervals. 
In particular we may write the expectation as

$$
\begin{array}{l}
{\tiny
\E{\left[ \begin{array}{c} X^{1}_{T_{3}} \\ X^{2}_{T_{3}} \\ X^{3}_{T_{3}}  \end{array} \right]}
}
 = 
{\tiny
\frac{1}{3}}
\left(
\tiny{
\left[ \begin{array}{c}
(e^{-\alpha T_{3}}+2e^{-\beta T_{3}})\E{X^{1}_{T_{2}}} +
2(e^{-\alpha T_{3}}-e^{-\beta T_{3}})\E{X^{2}_{T_{2}}} 
\\~\\
(e^{-\alpha T_{3}}-e^{-\beta T_{3}})\E{X^{1}_{T_{2}}} +
(2e^{-\alpha T_{3}}+e^{-\beta T_{3}})\E{X^{2}_{T_{2}}} 
\\~\\
(e^{-\alpha T_{3}}-e^{-\beta T_{3}})\E{X^{1}_{T_{2}}} +
(2e^{-\alpha T_{3}}+e^{-\beta T_{3}})\E{X^{2}_{T_{2}}} 
\end{array} \right]}
\right. \\~\\  \left.
+ 
{\tiny
\left[ \begin{array}{c}
(1-e^{-\alpha T_{3}}+2\frac{\alpha}{\beta}(1-e^{-\beta T_{3}}))\theta_{1} +  
(1-e^{-\alpha T_{3}}-\frac{\alpha}{\beta}(1-e^{-\beta T_{3}}))\theta_{2} + 
(1-e^{-\alpha T_{3}}-\frac{\alpha}{\beta}(1-e^{-\beta T_{3}}))\theta_{3}
\\
(1-e^{-\alpha T_{3}}-\frac{\alpha}{\beta}(1-e^{-\beta T_{3}}))\theta_{1} +  
(1-e^{-\alpha T_{3}}+2\frac{\alpha}{\beta}(1-e^{-\beta T_{3}}))\theta_{2} + 
(1-e^{-\alpha T_{3}}-\frac{\alpha}{\beta}(1-e^{-\beta T_{3}}))\theta_{3}
\\
(1-e^{-\alpha T_{3}}-\frac{\alpha}{\beta}(1-e^{-\beta T_{3}}))\theta_{1} +  
(1-e^{-\alpha T_{3}}-\frac{\alpha}{\beta}(1-e^{-\beta T_{3}}))\theta_{2} + 
(1-e^{-\alpha T_{3}}+2\frac{\alpha}{\beta}(1-e^{-\beta T_{3}}))\theta_{3}
\end{array} \right] }\right)
\end{array}
$$
where

$$
\begin{array}{rcl}
{\tiny
\E{\left[ \begin{array}{c} X^{1}_{T_{2}} \\ X^{2}_{T_{2}}  \end{array} \right]}}
& = &
{\tiny
\left[ \begin{array}{c} e^{-(\alpha T_{1}+\alpha T_{2})}X(0) \\  e^{-(\alpha T_{1}+\alpha T_{2})}X(0) \end{array} \right]
+ 
\left[ \begin{array}{c}
(e^{-\alpha T_{2}}-e^{-\alpha (T_{1}+T_{2})})\theta_{1}\\ (e^{-\alpha T_{2}}-e^{-\alpha (T_{1}+T_{2})})\theta_{1}
\end{array} \right]}
\\~\\ &&
+
{\tiny
\frac{1}{2}
\left[ \begin{array}{c}
(1-e^{-\alpha T_{2}} + \frac{\alpha}{\alpha +2m}(1-e^{-(\alpha +2m)T_{2}}))\theta_{1} 
+
(1-e^{-\alpha T_{2}} - \frac{\alpha}{\alpha +2m}(1-e^{-(\alpha +2m)T_{2}}))\theta_{2}
\\~\\ 
(1-e^{-\alpha T_{2}} - \frac{\alpha}{\alpha +2m}(1-e^{-(\alpha +2m)T_{2}}))\theta_{1} 
+ 
(1-e^{-\alpha T_{2}} + \frac{\alpha}{\alpha +2m}(1-e^{-(\alpha +2m)T_{2}}))\theta_{2}
\end{array} \right].}
\end{array}
$$
If we take $T_{2} \to \infty$, i.e. we assume that we have a pair of species exchanging gene flow forever without speciating 
we obtain the stationary expectation

$$
\begin{array}{rcl}
\E{\left[ \begin{array}{c} X^{1}_{\infty} \\ X^{2}_{\infty}  \end{array} \right]} 
& = &
\frac{1}{2}
\left[ \begin{array}{c}
(1 + \frac{\alpha}{\alpha +2m})\theta_{1} 
+
(1 - \frac{\alpha}{\alpha +2m})\theta_{2}
\\~\\ 
(1 - \frac{\alpha}{\alpha +2m})\theta_{1} 
+ 
(1 + \frac{\alpha}{\alpha +2m}1)\theta_{2}
\end{array} \right].
\end{array}
$$
If $m=0$ then we recover $\E{[ X^{1}_{\infty} ~~ X^{2}_{\infty} ]^{T}} = [\theta_{1}~~\theta_{2}]^{T}$
as expected. 
The variance 
$\var{\left[ \begin{array}{ccc} X^{1}_{T_{3}} & X^{2}_{T_{3}} &  X^{3}_{T_{3}} \end{array} \right]^{T}}$
can be written as,

$$
\begin{array}{l}
{\tiny\frac{1}{9}}
\left(
{\tiny
\left[ \begin{array}{cc}
(e^{-\alpha T_{3}}+2e^{-\beta T_{3}}) &
2(e^{-\alpha T_{3}}-e^{-\beta T_{3}}) 
\\
(e^{-\alpha T_{3}}-e^{-\beta T_{3}}) &
(2e^{-\alpha T_{3}}+e^{-\beta T_{3}}) 
\\
(e^{-\alpha T_{3}}-e^{-\beta T_{3}})&
(2e^{-\alpha T_{3}}+e^{-\beta T_{3}}) 
\end{array} \right]}
{\tiny
\var{\left[ \begin{array}{c} X^{1}_{T_{2}} \\ X^{2}_{T_{2}}  \end{array} \right]}}
{\tiny
\left[ \begin{array}{cc}
(e^{-\alpha T_{3}}+2e^{-\beta T_{3}}) &
2(e^{-\alpha T_{3}}-e^{-\beta T_{3}}) 
\\
(e^{-\alpha T_{3}}-e^{-\beta T_{3}}) &
(2e^{-\alpha T_{3}}+e^{-\beta T_{3}}) 
\\
(e^{-\alpha T_{3}}-e^{-\beta T_{3}})&
(2e^{-\alpha T_{3}}+e^{-\beta T_{3}}) 
\end{array} \right]^{T}}
\right. \\ \left.
+
{\tiny
\sigma^{2}\var{
\left[ \begin{array}{c}
\int\limits_{0}^{T_{3}} e^{-\alpha(T_{3}- s)}+2e^{-\beta(T_{3} -s)}  \ud B^{1}_{s} + 
\int\limits_{0}^{T_{3}} 
e^{-\alpha(T_{3}- s)}-e^{-\beta(T_{3} -s)} \ud B^{2}_{s} + 
\int\limits_{0}^{T_{3}} 
e^{-\alpha(T_{3}- s)}-e^{-\beta(T_{3} -s)} \ud B^{3}_{s} \\
\int\limits_{0}^{T_{3}} e^{-\alpha(T_{3}- s)}-e^{-\beta(T_{3} -s)}  \ud B^{1}_{s} + 
\int\limits_{0}^{T_{3}} 
e^{-\alpha(T_{3}- s)}+2e^{-\beta(T_{3} -s)} \ud B^{2}_{s} + 
\int\limits_{0}^{T_{3}} 
e^{-\alpha(T_{3}- s)}-e^{-\beta(T_{3} -s)} \ud B^{3}_{s} \\
\int\limits_{0}^{T_{3}} e^{-\alpha(T_{3}- s)}-e^{-\beta(T_{3} -s)}  \ud B^{1}_{s} + 
\int\limits_{0}^{T_{3}} 
e^{-\alpha(T_{3}- s)}-e^{-\beta(T_{3} -s)} \ud B^{2}_{s} + 
\int\limits_{0}^{T_{3}} 
e^{-\alpha(T_{3}- s)}+2e^{-\beta(T_{3} -s)} \ud B^{3}_{s} 
\end{array} \right]}}
\right)
\end{array}
$$
where

$$
\begin{array}{rcl}
\var{\left[ \begin{array}{c} X^{1}_{T_{2}} \\ X^{2}_{T_{2}}  \end{array} \right]}
&= &
\sigma^{2}
\var{
\left[ \begin{array}{c}
e^{-\alpha (T_{1}+T_{2})}\int\limits_{0}^{T_{1}} e^{\alpha s}\ud B^{1}_{s}\\
e^{-\alpha (T_{1}+T_{2})}\int\limits_{0}^{T_{1}} e^{\alpha s}\ud B^{1}_{s}
\end{array} \right]}
\\&&+
\frac{\sigma^{2}}{4}
\var{
\left[ \begin{array}{c}
\int\limits_{0}^{T_{2}} e^{\alpha (s-T_{2})} + e^{(\alpha +2m)(s-T_{2})} \ud B^{1}_{s} 
+ 
\int\limits_{0}^{T_{2}}e^{\alpha (s-T_{2})} - e^{(\alpha +2m)(s-T_{2})}\ud B^{2}_{s} 
\\ 
\int\limits_{0}^{T_{2}}e^{\alpha (s-T_{2})} - e^{(\alpha +2m)(s-T_{2})}\ud B^{1}_{s} 
+ 
\int\limits_{0}^{T_{2}}e^{\alpha (s-T_{2})} + e^{(\alpha +2m)(s-T_{2})}\ud B^{2}_{s} 
\end{array} \right]}.
\end{array}
$$
The above can be calculated explicitly.
First 

$$
\sigma^{2}\var{e^{-\alpha (T_{1}+T_{2})}\int\limits_{0}^{T_{1}} e^{\alpha s}\ud B^{1}_{s}}=
\frac{\sigma^{2}}{2\alpha}\left(e^{-2\alpha T_{2}}-e^{-2\alpha (T_{1}+T_{2})}\right) 
\xrightarrow{T_{2} \to \infty} 0.
$$
Now we use the It\=o isometry for the second part of the variance.

$$
\begin{array}{cl}
A_{T_{2}}=&\var{\int\limits_{0}^{T_{2}} e^{\alpha (s-T_{2})} + e^{(\alpha +2m)(s-T_{2})} \ud B^{1}_{s} } 
= e^{-2\alpha T_{2}}\var{\int\limits_{0}^{T_{2}} e^{\alpha s} + e^{-2mT_{2}}e^{(\alpha +2m)s} \ud B^{1}_{s} } 
\\ = &e^{-2\alpha T_{2}}\int\limits_{0}^{T_{2}}\left(e^{\alpha s} + e^{-2mT_{2}}e^{(\alpha +2m)s} \right)^{2} \ud s
= e^{-2\alpha T_{2}}\int\limits_{0}^{T_{2}}e^{2\alpha s} + 2e^{-2mT_{2}}e^{2(\alpha +m)s} 
\\ & + e^{-4mT_{2}}e^{2(\alpha +2m)s} \ud s
\\
=& e^{-2\alpha T_{2}}
\left(
\frac{1}{2\alpha}(e^{2\alpha T_{2}} -1) + \frac{e^{-2mT_{2}}}{\alpha +m}(e^{2(\alpha +m)T_{2}}-1) + 
\frac{e^{-4mT_{2}}}{2(\alpha +2m)}(e^{2(\alpha +2m)T_{2}}-1) 
\right)
\\ =&
\frac{1}{2\alpha}(1-e^{-2\alpha T_{2}}) + \frac{1}{\alpha +m}(1-e^{-2(\alpha +m)T_{2}}) + 
\frac{1}{2(\alpha +2m)}(1-e^{-2(\alpha +2m)T_{2}}) 
\\ \xrightarrow{T_{2} \to \infty} &
\frac{1}{2\alpha} + \frac{1}{\alpha +m} + \frac{1}{2(\alpha +2m)} 
\\
B_{T_{2}}=&\var{\int\limits_{0}^{T_{2}} e^{\alpha (s-T_{2})} - e^{(\alpha +2m)(s-T_{2})} \ud B^{1}_{s} } 
\\=&
\frac{1}{2\alpha}(1-e^{-2\alpha T_{2}}) - \frac{1}{\alpha +m}(1-e^{-2(\alpha +m)T_{2}}) + 
\frac{1}{2(\alpha +2m)}(1-e^{-2(\alpha +2m)T_{2}}) 
\\  \xrightarrow{T_{2} \to \infty} &
\frac{1}{2\alpha} - \frac{1}{\alpha +m} + \frac{1}{2(\alpha +2m)} 
\\
C_{T_{2}}=&\cov{\int\limits_{0}^{T_{2}} e^{\alpha (s-T_{2})} + e^{(\alpha +2m)(s-T_{2})} \ud B^{1}_{s} }{\int\limits_{0}^{T_{2}} e^{\alpha (s-T_{2})} - e^{(\alpha +2m)(s-T_{2})} \ud B^{1}_{s} }
\\=&
e^{-2\alpha T_{2}}
\int\limits_{0}^{T_{2}} e^{2\alpha s}  -  e^{-4mT_{2}}e^{2(\alpha +2m)s} \ud s 
=
e^{-2\alpha T_{2}}\left(
\frac{1}{2\alpha}(e^{2\alpha T_{2}}-1)  -  \frac{e^{-4mT_{2}}}{2(\alpha +2m)}(e^{2(\alpha +2m)T_{2}} -1)
\right)
\\=&
\frac{1}{2\alpha}(1-e^{-2\alpha T_{2}}) - \frac{1}{2(\alpha+2m)}(1-e^{-2(\alpha+2m) T_{2}})
  \xrightarrow{T_{2} \to \infty} 
\frac{1}{2\alpha} - \frac{1}{2(\alpha+2m)}
\end{array}
$$
Then

$$
\begin{array}{cl}
\var{\left[ \begin{array}{c} X^{1}_{T_{2}} \\ X^{2}_{T_{2}}  \end{array} \right]}
= &
\sigma^{2}\var{e^{-\alpha (T_{1}+T_{2})}\int\limits_{0}^{T_{1}} e^{\alpha s}\ud B^{1}_{s}}
\\ ~~~~~~~~~~~~~~~~~~~ = &
\frac{\sigma^{2}}{2\alpha}\left(e^{-2\alpha T_{2}}-e^{-2\alpha (T_{1}+T_{2})}\right) 
+
\frac{\sigma^{2}}{4}
\left[
\begin{array}{cc}
A_{T_{2}} + B_{T_{2}} & 2C_{T_{2}} \\
2C_{T_{2}} & A_{T_{2}} + B_{T_{2}}  
\end{array}
\right]
\\
 \xrightarrow{T_{2} \to \infty} &
\frac{\sigma^{2}}{4}
\left[
\begin{array}{cc}
\frac{1}{\alpha} + \frac{1}{(\alpha +2m)} 
&
\frac{1}{\alpha}-\frac{1}{\alpha+2m} 
\\
\frac{1}{\alpha}-\frac{1}{\alpha+2m} 
&
\frac{1}{\alpha} + \frac{1}{(\alpha +2m)} 
\end{array}
\right]
\end{array}
$$
We can see with no migration we recover that both species are independent with
stationary variance $\sigma^{2}/(2\alpha)$ as expected. The covariance between the
two species depends on the values of $\alpha$ and $m$ and as expected is 
always positive. It increases with the increase of the migration rate.

In the three dimensional case as $T_{3} \to \infty$ we can also calculate
the limits. 
We can immediately see that the mean value will be in the limit

$$
\begin{array}{l}
\E{\left[ \begin{array}{c} X^{1}_{T_{3}} \\ X^{2}_{T_{3}} \\ X^{3}_{T_{3}}  \end{array} \right]}
  \xrightarrow{T_{3} \to \infty} 
\left[ \begin{array}{c}

(1+2\frac{\alpha}{\beta})\theta_{1} +  
(1-\frac{\alpha}{\beta})\theta_{2} + 
(1-\frac{\alpha}{\beta})\theta_{3}
\\
(1-\frac{\alpha}{\beta})\theta_{1} +  
(1+2\frac{\alpha}{\beta})\theta_{2} + 
(1-\frac{\alpha}{\beta})\theta_{3}
\\
(1-\frac{\alpha}{\beta})\theta_{1} +  
(1-\frac{\alpha}{\beta})\theta_{2} + 
(1+2\frac{\alpha}{\beta})\theta_{3}
\end{array} \right].
\end{array}
$$
Using the same techniques as in the two dimensional case we calculate that the non--vanishing part
of the covariance matrix will equal

$$
{\tiny
\frac{\sigma^{2}}{9}\left[
\begin{array}{ccc}
\frac{3}{2\alpha}(1-e^{-2\alpha T_{3}})+\frac{6}{2\beta}(1-e^{-2\beta T_{3}})  &
\frac{3}{2\alpha}(1-e^{-2\alpha T_{3}})-\frac{3}{2\beta}(1-e^{-2\beta T_{3}})  &
\frac{3}{2\alpha}(1-e^{-2\alpha T_{3}})-\frac{3}{2\beta}(1-e^{-2\beta T_{3}})  
\\
\frac{3}{2\alpha}(1-e^{-2\alpha T_{3}})-\frac{3}{2\beta}(1-e^{-2\beta T_{3}})  &
\frac{3}{2\alpha}(1-e^{-2\alpha T_{3}})+\frac{6}{2\beta}(1-e^{-2\beta T_{3}})  &
\frac{3}{2\alpha}(1-e^{-2\alpha T_{3}})-\frac{3}{2\beta}(1-e^{-2\beta T_{3}})  
\\
\frac{3}{2\alpha}(1-e^{-2\alpha T_{3}})-\frac{3}{2\beta}(1-e^{-2\beta T_{3}})  &
\frac{3}{2\alpha}(1-e^{-2\alpha T_{3}})-\frac{3}{2\beta}(1-e^{-2\beta T_{3}})  &
\frac{3}{2\alpha}(1-e^{-2\alpha T_{3}})+\frac{6}{2\beta}(1-e^{-2\beta T_{3}}) 
\end{array}
\right]}
$$
becoming in the limit

$$
\sigma^{2}\left[
\begin{array}{ccc}
\frac{1}{3\cdot 2\alpha}+\frac{2}{3\cdot 2\beta}  &
\frac{1}{3\cdot2\alpha}-\frac{1}{3\cdot2\beta}  &
\frac{1}{3\cdot2\alpha}-\frac{1}{3\cdot2\beta}
\\
\frac{1}{3\cdot2\alpha}-\frac{1}{3\cdot2\beta}  &
\frac{1}{3\cdot2\alpha}+\frac{2}{3\cdot2\beta}  &
\frac{1}{3\cdot2\alpha}-\frac{1}{3\cdot2\beta}
\\
\frac{1}{3\cdot2\alpha}-\frac{1}{3\cdot2\beta}  &
\frac{1}{3\cdot2\alpha}-\frac{1}{3\cdot2\beta}  &
\frac{1}{3\cdot2\alpha}+\frac{2}{3\cdot2\beta}
\end{array}
\right].
$$
As $\beta = \alpha + 3m/2$ then when $m=0$ one obtains the limiting variance
as $\sigma^{2}/(2\alpha)$ and the limiting covariance $0$. Also for $m>0$ the covariance
is positive as expected.

\bibliographystyle{plainnat}
\bibliography{MigrationOU.bib}

\newpage
\includepdf[pages={-}]{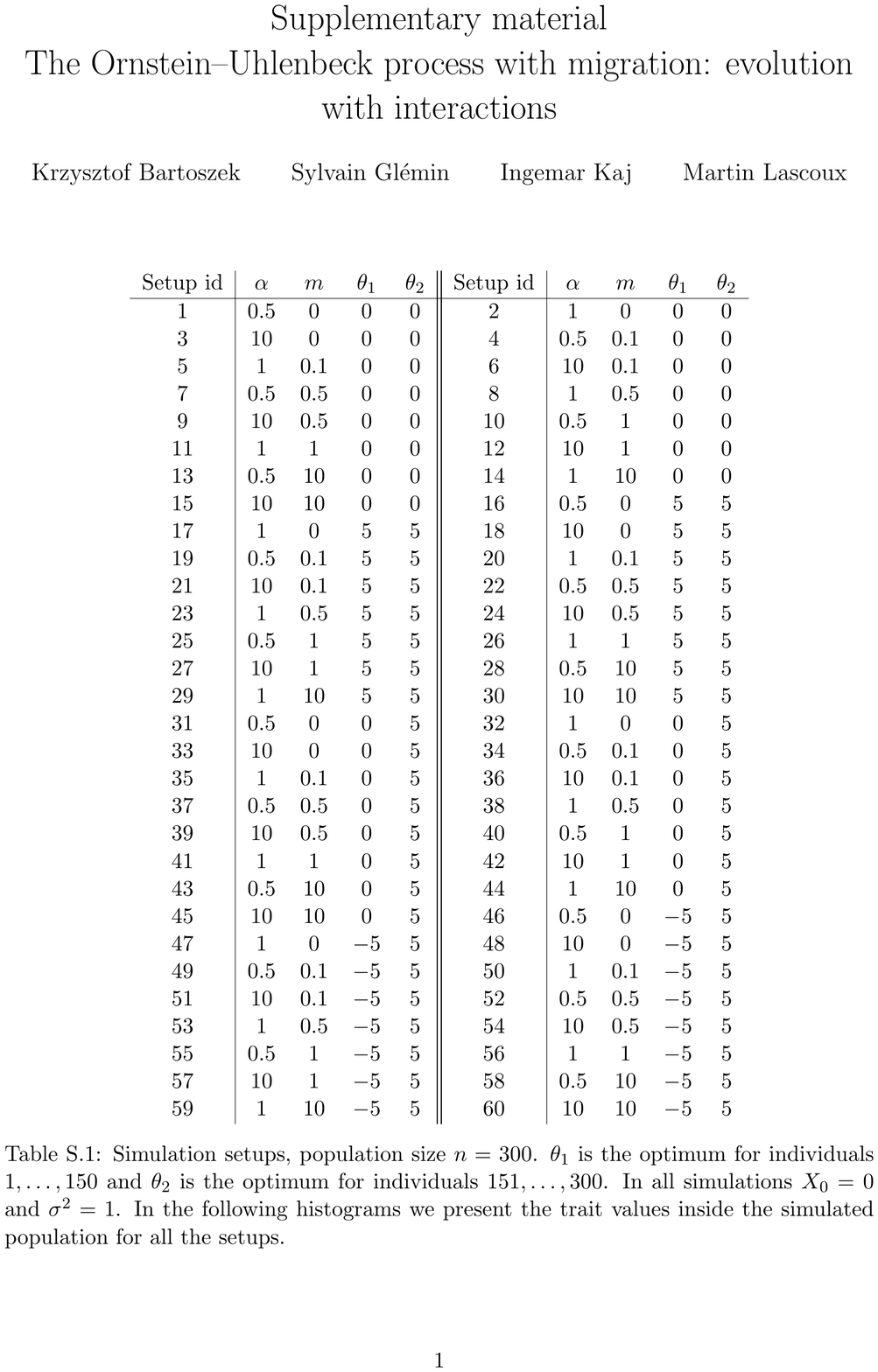}

\end{document}